\documentstyle[preprint,floats,prd,aps]{revtex}

\def\PRref#1&#2&#3(#4){\unskip\ #1~\bf #2\rm, #3 (#4)}

\def\NIMA{Nucl. Inst. and Meth. A}

\def\PLB{Phys. Lett. B}
\def\PRL{Phys. Rev. Lett.}
\def\PRD{Phys. Rev. D}
\def\ZPC{Z. Phys. C}
\def\etal{{\it et al.}}
\input psfig

\begin{document}

\tighten 
\preprint{\vbox{\hbox{CLNS 93/1261  \hfill}
                \hbox{UH-511-778-93 \hfill}
                \hbox{OHSTPY-HEP-E-93-018 \hfill}
                \hbox{HEPSY 93-10 \hfill}
                \hbox{\today        \hfill}}}

\title{A Review of Hadronic and Rare B Decays \cite{clns}}

\author{ Thomas E. Browder}
\address{University of Hawaii at Manoa, Honolulu, Hawaii 96822}
\author{Klaus Honscheid}
\address{Ohio State University, Columbus, Ohio 43210}
\author{Stephen Playfer}
\address{Syracuse University, Syracuse, New York 13244}

\bigskip
\bigskip
\bigskip
\author{(To appear in $B$ Decays, 2nd edition, Ed. by S. Stone, World
Scientific)}

\maketitle

\vfill

\begin{abstract}
We review recent experimental results on $B$ meson decays. These include
measurements of the inclusive production of charmed and non-charmed mesons and
baryons, the reconstruction of a large number of exclusive hadronic final
states with charmed mesons, the search for exclusive hadronic
final states without charmed mesons, and the first observation of the
decay $B\to K^*\gamma$ which is described by an electromagnetic penguin
diagram. The theoretical implications of these results will be considered.
\end{abstract}

\vspace{0.5cm}

\newpage

\tableofcontents

\newpage

\section{INTRODUCTION}

In the past two years there have been major advances in our knowledge
of the decays of $B$ mesons. This is primarily a result of the large
data sample of about $2 ~\rm{fb}^{-1}$ that has been collected on
the $\Upsilon (4S)$ by the CLEO~II collaboration at
the Cornell Electron Storage Ring (CESR). The CLEO~II detector has
excellent capabilities for measuring both charged tracks and neutral showers,
and has reconstructed a large number of exclusive hadronic $B$ decays.
We will discuss the results that have been obtained
on Cabibbo favored hadronic $B$ decays\cite{SixthB},
as well as the new results on rare $B$ decays\cite{PRLkpi,PRLbsg}.
Older results from the ARGUS experiment, which operated at the DORIS storage
ring
and from the CLEO~1.5 experiment, which preceded the CLEO~II detector are
also discussed.
We note that the LEP experiments and the CDF experiment at the Tevatron
Collider have recently observed exclusive hadronic decays of both $B$ and $B_s$
mesons.
This demonstration of the feasibility of reconstructing $B$ decays in $e^+e^-$
collisions at the $Z^0$, and in high energy $p\bar{p}$ collisions, is
an indication of the possibilities for future studies of $B$ mesons.

\begin{figure}[htb]
\begin{center}
\unitlength 1.0in
\begin{picture}(3.,1.)(0,0)
\end{picture}
\bigskip
\vskip 10 mm
\caption{$B$ meson decay diagrams: (a)
external spectator and (b) color suppressed spectator.}
\label{Fdiag}
\end{center}
\end{figure}

Since the top quark mass is large, $B$ mesons are expected to be the only
weakly decaying mesons containing quarks of the third generation.
This means that their decays are a unique window on the
Cabibbo-Kobayashi-Maskawa (CKM) matrix elements $V_{cb}$, $V_{ub}$, $V_{ts}$
and $V_{td}$, describing the couplings of the third generation of quarks
to the lighter quarks. Hadronic $B$ meson decays occur primarily
through the Cabibbo favored $b\to c$ transition. In such decays
the dominant weak decay diagram is the spectator diagram, shown in
Fig.~\ref{Fdiag}(a), where the virtual $W^-$ materializes into either a
$\bar u d$ or a $\bar c s$ pair. This pair becomes one of the final state
hadrons while the $c$ quark pairs with the spectator anti-quark to form the
other hadron.

The spectator diagram is modified by hard gluon exchanges
between the initial and final quark lines.
This leads to the ``color suppressed'' diagram shown in Fig.~\ref{Fdiag}(b),
which has a different set of quark pairings.
Observation of $B \to \psi X_s$ decays, where $X_s$ is a strange meson, gives
experimental evidence for the presence of this diagram.
Further information on the size of the color suppressed contribution can be
obtained from $\bar{B^0} \to D^0$ (or $D^{*0} ) X^0$ transitions,
where $X^0$ is a neutral meson. In $B^-$ decays, both types of
diagrams are present and can interfere.
By comparing the rates for $B^-$ and $\bar{B^0}$ decays, the size and the sign
of the color suppressed amplitude can be determined.

It has been suggested by Bjorken\cite{Bjorken} that, in
analogy to semileptonic decays, two body decays of $B$ mesons can be expressed
as the product of two independent hadronic currents, one
describing the formation of a charm meson and the other the hadronization of
the $\bar{u} d$ (or $\bar c s$) system from the virtual $W^-$. Qualitatively,
he argues that for a $B$ decay with a large energy release the
$\bar{u} d$ pair, which is produced as a color singlet, travels fast enough to
leave the interaction region without influencing the
second hadron formed from the $c$ quark and the spectator anti-quark.
The  assumption that the amplitude can be expressed
as the product of two hadronic currents is called ``factorization'' in this
paper. It is expected that the simple approximation of the strong interaction
effects by the factorization hypothesis will be more reliable in $B$ meson
decays than in the equivalent $D$ meson decays due to the larger
characteristic energy transfers and the consequent suppression
of final state interactions.
We will discuss several tests of the factorization hypothesis based on
the comparison of semileptonic and hadronic $B$ meson decays.

All $B$ meson decays that do not occur through the usual $b\to c$ transition
are known as rare $B$ decays. The simplest diagram for a rare $B$ decay is
obtained by replacing the $b\to c$ transition by
a Cabibbo suppressed $b\to u$ transition. These decays probe the small
CKM matrix element $V_{ub}$, the magnitude of which sets bounds on
the combination $\rho^2 + \eta^2$ in the Wolfenstein
parameterization of the CKM matrix. So far the only
measurement of the magnitude of $V_{ub}$ has been obtained from semileptonic
$B$ decays \cite{btoulnu}. We discuss the status of the search for rare
hadronic $B$ decays, and in particular the possibility of measuring the decay
$B^0\to\pi^+\pi^-$ which is important for the study of CP violation in
$B$ decays.

\begin{figure}[htb]
\begin{center}
\unitlength 1.0in
\begin{picture}(3.,1.0)(0,0)
\end{picture}
\bigskip
\vskip 10 mm
\caption{Diagram for the
 electromagnetic penguin in $B$ meson decay.}
\label{empeng}
\end{center}
\end{figure}

Since the contribution of the spectator diagram to rare $B$ decays is
suppressed it is expected that additional diagrams will make a large
contribution to some decay modes. The most significant of these diagrams
is the one-loop flavor-changing neutral current diagram known as the
``penguin'' diagram (Fig. \ref{empeng}). The Cabibbo allowed part of
this diagram, corresponding to a $b\to s$ transition, is expected to dominate
the amplitude of rare decays to final states with one or three $s$-quarks.
There is also a Cabibbo suppressed $b\to d$ amplitude which may not be
negligible in decays to final states with no $c$ or $s$ quarks.
It should be noted that the loop diagram is much more significant in $B$ decays
than in $D$ decays because the $b\to s$ loop contains the heavy top quark
with large couplings $V_{tb}$ and $V_{ts}$, whereas contributions to
the equivalent $c\to u$ loop are suppressed either by the small couplings
$V_{cb}$ and $V_{ub}$, or by the small $s$ and $d$ quark masses.

The observation of the decay $B\to K^*(892) \gamma$,
recently reported by the CLEO~II experiment, is the first direct evidence for
the penguin diagram.

This decay is described by the electromagnetic transition
$b\to s\gamma$, which is a $b\to s$ penguin loop
accompanied by the radiation of a photon from either the loop, or the initial
or final state quarks.
This important new result will be discussed in some detail.
We will also comment on the recent discussion about the sensitivity of
the $b\to s\gamma$ process to non-standard model contributions within the
loop\cite{Hewett}.
In many extensions of the standard model an additional
contribution to $b\to s\gamma$ is expected to come from a charged Higgs.
We will discuss the extent to which the data from the CLEO~II experiment
allow bounds to be set on such non-standard model contributions.

\section{THE EXPERIMENTAL STUDY OF B DECAY}

\subsection{$\Upsilon (4S)$ Experiments}
\label{y4sexp}

The first fully reconstructed $B$ mesons were reported in 1983 by the
CLEO~I collaboration \cite{FirstB}. Since then the CLEO~1.5 experiment
\cite{anotherB,SecondB}
has collected $212 ~\rm{pb}^{-1}$ \cite{only4s}, the ARGUS experiment
\cite{ThirdB,FourthB,FifthB}
has collected $246 ~\rm{pb}^{-1}$,
and the CLEO~II experiment
has collected about $2 ~\rm{fb}^{-1}$, of which between
$0.9$ and $1.4 ~\rm{fb}^{-1}$ have been used
to obtain the results
described in this review
\cite{SixthB,PRLkpi,PRLbsg}.
All these experiments take data on the $\Upsilon ( 4 S )$ resonance
at $e^+ e^-$ colliders. The techniques used by experiments which operate
at the $\Upsilon (4S)$ resonance are discussed in detail in the
review by M. Artuso in this volume. It is assumed here that
the $\Upsilon (4 S)$ resonance always decays to pairs of $B$ mesons, and
that $f_+$, the fraction of $B^+ B^-$ pairs produced in
$\Upsilon (4 S)$ decay, is equal to $f_0$,
the fraction of $B^0 \bar{B^0}$ pairs produced in these decays.
Older results which assumed other values of $f_+$ and $f_0$ have been
rescaled.

The $\Upsilon (4 S)$ resonance sits on a continuum background
consisting of $e^+ e^- \to q\bar{q}$, where $q$ can be any of $u,d,s,c$.
This continuum background is studied by taking a significant amount of data at
an energy just below the $\Upsilon (4 S)$ resonance, e.g. CLEO~II records a
third of its data at an energy 55~MeV below the resonance.
Using this data sample, and Monte Carlo simulations of $q\bar{q}$ jets,
cuts have been devised to suppress the continuum background.
In $\Upsilon (4S)$ production of $B\bar{B}$ pairs, the $B$ mesons are produced
almost at rest, and their decay axes are uncorrelated. These events are rather
spherical in shape, and can be distinguished from jetlike continuum events
using a variety of event shape variables. For the study of inclusive production
in $B$ decays a particularly useful variable is the normalized second
Fox-Wolfram moment\cite{fw}, $R_2$,  which is 0 for a perfectly spherical
event, and 1 for an event completely collimated around the jet axis.
For the study of exclusive $B$ decay modes it is more useful to compare the
axis of the reconstructed $B$ candidate with the axis of the
rest of the event.
Examples of variables used are the direction of the sphericity axis or the
thrust axis of the rest of the event with respect to the $B$ candidate,
$\theta_S$ or $\theta_T$, and the sum of the momenta transverse
to the axis of the $B$ candidate, known as $s_{\perp}$ \cite{Artuso}.
There is also some information in the direction of flight of the $B$
meson, which is expected to be distributed like $\sin^2 \theta_B$, whereas the
continuum background is flat. We will discuss the
use of these cuts, and their effectiveness for particular analyses, but refer
the reader to the article by M. Artuso for a more detailed discussion
of the shape variables.

\subsection {High Energy Collider Experiments}

In the past, evidence for the production of $b$ quarks
in high energy experiments has been
deduced from the presence of high $p_{T}$ leptons.
Recently, significant progress in the isolation of events
containing $b$ quarks
has been made possible by the installation of silicon vertex detectors
near the interaction point at several collider experiments.
These $b$ quarks hadronize as $B_d$, $B_u$, $B_s$ mesons and
baryons containing $b$ quarks. Evidence for $B_s$ and
$\Lambda_b$ production has been reported but the relative
production fractions are not well known \cite{Sharma}.
 With the improvement in background suppression provided by these
solid state detectors, signals for exclusive hadronic
$B_d$, $B_u$ and $B_s$
 meson decays have been isolated in the invariant mass spectra for low
multiplicity final states (e.g. $B_s \to \psi \phi$).
However, the resolution in invariant mass (O(20 MeV))
is poorer than the resolution in beam constrained mass in threshold
experiments and is  not sufficient
to clearly separate
 modes with an additional photon or modes where one kaon
is replaced with a pion.
Although collider experiments cannot determine absolute branching
fractions without
 making further assumptions or using information from
experiments at the $\Upsilon (4S)$, they can measure relative
branching fractions.
Some high energy
experiments have also obtained inclusive signals
for $D^0, D^{*+}, \psi$ mesons in $B$ decay.
However, it is difficult
to distinguish the contribution of $B_d$, $B_u$ and $B_s$ mesons.

\subsection {Determination of $B$ Meson Branching Fractions}
\label{thatsit}
To extract $B$ meson branching ratios, the detection efficiencies are
determined from a Monte Carlo simulation and the yields are corrected
for the charmed meson branching fractions.
In order to determine new average branching ratios for $B$ meson
decays the results from individual experiments must be normalized
with respect to a common set of charm meson and baryon absolute branching
fractions.
The branching fractions for
the $D^0$ and $D^+$ modes used to calculate the $B$ branching fractions
are given in the Tables below. We have chosen the precise value
of the $D^0 \to K^- \pi^+$ branching fraction recently reported by
CLEO~II to normalize the results \cite{DKpi}.
The branching fractions of other $D^0$
decay modes relative to $D^0 \to K^- \pi^+$ are taken from the PDG
compilation\cite{PDG}. The $D^+$ branching ratio is taken
from the Mark~III experiment \cite{AbsoD}.

\begin{table}[htb]
\caption{$D^0$ branching fractions [\%] used in previous publications and
this review.}
\label{Tbd0br}
\begin{tabular}{lllll}
Mode & ARGUS, CLEO 1.5 \cite{SecondB},\cite{FourthB}
 & ARGUS ($DD_s$) \cite{ARGUSDDs} & CLEO II \cite{SixthB} & This review\\
\hline
$K^-\pi^+$ & $4.2 \pm 0.6 $ & $ 3.7 \pm 0.3 $ & $ 3.91 \pm 0.19 $ & $3.9 \pm
0.2 $\\
$K^-\pi^+ \pi^- \pi^+$ & $ 9.1 \pm 1.1 $ & $ 7.5 \pm 0.5 $ & $ 8.0 \pm 0.5 $ &
$ 8.0 \pm 0.5$\\
$K^-\pi^+ \pi^0$ & $ 13.3 \pm 1.8 $ & $ 11.3 \pm 1.1 $ & $ 12.1 \pm 1.1 $&
$12.1 \pm 1.1$\\
$K^0\pi^+ \pi^-$ & $ 6.4 \pm 1.1 $ & $ 5.4 \pm 0.5 $ &         & $5.8 \pm 0.5$
\end{tabular}
\end{table}

\begin{table}[htb]
\caption{$D^+$ branching fractions [\%] used in previous publications and
this review.}
\label{Tbdpbr}
\begin{tabular}{lllll}
Mode & ARGUS, CLEO 1.5 \cite{SecondB},\cite{FourthB}
 & ARGUS ($DD_s$) \cite{ARGUSDDs} & CLEO II \cite{SixthB} & This review\\
\hline
$K^-\pi^+ \pi^+$ & $9.1 \pm 1.4 $ & $ 7.7 \pm 1.0 $ & $ 9.1 \pm 1.4 $ & $9.1
\pm 1.4$\\
$K^0\pi^+ $& $ 3.2 \pm 0.5 $ & $ 2.6 \pm 0.4 $ & & \\
$K^0\pi^+ \pi^+ \pi^-$ &  & $ 6.9 \pm 1.1 $ &  & \\
\end{tabular}
\end{table}

Branching ratios for all $D_s$ decay modes are normalized relative to
${\cal{B}}(D_s\rightarrow \phi \pi)$. There are no model-independent
measurements of the absolute branching fraction for
$D_s\rightarrow \phi \pi$.
The currently favored method uses measurements of
$\Gamma(D_s\rightarrow \phi l\nu)$/$\Gamma(D_s\rightarrow \phi \pi)$.
The rate $\Gamma(D_s\rightarrow \phi l\nu)$
is determined from measurements of $\tau_{D_s}/\tau_{D^+}$,
$\Gamma(D^+\rightarrow K^* l\nu)$, and using
$\Gamma(D^+\rightarrow K^* l\nu)$/
$\Gamma(D_s\rightarrow \phi l\nu)$ obtained from theory.
We use the value of ${\cal{B}}(D_s\rightarrow \phi \pi)$
derived in reference \cite{CLNS9314} .
\begin{table}[htb]
\caption{$D_s$ branching fraction [\%] used in previous publications and
this review.}
\label{Tbdsbr}
\begin{tabular}{llll}
Mode & CLEO 1.5 \cite{SecondB}
 & ARGUS ($DD_s$) \cite{ARGUSDDs} & This review \\ \hline
$\phi \pi^+ $& $2.7 \pm 0.7 $ & $ 3.0 \pm 1.1 $&  $ 3.7 \pm 0.9 $
\end{tabular}
\end{table}

Since the publication of the
ARGUS and CLEO~1.5 papers on hadronic
decays, the branching fractions for the $D^{*} \to D \pi (\gamma)$
modes have been significantly improved by more
precise measurements from CLEO~II \cite{CLEODSTAR}.
For modes which contain $D^{*}$ mesons
we have recalculated the branching ratios using the CLEO~II measurements.

\begin{table}[htb]
\caption{$D^{*}$ branching fractions [\%] used in previous publications and
this review.}
\label{Tbdstarbr}
\begin{tabular}{llll}
Mode & ARGUS, CLEO 1.5 \cite{SecondB},\cite{FourthB}
 & CLEO II \cite{SixthB} & This review\\ \hline
$D^{*0}\rightarrow D^0\pi^0 $& $55.0 \pm 6 $ & $ 63.6 \pm 4.0 $ & $63.6 \pm
4.0$\\
$D^{*0}\rightarrow D^0\gamma $& $45.0 \pm 6 $ & $ 36.4 \pm 4.0 $ & $36.4 \pm
4.0$\\
$D^{*+}\rightarrow D^0\pi^+ $& $57.0 \pm 6 $ & $ 68.1 \pm 1.6 $ & $68.1 \pm
1.6$
\end{tabular}
\end{table}

\begin{table}[htb]
\caption{Charmonium branching fractions [\%] used in previous publications and
this review.}
\label{Tbccbr}
\begin{tabular}{llll}
Mode & ARGUS, CLEO 1.5 \cite{SecondB},\cite{FourthB}
 & CLEO II \cite{SixthB} & This review \cite{BRpsi}\\ \hline
$\psi \rightarrow e^+e^- $& $6.9 \pm 0.9$ & $ 6.3 \pm 0.2 $ & $ 5.9 \pm 0.25
$\\
$\psi \rightarrow \mu^+\mu^- $& $6.9 \pm 0.9$ & $ 6.0 \pm 0.25 $ & $ 5.9 \pm
0.25 $\\
$\psi' \rightarrow e^+e^-$ and $\mu^+\mu^- $& $1.7 \pm 0.3$ & $ 1.7 \pm 0.3 $ &
$ 1.7 \pm 0.3 $\\
$\psi' \rightarrow \psi \pi+\pi-$& $32.4 \pm 2.6$ & $ 32.4 \pm 2.6 $ &
$ 32.4 \pm 2.6 $\\
$\chi_{c1} \rightarrow \psi \gamma $& $ 27.3 \pm 1.6$ & $ 27.3 \pm 1.6 $ &
$ 27.3 \pm 1.6 $
\end{tabular}
\end{table}

We also give the old and new values assumed for the decays
$\psi\to e^+e^-$ and $\psi\to \mu^+\mu^-$.
We have chosen to use the precise measurement of these decays
recently performed by the MARK III collaboration \cite{BRpsi}.
The modes $\psi^{'}\to \ell ^+ \ell ^-$ and
$\psi^{'}\to \psi \pi^+ \pi^-$ are used to form $B$ meson candidates in modes
involving
$\psi'$ mesons.
B meson decays into final states containing $\chi_{c}$ mesons are reconstructed
using the channel $\chi_{c1}\to \psi \gamma$.
Product branching ratios for all modes containing $\psi$ mesons have been
rescaled
to account for the improved  $\psi$ branching fractions.

In the cases where only one $D^0$ decay mode was used to reconstruct the $B$
meson
the published branching ratio is simply rescaled.
The procedure for recalculating the branching ratios becomes more complicated
when more than one $D$ decay channel is used.
All experiments used the following procedure to obtain their results
\[\displaystyle
{\cal{B}}(B)\; = \; \frac{N_{observed}}{\epsilon \times N_B \times
({\cal{B}}(D^*)) \times
\sum{{\cal{B}}_i(D^0)}}\]
where $N_B$ is the number of $B$ mesons.
The efficiency $\epsilon$ is defined as
\[\epsilon \; = \; \frac{\sum{{\cal{B}}_i(D^0)\epsilon
_i}}{\sum{{\cal{B}}_i(D^0)}}\]
The index $i$ refers to the $D$ meson decay channel.
Therefore the rescaled branching ratio is given by
\[{\cal{B}}\; = \; \frac{N_{observed}}{N_B \times ({\cal{B}}(D^*))\times
\sum{{\cal{B}}_i(D^0)\epsilon _i}}\]

The CLEO collaboration published enough information, including the
yields and the efficiencies for the individual $D^0$ decay channels, so that
rescaling their $B$ branching ratios is straightforward.

Although the $D^0$ reconstruction efficiencies depend slightly on the $B$ meson
decay channel under study, the only information available from the ARGUS
collaboration are average $D^0$ reconstruction efficiencies $<\epsilon >_i$.
Therefore we had to  make the assumption that the correct way to
renormalize the ARGUS results is to multiply their branching ratios by
the scale factor $F$
where
\[
F\;=\;\frac{\sum{<\epsilon >_i \times {\cal{B}}_i(D^0)_{old}}}
{\sum{<\epsilon > _i \times {\cal{B}}_i(D^0)_{new}}}
\]
The validity of this assumption has been checked using CLEO~1.5 data.

Statistical errors are recalculated in the same way as the branching ratios.
For the results from individual experiments on $B$ decays to final states with
$D$
mesons two systematic errors are quoted.
The second systematic error contains the contribution
due to the uncertainties in the $D^0\to K^-\pi^+$ or $D^+\to K^-\pi^+\pi^+$
branching fractions.
This will allow easier rescaling when these
branching ratios  are measured more precisely.
The first systematic error includes the experimental uncertainties and when
relevant the uncertainties in the
ratios of charm branching ratios, e.g. $\Gamma(D^0 \rightarrow
K^- \pi^+ \pi^+ \pi^-)/\Gamma(D^0 \rightarrow K^- \pi^+)$, the error in the
$D^*$ branching fractions and the error in
${\cal{B}}(D_s\to\phi\pi^+)$.
For all other modes only one systematic error is given.
For the world averages the
 statistical and the first systematic error are combined
in quadrature while the errors due to the $D^0$ and $D^+$ scales
are still listed separately.
With the improvement in the precision
of the $D^0$ and $D^*$ branching fractions these are no longer the
dominant source of systematic error in the study of hadronic $B$
meson decay.

\section{INCLUSIVE B DECAY}

\subsection{Motivation}

Due to the large mass of the $b$ quark $B$ meson decays
give rise to a large
number of secondary decay products. For instance,
CLEO finds that the charged and photon multiplicities at the
$\Upsilon (4 S)$ are:
$n_{\rm charged}=10.99 \pm 0.06 \pm 0.29$,
$n_{ \gamma}=10.00\pm 0.53 \pm 0.50$, respectively
\cite{multi}.
Similarly, ARGUS \cite{multiARG} finds $n_{\rm charged}=10.74 \pm 0.02$.
The high multiplicity of final state particles leads to a large
number of possible exclusive final states. Even with a
detector that has a large acceptance
for both charged tracks and showers, it is difficult to reconstruct
many exclusive final states because of the combinatorial backgrounds.
Furthermore the detection efficiency drops for high multiplicity final states.
Thus, to get a complete picture of $B$ meson decay, it is important to
study inclusive decay rates.

A number of theoretical calculations of inclusive $B$ decay rates have been
made using the parton model.
It is believed that measurements
of such inclusive rates can be more reliably compared to the theoretical
calculations than can measurements of exclusive decays (e.g. see the
contribution
by Bigi in this volume).
While this is sufficient motivation for studying the inclusive rates,
there is also a need for accurate measurements in order to model the
decays of $B$ mesons both for high energy collider experiments, and for
experiments at the $\Upsilon (4S)$. As a specific example, the inclusive
rate for $B\to\psi$ has been used to determine the $B$ meson production
cross-section at the Tevatron \cite{pppsi}.

The branching ratios for inclusive $B$ decays to particular final state
particles are determined by measuring the inclusive yields of these
particles in data taken
on the $\Upsilon (4S)$ resonance, and subtracting the non-resonant background
using data taken at energies below the
$\Upsilon (4S)$ resonance. The off-resonance data are scaled to correct
for the energy dependence of the continuum cross-section.
Results on inclusive production at the $\Upsilon (4S)$ are usually presented
as a function of the variable $x$, which is the fraction of the maximum
possible momentum carried by the particle, $p_{max}=\sqrt{E_{beam}^2 - M^2}$.
The endpoint for production in $B$ decays is at $x=0.5$.

\subsection{Inclusive $B$ Decay to Mesons}

CLEO~1.5 \cite{CLEOK} has measured the branching fractions of inclusive $B$
decays
to light mesons, while
ARGUS has determined the average multiplicities of light mesons
in $B$ decay. If more than one meson of the particle type under study is
produced in a $B\bar{B}$ decay, then the branching fraction and the
multiplicity
will differ. Unless otherwise noted,
the results reported in Table \ref{Tbmulti} are averaged over $B$ and
$\bar{B}$ decay.

\begin{table}[htb]
\caption{Multiplicities or branching fractions of light mesons per $B$ meson
decay.}
\label{Tbmulti}
\begin{tabular}{lll}
Mode & CLEO 1.5 \cite{CLEOK} & ARGUS \cite{ARGUSK} \\
     & (Branching Ratio) & (Multiplicity) \\ \hline
$ B/\bar{B}\to \pi^{\pm} $ &      & $ 3.59\pm 0.03\pm0.07$   \\
(not from $K_s,\Lambda$) & & \\
$ B/\bar{B}\to \pi^{\pm} $ &      & $ 4.11\pm 0.03\pm0.08$   \\
(incl. $K_s,\Lambda$) & & \\
$ B/\bar{B}\to K^{\pm} $ & $ 0.85\pm 0.07\pm 0.09$ & $0.78\pm 0.02\pm 0.03$ \\
$ \bar{B}\to K^{-} $  & $ 0.66\pm 0.05\pm 0.07$ &  \\
$ \bar{B}\to K^{+} $  & $ 0.19\pm 0.05\pm 0.02$ &  \\
$ B/\bar{B}\to K^0/\bar{K}^0 $ & $ 0.63 \pm 0.06\pm0.06$ & $0.64\pm 0.01 \pm
0.04$   \\
$ B/\bar{B}\to K^{*0} $    &    & $0.146\pm 0.016\pm 0.020$ \\
$ B/\bar{B}\to K^{*+} $    &    & $0.182\pm 0.054\pm 0.024$  \\
$ B/\bar{B}\to \rho^0 $    &    & $0.209\pm 0.042 \pm 0.033$ \\
$ B/\bar{B}\to \omega $    &    & $< 0.41$ (90\% C.L.)       \\
$ B/\bar{B}\to f_0(975) $  &    & $<0.025$ (90\% C.L.) \\
$ B/\bar{B}\to \eta ' $  &    & $<0.15$ (90\% C.L.) \\
$ B/\bar{B}\to \phi $      &    & $0.039\pm 0.003\pm 0.004 $ \\
\end{tabular}
\end{table}

In the decay $b \to c \to s$ the
charge of the kaon can be used to determine the
flavor of the $b$ quark. A first attempt to measure the tagging efficiency and
misidentification probability for this method
has been performed by ARGUS \cite{ARGUSK}.
With the large sample of
reconstructed $B^0$ and $B^+$ decays from CLEO~II it should
be possible to measure these quantities directly.
The experiments also measure the momentum spectra for the particles listed
in Table \ref{Tbmulti}. These results provide important information needed
to improve Monte Carlo generators for future $B$ experiments.

\begin{figure}[htb]
\begin{center}
\unitlength 1.0in
\begin{picture}(3.,3.)(0,0)
\end{picture}
\bigskip
\vskip 10 mm
\caption{$B\to D^0 X$, $D^+ X$, and $D^{*+}X$
momentum spectra in CLEO 1.5 data. The dashed curve is
the prediction of the phenomenological model of Wirbel and Wu while
  the solid histogram is the prediction of their free quark model}
\label{Fdmomdata}
\end{center}
\end{figure}
\begin{figure}[htb]
\begin{center}
\vskip 10mm
\unitlength 1.0in
\begin{picture}(3.,2.8)(0,0)
\end{picture}
\bigskip
\vskip 5 mm
\caption{$B\to D_s X$ momentum spectrum in ARGUS data.
The solid line is the sum of the two components. The two dotted lines
indicate the two body component and
three body process.}
\label{Fdsmomdata}
\end{center}
\end{figure}
The inclusive production of $D^0, D^+, D_{s}^+$ and $D^{*+}$ mesons
in $B$ decay has been measured
by ARGUS \cite{ARGUSD} and CLEO~1.5 \cite{CLEOD}.
To improve signal to background, only the $D^0 \to K^- \pi^+$,
$D^+ \to K^- \pi^+ \pi^+$ and $D_{s}^{+} \to \phi \pi^+$ decay modes are used.
The results, rescaled for the charm branching ratios,
are given in Tables ~\ref{Tbhad_inc}--\ref{Tbincav}.
The momentum spectrum for the inclusive decay of $B$ mesons
to $D^0$, $D^+$, and $D^{*+}$ as measured by
CLEO~1.5 are shown in Fig. \ref{Fdmomdata}.
The $D^{*+}$ spectrum is not measured for $x <0.1$ due to poor
reconstruction efficiency for slow tracks. The shape of the $D_s$ momentum
spectrum (Fig. \ref{Fdsmomdata})
indicates that there is a substantial two body component. In model
dependent fits the ARGUS and CLEO~1.5
experiments find two body fractions of $(58 \pm 7 \pm 9)$\%
\cite{ARGUSD} and $(56 \pm 10)$\% \cite{CLEOD}, respectively.

The polarization as a function of $x$ for $B \to D^{*+}$ has also been
measured and found to agree with the predictions of Wirbel
and Wu\cite{D*pol} and of Pietschmann and Rupertsberger\cite{PR}.

Results on inclusive $B$ decay to final states with $\psi$ and $\psi '$ mesons
have been reported by CLEO~1.5\cite{SecondB}, ARGUS\cite{ARGUSpsinc}, and
CLEO~II\cite{CLEOpsiinc}.
In the most recent high statistics analysis from CLEO~II, the effect of final
state radiation has been taken into account \cite{dmc}.
This effect leads
to a significant tail on the low side of the
$\psi \to e^+ e^-$ mass peak and a smaller
effect in the $\mu^+ \mu^-$ spectrum. Even with a large mass window that
extends from $2.50$ to $3.05$ GeV$/c^2$, this effect
can modify the calculated detection efficiency by more than $10\%$.
Corrections are also made for non-resonant $\psi$ production in the
CLEO~II analysis \cite{fastpsi}.
The resulting invariant dielectron and dimuon mass distributions are
shown in Fig. \ref{Fpsi}.

\begin{figure}[htb]
\begin{center}
\unitlength 1.0in
\begin{picture}(3.,3.)(0,0)
\end{picture}
\bigskip
\bigskip
\vskip 2mm
\caption{$B\to \psi X$ mass spectra from CLEO II: (a) $\psi \to e^+ e^-$
channel and (b) $\psi \to \mu^+ \mu^-$ channel.}
\label{Fpsi}
\end{center}
\end{figure}

\begin{figure}[htb]
\begin{center}
\unitlength 1.0in
\begin{picture}(3.,3.)(0,0)
\end{picture}
\bigskip
\vskip 10 mm
\caption{$B\to \psi X$ momentum spectra in CLEO~II data.}
\label{Fpsimomdata}
\end{center}
\end{figure}

The momentum spectrum  for $B \to \psi, \psi^{'}$ transitions
has been measured (Fig.
\ref{Fpsimomdata} shows the $\psi$ momentum spectrum).
The two body component due to $B\to \psi K$ and $B\to \psi K^*$
saturates the spectrum in the momentum range between 1.4 and 2.0 GeV.
ARGUS has determined ${\cal{B}}(B \to \psi$, where $\psi$ not from $\psi ')
\; = \; (0.95 \pm 0.27)\%$.
The two body component constitutes about 1/3 of direct $\psi$ production.

\begin{table}[htb]
\caption{$\psi$ polarization $\Gamma_L/\Gamma $
in inclusive $B$ meson decays.}
\label{Tbpsipol}
\begin{tabular}{lll}
$\psi$ momentum & CLEO II \cite{psipol} & ARGUS \cite{argpol} \\ \hline
$p_{\psi}< 0.8$ GeV/c & $0.55 \pm 0.35 $ & \\
0.8 GeV/c $<p_{\psi}< 1.4$ GeV/c & $0.49 \pm 0.32 $ & \\
1.4 GeV/c $<p_{\psi}< 2.0$ GeV/c & $0.78 \pm 0.17 $ & $1.17\pm 0.17$ \\
all $p_{\psi}< 2.0$ GeV/c & $0.59 \pm 0.15 $ &
\end{tabular}
\end{table}

The polarization $\Gamma_L/\Gamma$
as a function of momentum for $B\to \psi$ transitions
 has also been determined (see Table~\ref{Tbpsipol}).
According to ARGUS, the $\psi$ mesons in the highest momentum bin are
completely longitudinally polarized.
Since the highest momentum
bin is dominated by two body $B$ decay, the polarization measured in this bin
can be used to estimate the polarization of $B \to \psi K^*$ after
correcting for the contribution of $B \to \psi K$.
Therefore the ARGUS result indicates that the $B \to \psi K^*$
mode is dominated by a single orbital angular momentum state and hence
by a single CP eigenstate.

CLEO~II \cite{fastpsi} and ARGUS \cite{arguspsi} have reported results on
inclusive $B \to \chi_c X, \chi_c \to \gamma \psi$
decays. ARGUS assumes there is no $\chi_{c2}$ production.
CLEO~II has a 4.5 times better $\chi_c$ mass resolution than ARGUS and allows
for both possibilities. The branching ratio for $\chi_{c0} \to \gamma \psi$ is
$(6.6\pm 1.8) \times 10^{-3}$ so the contribution
of the $\chi_{c0}$ meson can be neglected.

\begin{table}[htb]
\caption{Branching fractions for inclusive $B$ decays to charm mesons.}
\label{Tbhad_inc}
\begin{tabular}{lllll}
Particle & Signature & ARGUS & CLEO 1.5 & CLEO II \\ \hline
 $ D^0$ & $ K^- \pi ^+$ &
 $ 49.5 \pm 3.8 \pm 6.4 \pm 2.0 $ &
 $ 59.7 \pm 3.2 \pm 3.6 \pm 2.4 $ & \\
 $ D^+$ & $ K^- \pi ^+ \pi ^+$ &
 $ 23.9 \pm 2.9 \pm 4.4 \pm 2.4 $ &
 $ 24.9 \pm 3.3 \pm 2.0 \pm 2.5 $ & \\
 $ D^{*+}$ & $ D^0 \pi ^+ $ &
 $ 27.6 \pm 2.3 \pm 4.7 \pm 1.1 $ &
 $ 22.7 \pm 1.3 \pm 2.3 \pm 0.9 $ & \\
 $ D_s^+$ & $ \phi \pi ^+$ &
 $ 7.9 \pm 1.1 \pm 0.8 \pm 2.6$ &
 $ 8.3 \pm 1.2 \pm 0.8 \pm 2.7$ & \\
 $ \psi$ & $ e^+e^-,\mu ^+ \mu^- $ &
 $ 1.25 \pm 0.19 \pm 0.26 $ &
 $ 1.31 \pm 0.12 \pm 0.27 $ &
 $ 1.09 \pm 0.04 \pm 0.07 $ \\
 $ \psi'$ & $ \ell ^+ \ell ^-,\psi \pi^+ \pi^- $ &
 $ 0.50 \pm 0.19 \pm 0.12 $ &
 $ 0.36 \pm 0.09 \pm 0.13 $ &
 $ 0.30 \pm 0.05 \pm 0.03 $ \\
 $ \chi_{c1}$ & $ \psi \gamma $ &
 $ 1.23 \pm 0.41 \pm 0.29 $ &
 &  $ 0.56 \pm 0.16 \pm 0.14 $
\end{tabular}
\end{table}

\begin{table}[htb]
\caption{World averages for branching fractions of
inclusive $B$ decays to charm mesons.}
\label{Tbincav}
\begin{tabular}{lll}
Particle & Signature & Branching Ratio \\ \hline
 $ D^{0}$ & $ K^- \pi ^+$ &
 $ 56.7 \pm 4.0 \pm 2.3 $ \\
 $ D^+$ & $ K^- \pi ^+ \pi ^+$ &
 $ 24.6 \pm 3.1 \pm 2.5 $ \\
 $ D^{*+}$ & $ D^0 \pi ^+$ &
 $ 23.7 \pm 2.3 \pm 0.9 $ \\
 $ D_s^+$ & $ \phi \pi ^+$ &
 $ 8.1 \pm 0.9 \pm 0.1 $ \\
 $ \psi$ & $ e^+e^-$, $ \mu ^+ \mu ^-$ &
 $ 1.11 \pm 0.08 $ \\
 $ \psi' $& $ \ell ^+\ell ^-$, $\psi \pi ^+ \pi ^-$ &
 $ 0.32 \pm 0.05 $ \\
 $ \chi_{c1}$ & $ \psi \gamma $ &
 $ 0.66 \pm 0.20 $
\end{tabular}
\end{table}

In a parton level calculation, Palmer and Stech \cite{palmstech}
find that ${\cal{B}}(B \to X_{c \bar{c}}) = 19 \pm 1 \%$
where the theoretical error is the uncertainty due to the choice
of quark masses. This can be
compared to the sum of the experimental measurements
${\cal{B}}(B \to D_s X) + 4*{\cal{B}}( B \to \psi X) + {\cal{B}}(B \to \psi' X)
 = 12.9 \pm 1.0 \%$ where the factor of 4
 which multiplies ${\cal{B}}(B\to \psi X)$ includes a factor of 2 for the
two charm quarks, and an additional
factor of 2 which is a generous estimate
of the unobserved contributions from $B$ decays to
 $\eta_c$ and $\chi$ states.

\subsection{Inclusive $B$ Decay to Baryons }

ARGUS\cite{argusbary}
and CLEO~1.5\cite{crawbary} have observed inclusive yields of $\bar{p}$,
$\Lambda$, $\Xi$ and the charmed $\Lambda_c$ baryon, and recently
CLEO~II has reported the observation
of $B \to \Sigma_c X$\cite{sigmamz}.
The measured branching ratios for these decays and the world
averages can be found in Table \ref{Tbbaryon}.
The determination of branching ratios for inclusive $B$ decays to the charmed
baryons $\Lambda_c$ and $\Sigma_c$ requires knowledge of
${\cal{B}}(\Lambda_c \to pK^-\pi^+)$.
However, the uncertainty in this quantity is still large and it can only
be determined by indirect methods. The results given in this review use
${\cal{B}}(\Lambda_c \to pK^-\pi^+)\:=\:(4.3 \pm 1.0 \pm 0.8)$\%
\cite{crawbary}.
For modes involving $\Lambda_c$ baryons the uncertainty due to
the $\Lambda_c$ branching ratio scale is listed as a separate error.

\begin{table}[htb]
\caption{Branching fractions [\%] of inclusive $B$ decays to baryons.}
\label{Tbbaryon}
\begin{tabular}{lllll}
Mode          &       CLEO~1.5 & ARGUS  & CLEO~II & Average \\ \hline
$B\to \bar{p} ~X$
& $ 8.0\pm 0.5 \pm 0.3   $       & $8.2\pm 0.5^{+1.3}_{-1.0}$& & $8.0\pm 0.5$
\\
(incl. $\Lambda$) & & & & \\
$ B\to \bar{p} ~X $
& $ 5.6 \pm 0.6 \pm 0.5 $ & $ 5.5\pm 1.6$ & & $5.6 \pm 0.7$ \\
(not from $\Lambda$) & & & & \\
$B\to \Lambda ~X$ & $3.8\pm 0.4 \pm 0.6 $  & $4.2\pm 0.5\pm 0.6$& & $4.0\pm
0.5$ \\
$B\to \Xi^- ~X$ & $0.27 \pm 0.05 \pm 0.04  $ & $0.28\pm 0.14$ & & $0.27\pm
0.06$ \\
$\bar{B}\to \Lambda_c ~X$ & $6.3\pm 1.2\pm 0.9\pm 1.9  $  & $7.0\pm 2.8\pm
1.4\pm 2.1$ &
 & $6.4 \pm 1.3 \pm 1.9$ \\
$\bar{B}\to \Sigma_c^0 ~X$ &           & $ $ & $0.53\pm 0.19\pm 0.16\pm 0.16 $
&\\
$\bar{B}\to \Sigma_c^{++} ~X$ &        & $ $ & $0.50\pm 0.18\pm 0.15 \pm 0.15 $
& \\
$\bar{B}\to \Sigma_c^{0} \bar{N}$ &   &  & $< 0.17$ (90\% C.L.) & \\
$\bar{B}\to \Sigma_c^{++} \bar{\Delta}^{--}$ &  &  & $< 0.12 $ (90\% C.L.) & \\
\end{tabular}
\end{table}

In a parton level calculation with diquark correlation taken into
account, Palmer and Stech \cite{palmstech}
have performed a calculation of the total rate
for inclusive $B$ decay to charmed baryons.
They find ${\cal{B}}(B \to$ charmed baryons) $\approx 6\%$.
Neglecting small contributions from $B \to \Xi_c$ transitions, we
assume all $B$ to charmed baryon decays proceed through a $\Lambda_c$
baryon so we can
compare this prediction to the experimental result ${\cal{B}}(B\to \Lambda_c X)
 = (6.4 \pm 1.3 \pm 1.9)$\%.

The momentum spectrum of $B\to \Lambda_c$ transitions has been measured
by CLEO~1.5 \cite{crawbary} and ARGUS \cite{argusbary}.
No significant two body component is found. Similarly,
CLEO~II has found that $B\to \Sigma_c^0 X$ and $B\to \Sigma_c^{++} X$ have
no two body contribution. To date, no exclusive $B\to $ baryon decays have
been reconstructed.

In addition to the inclusive branching ratios given above, CLEO~1.5
\cite{crawbary}
and ARGUS \cite{argusbary} have investigated baryon correlations in  $B$ decay
in order
to elucidate the underlying decay process.
\begin{table}[htb]
\caption{Branching fractions [\%] of
inclusive $B$ decays to baryon pairs.}
\label{Tbbaryonp}
\begin{tabular}{lll}
Mode          &       CLEO 1.5 & ARGUS                     \\ \hline
$B\to p \bar{p}~X$ & $2.4\pm 0.1\pm 0.4   $ & $ 2.5\pm 0.2\pm 0.2$  \\
$B\to \Lambda \bar{\Lambda}~X$ & $<0.5$ (90\% C.L.)  & $<0.88$ (90\% C.L.) \\
$B\to \Lambda \bar{p} ~X$ & $2.9\pm 0.5\pm 0.5  $ & $2.3\pm 0.4\pm 0.3$  \\
$B\to D^{*+} p \bar{p} ~X$ & $< 0.35 $ (90\% C.L.) & $       $       \\
$B\to D N \bar{N} ~X$ & $<5.2$ (90\% C.L.) & $       $
\end{tabular}
\end{table}
We follow the notation of Reference\cite{crawbary} . Let N
denote baryons with $S=C=0$ (e.g. p, n, $\Delta$, $N^*$). Let Y
refer to baryons with $S=-1, C=0$ (e.g. $\Lambda$, $\Sigma^0$, $\Sigma^+$).
Let $Y_c$ refer to baryons with $S=0, C=1$ [e.g. $\Lambda_{c}^{+}$,
$\Sigma_{c}^{(+,0,++)}$] . Then the following final states can be used
to distinguish possible mechanisms for baryon production in $B$ decay
(Fig. \ref{btobaryon}).
\begin{figure}[htb]
\begin{center}
\unitlength 1.0in
\begin{picture}(3.,2.5)(0,0)
\end{picture}
\vskip 10 mm
\caption{
Decay diagrams for $B$ meson decays to baryons: (a) External Spectator
Diagram (b) W Exchange Diagram
(c) External Spectator Diagram producing $D N \bar{N} X$
and $D Y \bar{Y} X$ (d) Internal Spectator Diagram producing
 $DN\bar{N}X$ and $DY\bar{Y}X$.}
\label{btobaryon}
\end{center}
\end{figure}

\begin{enumerate}
\item {$\bar{B} \to Y_c \bar{N} X$, $\bar{B} \to \Xi_c \bar{Y} X$}\\
These final states are produced by the usual $b \to c W^-$ coupling
in a spectator or exchange diagram in conjunction with the popping of
two quark pairs from the vacuum (as shown in Figs.~\ref{btobaryon}(a),(b)).
 It should be noted that the two
mechanisms can be distinguished by examination of the $Y_c$ momentum
spectrum,  since the exchange diagram will
produce two body final states (e.g. $\Lambda_c \bar{p}$ or
$\Sigma_c^{++} \bar{\Delta}^{--}$).

\item {$\bar{B}\to D N \bar{N} X$, $\bar{B} \to D Y \bar{Y} X$}\\
 The noncharmed baryon-antibaryon is produced from W fragmentation
after hadronization with two quark-antiquark pairs popped from
the vacuum (as shown in Figs.~\ref{btobaryon}(c),(d)).
The $D$ meson is formed from the charm spectator quark system.
If this mechanism is significant, inclusive production of
charmless baryon-antibaryon pairs should be
observed in $B$ decay.

\item{$\bar{B} \to Y_c \bar{Y} X$,  $\bar{B} \to \Xi_c \bar{Y_c} X$}\\
 These states are produced by the internal spectator graph
with $W^- \to \bar{c} s$ in conjunction with the popping of two quark
antiquark pairs. Since ${\cal B} (W^- \to \bar{c} s)/ {\cal B} (W^- \to all)$
is about $0.15$, this mechanism may be suppressed.

\item {$\bar{B}\to D_{s}^{-} Y_c \bar{N} X$,
 $\bar{B}\to D_{s}^{-} \Xi_c \bar{Y} X$}\\
This is the same as mechanism (1) with $W^- \to \bar{c} s$.
\end{enumerate}

The low rates for $B\to \Lambda \bar{\Lambda} X$, $\Lambda \bar{p} X$ and
$D^* p \bar{p} X$ suggest that mechanism (2) is small (Table \ref{Tbbaryonp}).
The absence
of a two body component in the momenta spectra
 of $B\to \Lambda_c X$, $\Sigma_c X$ indicates that the W-exchange
and internal spectator mechanisms are small. Thus it is
reasonable to assume that $\bar{B}\to Y_c \bar{N} X$ with
an external spectator  $b\to c W^-$
coupling (Fig.~\ref{btobaryon}(a))
 is the principal mechanism in $B$ to baryon transitions.

If $B$ decays to baryons are dominated by $\bar{B} \to \Lambda_c \bar{p} X$
and $\bar{B} \to \Lambda_c \bar{n} X$ then
measurements of the branching ratios for
 $B \to \bar{p} X$,
$B \to p \bar{p} X$ can be used to extract the absolute $\Lambda_c \to
p K^- \pi^+$ branching ratio.  The CLEO~1.5 measurements give
$B (\Lambda_c \to p K^- \pi^+) = 4.3 \pm 1.0 \pm 0.8 \%$ which can be
used to normalize all other measured $\Lambda_c$ branching ratios. In a similar
analysis ARGUS finds $(4.1\pm 2.4)$\% for this branching ratio.

\subsection{Charm Production in $B$ Decay}

The measurements of inclusive decay rates can be used to test the parton level
expectation that almost all $B$ decays proceed via a $b\to c$ transition.
This means ${\cal B}(B \to D^0 X) + {\cal B}(B \to D^+ X) +
{\cal B}(B \to D_s X) + {\cal B}(B \to \Lambda_c X) + 2
f\times{\cal B}(B\to \psi X)$
should be about 115\%, where the extra $15\%$ is due
to the fact that the virtual W should form a $s \bar{c}$ quark pair with
a probability of 0.15 . The small contributions from $b \to u$, penguin
transitions and
contributions from $B \to \Xi_c X$ not proceeding through a $\Lambda_c$
have been neglected.
The factor of 2 which multiplies ${\cal{B}}(B\to \psi X)$ accounts for the
two charm quarks, while
the factor $f=2$ is a generous estimate of the contribution
from charmonium states that do not decay via $\psi$.
The CLEO~1.5 result of $104.4 \pm 7.7 \%$ and
the ARGUS result of $93.3 \pm 10.5 \%$ can be combined to give a world average
of $100.2\pm 6.6\%$ for the average number of charm quarks produced in a $B$
decay.
While the existing data are not yet
sufficiently precise to make a convincing case for a charm deficit in $B$
decay,
there are several possible explanations for a deviation from
the parton level expectation.
There may be systematic biases in the method of determining inclusive
production rates at the $\Upsilon (4S)$.
The charm meson absolute branching
fractions can also contribute a systematic uncertainty, although this
error has been significantly reduced by the new determination
of ${\cal B}(D^0\to K^-\pi^+)$\cite{DKpi}.
There could be a breakdown of the parton level approximation in $B$ decay,
or there could be a large contribution to the inclusive rate that has not been
included. It has been suggested by Palmer and Stech\cite{palmstech},
that $b \to c \bar{c} s$ followed by $c \bar{c} \to \rm{gluons}$,
 which in turn hadronize into a final state with no charm, has a large
branching ratio. Another suggestion is that the rate for the hadronic penguin
diagram $b\to sg$ is larger than expected.

\begin{figure}[htb]
\begin{center}
\unitlength 1.0in
\vskip 12 mm
\begin{picture}(3.,3.)(0,0)
\end{picture}
\bigskip
\bigskip
\bigskip
\bigskip
\vskip 15 mm
\caption[]{The beam constrained mass distributions
from CLEO~II for:
 (a) $ B^- \to D^0 \pi^-$ decays.
(b) $ B^- \to D^0\rho^-$ decays for $|\cos\Theta_{\rho}|>0.4$.
 (c) $ \bar{B^0} \to D^+\pi^-$ decays .
(d) $ \bar{B}^0 \to D^+ \rho^-$
decays for
$|\cos\Theta_{\rho}|>0.4$.}
\label{dpi}
\end{center}
\end{figure}

\begin{figure}[hbt]
\begin{center}
\unitlength 1.0in
\vskip 12 mm
\begin{picture}(3.,3.)(0,0)
\end{picture}
\bigskip
\bigskip
\bigskip
\bigskip
\vskip 15 mm
\caption[]{Beam constrained mass
 distributions from CLEO~II for:
(a) $B^- \to D^{*0} \pi^-$ decays,
 (b) $B^- \to D^{*0} \rho^-$ decays,
 (c) $\bar{B}^0 \to
D^{*+} \pi^-$ decays,  and
  (d) $\bar{B}^0
\to D^{*+} \rho^-$ decays.}
\label{dspi}
\end{center}
\end{figure}

\begin{figure}[htb]
\begin{center}
\unitlength 1.0in
\begin{picture}(2.2,2.4)(0.0,0.0)
\end{picture}
\vskip 15 mm
\caption[]{Resonant substructure for $B\to D^* \rho^-$ from CLEO~II for:
 (a) the
$\pi^0\pi^-$ invariant mass spectrum for the
$ \bar{B}^0 \to D^{*+} \pi^0\pi^-$ decay mode in data.
 (b) the
$\pi^0\pi^-$ invariant mass spectrum for the
$ \bar{B}^0 \to D^{*+} \pi^0\pi^-$ decay mode in data.}
\label{subs}
\end{center}
\end{figure}


\begin{figure}[htb]
\unitlength 1.0in
\vskip 10 mm
\begin{picture}(3.,2.0)(0,0)
\end{picture}
\bigskip
\caption[]{Beam constrained mass
 distributions from CLEO~II for:
(a) $B^- \to D^{*0} a_{1}^{-}$ and
 (b) $\bar{B}^0 \to D^{*+} a_{1}^{-}$.}
\label{FBmaone}
\end{figure}

\begin{figure}[htb]
\unitlength 1.0in
\vskip 10 mm
\begin{center}
\begin{picture}(2.2,2.4)(0.0,0.0)
\end{picture}
\vskip 15 mm
\caption[]{Resonant substructure of $\bar{B^0}\to D^{*+} a_1$ from
CLEO~II:
(a) The $\pi ^- \pi ^- \pi ^+$ invariant mass spectrum from
a Monte Carlo simulation of $\bar {B}^0 \to D^{*+} a_1^-$
(b) The $\pi ^- \pi ^- \pi ^+$ invariant mass spectrum from
Monte Carlo simulation for $\bar {B}^0 \to D^{*+} (\pi ^- \rho ^0)_{NR}$
(c) The $\pi ^- \pi ^- \pi ^+$ mass spectrum from data
after $B$ mass sideband subtraction. The fit
to the sum of (a) and (b) is superimposed.}
\label{Fmaonea}
\end{center}
\end{figure}

\begin{figure}[htb]
\unitlength 1.0in
\vskip 10 mm
\begin{center}
\begin{picture}(3.0,3.5)(0.0,0.0)
\end{picture}
\vskip 15 mm
\caption[]{Angular distributions (efficiency corrected)
from CLEO~II for
(a) the helicity angle from $D^{*+} \to D^0 \pi ^+$
in  $\bar {B^0} \to D^{*+} \rho ^-$ and
(b)the helicity angle from $\rho ^- \to \pi ^- \pi ^0$
in  $\bar {B^0} \to D^{*+} \rho ^-$  (c) the helicity angle from
$D^{*+} \to D^0 \pi ^+$ in  $\bar {B^0} \to D^{*+} \pi ^-$}
\label{helrho}
\end{center}
\end{figure}

\begin{figure}[htb]
\unitlength 1.0in
\vskip 10 mm
\begin{center}
\begin{picture}(3.0,3.4)(0.0,0.0)
\end{picture}
\vskip 15 mm
\caption[]{
Beam constrained mass distributions from CLEO~II
 for:  (a) $B^- \to D^{**0}(2420)
\pi^-$ where $D^{**0}(2420) \to D^{*+} \pi^-$, (b) $B^- \to D^{**0}(2460)
\pi^-$ where $D^{**0}(2460) \to D^{*+} \pi^-$,  (c) $B^- \to
D^{**0}(2420) \pi^-\pi^0$ where
$D^{**0}(2420) \to D^{*+}\pi^-$,  (d) $B^- \to D^{**0}(2460) \pi^-\pi^0$
where $D^{**0}(2460) \to D^{*+} \pi^-$}
\label{dsspi}
\end{center}
\end{figure}

\begin{figure}[htb]
\unitlength 1.0in
\begin{center}
\begin{picture}(2.2,1.6)(0.0,0.0)
\end{picture}
\bigskip
\caption[]{$B$ meson decay diagrams
with emission of $c$ $\bar{s}$ quarks: (a)
external spectator and (b) color suppressed.}
\label{Fcsdiag}
\end{center}
\end{figure}

\begin{figure}[htb]
\unitlength 1.0in
\begin{center}
\begin{picture}(3.0,2.6)(0.0,0.0)
\end{picture}
\vskip 7 mm
\caption[]{The
beam constrained mass distributions from ARGUS for:
(a) the sum of $\bar{B}^0\to D_s^-{D}^+$,
    $\bar{B}^0\to D_s^-{D}^{*+}$,
   $\bar{B}^0\to D_s^{*-}{D}^+$, and
 $\bar{B}^0\to D_s^{*-}{D}^{*+}$ and
(b) the sum of $B^+\to D_s^+\bar{D}^0$,
 $B^+\to D_s^+ \bar{D}^{*0}$,
 $B^+\to D_s^{*+} \bar{D}^0$, and
 $B^+\to D_s^{*+} \bar{D}^{*0}$.}
\label{Fargusdd}
\end{center}
\end{figure}

\begin{figure}[htb]
\vskip 10 mm
\unitlength 1.0in
\begin{center}
\begin{picture}(3.0,3.3)(0.0,0.0)
\end{picture}
\vskip 15 mm
\caption[]{
Beam-constrained mass from CLEO~II for: (a) $B^-\to\psi K^-$,  (b)
$\bar{B^0}\to\psi\bar{K^0}$,  (c) $B^- \to\psi\bar{K}^{*-}$, and  (d)
$\bar{B^0}\to\psi K^{*0}$.}\label{bpsik}
\end{center}
\end{figure}

\begin{figure}[htb]
\unitlength 1.0in
\vskip 10 mm
\begin{center}
\begin{picture}(3.0,3.3)(0.0,0.0)
\end{picture}
\vskip 15 mm
\caption[]{
Beam-constrained mass from CLEO~II for:  (a) $B^-\to\psi' K^-$,  (b)
$\bar{B^0}\to\psi'\bar{K^0}$,  (c) $B^-\to\psi'\bar{K}^{*-}$, and
(d) $\bar{B}^0\to\psi' K^{*0}$.}\label{bpsipk}
\end{center}
\end{figure}

\section{B MESON RECONSTRUCTION}
\label{B-recon}

\subsection{Selection of $B$ Candidates}

As an example of the techniques of $B$ reconstruction we will briefly describe
the procedure used by the CLEO~II experiment to reconstruct the decay modes
$B\to D^{(*)}(n\pi)^-$.
The CLEO~II detector is described in detail elsewhere \cite{TRA}.
It has a momentum resolution for charged tracks given by
$(\delta p/p)^2 = (0.0015p)^2 + (0.005)^2$,
and an energy resolution for isolated photons
from the CsI barrel calorimeter of
$\delta E/E [\%] = 0.35/E^{0.75} + 1.9 - 0.1E$, where $p$ and $E$ are in GeV.
Charged tracks are identified as pions or kaons if they have
ionization loss information $(dE/dx)$, and/or time-of-flight information (ToF),
consistent with the correct particle hypothesis.
Photon candidates are selected from showers in the calorimeter barrel
with a minimum energy of 30~MeV, which are not matched to charged tracks, and
which have a lateral energy distribution consistent with that expected for a
photon. Neutral pions are selected from pairs of photons with an invariant
mass within $2.5\sigma$ of the known $\pi^0$ mass.

Candidate $D^0$ mesons are identified in the decay modes $D^0\to K^-\pi^+$,
$D^0\to K^-\pi^+\pi^0$, and $D^0\to K^-\pi^+\pi^+\pi^-$, while $D^+$ mesons
are identified in the decay mode $D^+\to K^-\pi^+\pi^+$.
Charged $D^*$ candidates are found using the decay $D^{*+}\to D^0\pi^+$,
while neutral $D^*$ candidates are found using the decay $D^{*0}\to D^0\pi^0$.
Other $D$ and $D^*$ decay modes are not used because of poorer signal to
background ratios, or because of lower yields\cite{xdfeg}.
The reconstructed $D$ masses and $D^*-D^0$ mass differences are required to
be within $2.5\sigma$ of the known values.

The $D$ meson candidates are combined with one or more additional pions
to form a $B$ candidate.
Cuts on the topology of the rest of the event are made in order
to distinguish
$B \bar{B}$ events from continuum background, as discussed in Section
\ref{y4sexp}.
The following requirements are imposed:
$R_2<0.5$, and $|\cos(\theta_S)|<0.9(0.8,0.7)$ depending on
whether there are one(two,three) pions added to the $D$ meson.
The cosine of the sphericity angle $\theta_S$
is uniformly distributed for signal, but
peaks near $\pm 1$ for continuum background.
Requiring that  $|\cos(\theta_S)|$ be less than 0.7 typically
removes  80\% of the continuum background, while retaining
70\% of the $B$ decays.

The measured sum of charged and neutral energies, $E_{meas}$,
of correctly reconstructed $B$ mesons produced at the
$\Upsilon (4S)$, must equal the
beam energy, $E_{beam}$, to within the experimental resolution.
Depending on the $B$ decay mode,
$\sigma_{\Delta E}$, the resolution on the energy difference
$\Delta E\; = \; E_{beam} - E_{meas}$
varies between 14 and 46~MeV.
Note that this resolution is usually sufficient to distinguish
the correct $B$ decay mode from a mode that differs by one pion.
For final states with a fast $\rho^-$ the energy resolution
depends on the momenta of the final state pions from the $\rho$ meson.
This dependence is conveniently parameterized as a function
of the angle between the $\pi^-$ direction in the $\rho^-$ rest frame
and the $\rho^-$ direction in the lab frame, which we
denote as the $\rho$ helicity angle, $\Theta_{\rho}$.
When $\cos\Theta_{\rho} = +1$, the error in the energy measurement
is dominated by the momentum resolution on the fast $\pi^-$,
whereas at $\cos\Theta_{\rho}= -1$
the largest contribution to the error in the energy measurement comes from
the calorimeter energy resolution on the fast $\pi^0$.

To determine the signal yield and
display the data the beam constrained mass is formed
\begin{equation} M_B^2=E_{beam}^2 - \left(\sum_i{\vec{p_i}}\right)^2,
\label{EBmass}
\end{equation}
where $\vec{p_i}$ is the momentum of the $i$-th daughter of the $B$ candidate.
The resolution in this variable is determined by the beam energy spread,
and is about 2.7~MeV for CLEO~II, and about 4.0~MeV for ARGUS.
\cite{mbrange}
These resolutions are  a factor of ten better
than the resolution in invariant mass obtained without the beam energy
constraint.

For a specific $B$ decay chain,
such as $B^- \to D^0 \pi^-, D^0 \to K^- \pi^+ \pi^0$ there may be
multiple combinations in a given decay chain.
In the CLEO~II analysis,
if there are multiple candidates only the entry with the smallest absolute
value of $\Delta E$ is selected for events with $M_{B} > 5.2 $ GeV.
An alternative method is to select the candidate with the highest total
probability as calculated from the sum of all $\chi^2$ contributions
from particle identification, kinematical fits and the beam energy
constraint\cite{FifthB}.

\subsection{Background Studies}
\label{bkg-studies}

In order to extract the number of signal events
it is crucial to understand
the shape of the background in the $M_B$ distribution.
There are two contributions to this background, continuum
and other $B \bar{B}$ decays. The fraction of continuum
background varies between $58 \%$ and $91\%$ depending on the
B decay mode\cite{cfrac}. The shape of the continuum background is
well understood since it depends primarily
on the transverse momentum distributions of
the final state particles relative to the jet axis. This has been studied using
the off-resonance data sample, and using Monte Carlo techniques.

The shape of the $B \bar{B}$ background is more
difficult to understand since it is mode dependent.
It also has a tendency to peak in the signal region, since the combinatorial
background comes mostly from combinations in which the true final state is
altered by one low energy particle. A particularly troublesome background
occurs when the decay $D^{*0}\to D^0\gamma$ is replaced by the decay
$D^{*0}\to D^0\pi^0$. To determine the correct background shape
for each $B$ decay mode, CLEO~II has studied the
$M_B$ distributions for $\Delta E$
sidebands, and for combinations in which the charged particles have the wrong
charges for the expected spectator decay diagram, e.g. $D^+\pi^+$ and
$\bar{D^0}\pi^+$.

It is found that all of the background distributions can be
fitted with a linear background below M$_B$=5.282~GeV,
and a smooth kinematical cutoff at the endpoint, which is chosen to be
parabolic. For each $B$ decay mode CLEO~II
uses this background function
and a Gaussian signal with a  fixed width of $2.64$ MeV to determine the
yield of signal events. In the ARGUS and CLEO~1.5 experiments slightly
different background parameterizations were used \cite{ARGUSback}.

\section{EXCLUSIVE B DECAY TO D MESONS}
\label{BDpiDrho}

\subsection{Measurements of $D (n \pi)^-$ Final States}

The decay modes
$ \bar{B^0}\to D^+ \pi^-$, $ \bar{B^0} \to D^+ \rho^-$,
$ B^-\to D^0 \pi^-$, and $ B^- \to D^0 \rho^-$ are reconstructed
following the procedures outlined in Section \ref{B-recon}.
The beam constrained mass distributions from CLEO II
are shown in Fig.~\ref{dpi}, while
the experimental branching ratios are given in Tables ~\ref{kh1}  and
\ref{kh2} .

To select $ \bar{B} \to D \rho^-$ candidates additional requirements are
imposed on the $\pi^-\pi^0$ invariant mass and the $\rho$ helicity angle.
The CLEO~II analysis requires $ |m(\pi^- \pi^0) - 770|< 150$~MeV and
$|\cos\Theta_{\rho}|>0.4$.
For the $B \to D \rho^-$ modes there are
events which are consistent with both $B \to D \rho^-$ and with
$ B \to D^{*} \pi^-$, followed by $ D^{*} \to D \pi^0$. These
events are removed from the $B \to D \rho^-$ sample using a cut on
the $D^{*} - D$ mass difference.
By fitting the $\pi^- \pi^0$ mass spectrum and the helicity angle distribution,
CLEO~II finds that
at least 97.5\% of the $B \to D \pi^-\pi^0$ rate is described by the
decay $B \to D \rho^-$\cite{mcdd}.
ARGUS\cite{ThirdB} also finds that the $\pi^- \pi^0$ mass spectrum
is consistent
with the dominance of $\rho$ production.

\subsection{Measurements of $D^*(n\pi)^-$ Final States}

We now consider final states containing a $D^*$ meson and one, two or
three pions. These include the $B \to D^* \pi^-$ ,
$B \to D^* \rho^-$, and $B \to D^* a_1^-$ decay channels.
The results for the decays $\bar{B^0} \to D^{*+} \pi^-$,
$\bar{B^0} \to D^{*+} \rho^-$ and $\bar{B^0} \to D^{*+} \pi^-\pi^-\pi^+$
are listed in Table~\ref{kh2}, and the results for
$B^- \to D^{*0} \pi^-$, $B^- \to D^{*0} \rho^-$ and
$B^- \to D^{*0} \pi^-\pi^-\pi^+$ are given in Table ~\ref{kh1}.

The CLEO II $B^-$ and $\bar{B}^0$ signals in the $D^* \pi$ and $D^* \rho$
decay channels are shown in Fig. \ref{dspi}.
They find that $B \to D^* \pi^-\pi^0$ is saturated by the
decay $B \to D^* \rho^-$ (Fig. \ref{subs}) and
set a tight upper limit of $<9$\% at
90\% C.L. on a possible non-resonant contribution \cite{mcdrho}.
This disagrees with an ARGUS analysis that finds about 50\%
of $\bar{B}^0 \to D^{*+} \pi^- \pi^0$ decays
do not contain a $\rho^-$ meson \cite{FifthB}.

The CLEO~II data suggest that the signal in
$B\to D^{*}\pi^-\pi^-\pi^+$ arises dominantly
from $B\to D^{*} a_1^-$.
Taking into account the $a_1 \to \pi^-\pi^-\pi^+$ branching fractions it
follows
that ${\cal{B}}(B\to D^{*} a_1^-) = 2 \times
{\cal{B}}(B\to D^{*}\pi^-\pi^-\pi^+)$.
In Fig.~\ref{FBmaone} we show the $M_B$ distributions
when the $\pi^-\pi^-\pi^+$ invariant mass is required to be in the interval
$1.0 <\pi^-\pi^-\pi^+ < 1.6$ GeV.
Fig.~\ref{Fmaonea} shows a fit to the
$\pi^-\pi^-\pi^+$ mass distributions with contributions from
$B \to D^{*+} a_1^-$ and a $B \to D^{*+}\pi^- \rho^0$ non-resonant background.
The $a_1$ meson has been parameterized
as a Breit-Wigner resonance shape
with $m_{a_1} = 1182 $ MeV and $\Gamma_{a1} = 466$ MeV.
This fit gives an upper limit of 13\% on
the non-resonant component in this decay.
This conclusion differs from
CLEO~1.5 which attributed $(35\pm 15\pm 8)$\% of their
$\bar{B^0} \to D^{*+} \pi^-\pi^-\pi^+$ signal to non-resonant
$\bar{B^0} \to D^{*+} \pi^-\rho^0$ decays \cite{anotherB}.
ARGUS also finds a significant non-$a_1$ component in this decay
but does not quote a quantitative result \cite{FifthB}.

\subsection{Polarization in $B \to D^{*+}\rho^-$ Decays}
\label{pol-D*-rho}

The sample of fully
reconstructed $ \bar{B^0} \to D^{*+}\rho^-$ decays from CLEO~II
has been used to  measure the $D^{*+}$ and $\rho^-$ polarizations.
By comparing the measured polarizations in $\bar{B^0} \to D^{*+}\rho^-$
with the expectation from the corresponding semileptonic
B decay a test of the factorization hypothesis
can be performed (see Sec.~\ref{fac-ang-cor}).
The polarization is obtained from the distributions of the helicity angles
$\Theta_{\rho}$ and $\Theta_{D^*}$. The $D^{*+}$ helicity angle,
$\Theta_{D^*}$, is the angle between the $D^0$ direction
in the $D^{*+}$ rest frame and the $D^{*+}$ direction
in the rest frame of the $B$ meson.
After integration over $\chi$, the angle between the
normals to the $D^{*+}$ and the
$\rho^-$ decay planes, the helicity angle distribution can be expressed
as:
\begin{equation}
{d^2\Gamma\over{d\cos\Theta_{D^*}d\cos\Theta_{\rho}}}
\propto
{1\over{4}}\sin^2\Theta_{D^*}\sin^2\Theta_{\rho}(|H_{+1}|^2+|H_{-1}|^2)
+\cos^2\Theta_{D^*}\cos^2\Theta_{\rho}|H_{0}|^2 \label{polar3d}
\end{equation}
The fraction of  longitudinal polarization is defined by
\begin{equation}
 {{\Gamma_L}\over{\Gamma}}
 ~ = ~ {{|H_0|^2}\over{|H_{+1}|^2 + |H_{-1}|^2 + |H_{0}|^2}} \label{ratiohel}
\end{equation}
If $\Gamma_L$ is large both
the $D^{*+}$ and the $\rho^{-}$ helicity angles will
follow a $\cos^{2}\Theta$ distribution, whereas a large transverse
polarization, $\Gamma_T$, gives a $\sin^2\Theta$ distribution for both
helicity angles.

To measure the polarization the helicity angle distributions in the $B$
signal region are corrected by
subtracting the distributions from a properly scaled mass sideband.
The resulting helicity angle distributions, corrected for efficiency,
are fitted to the functional form:
\begin{equation}
{d\Gamma\over{d\cos\Theta}} =
 N ~  \left[ \cos^2 \Theta ~ + ~{{1}\over{2}} {{\Gamma_{T}}\over{\Gamma}}
 (1 -3 \cos^2 \Theta) \right]. \label{fithel}
\end{equation}
This form is derived from the angular distribution given above. It
is well behaved for large longitudinal polarization.
{}From the fit to the $D^{*+}$ helicity angle distribution, they find
$\Gamma_{L}/\Gamma =(88 \pm 10) \% $, while a fit to
the $\rho$ helicity angle distribution gives
$\Gamma_{L}/\Gamma = (91\pm 10)\%$. The results of the
fit are shown in Fig.~\ref{helrho}(a) and (b).
As a consistency check they have verified that the $D^{*+}$ mesons in
$\bar{B}^0 \to D^{*+}\pi^-$ are completely longitudinally polarized, as
expected from angular momentum conservation (Fig. \ref{helrho}(c)).

The statistical errors can be reduced by taking
advantage of the correlation between the two helicity angles.
An unbinned two dimensional likelihood fit
to the joint $(\cos\Theta_{D^*}, \cos\Theta_{\rho})$ distribution gives
\begin{equation}
(\Gamma_{L}/\Gamma)_{\bar{B^0} \to D^{*+} \rho^-}\; =\; 90 \pm 7 \pm 5 \%
\end{equation}

\subsection{Measurements of $D^{**}$ Final States}
\label{B->D**}

In addition to the production of $D$ and $D^*$ mesons,
the charm quark and spectator antiquark can hadronize as a $D^{**}$ meson.
The $D^{**0}(2460)$ has been observed experimentally and identified
as the J$^P=2^+$ state, while the
$D^{**0}(2420)$ has been identified as the $1^+$ state. These states have
full widths of approximately 20 MeV. Two other states, a $0^+$ and another
$1^+$ are predicted but have not yet been observed, presumably because of their
large intrinsic widths.
There is evidence for $D^{**}$ production in semileptonic $B$
decays\cite{Dssin}, and $D^{**}$ mesons have also been seen in hadronic
decays. However, early experiments did not have sufficient data to
separate the two narrow $D^{**}$ states and hence reported branching
ratios only for the combination of the two (see results listed under
$B \to D_J^{(*)0}$ in Tables~\ref{kh1} -- \ref{kh4}).

In order to search for $D^{**}$ mesons from $B$ decays the
final states $B^- \to D^{*+} \pi^- \pi^-$ and
$B^- \to D^{*+} \pi^- \pi^- \pi^0$ are studied.
These decay modes are not expected to occur via
a spectator diagram in which the $c$ quark and the spectator
antiquark form a $D^*$ rather than a $D^{**}$ meson.
The $D^{*+}$ is combined with a $\pi ^-$ to form a $D^{**}$ candidate.
If the $D^{**}$ candidate is within one full width of the nominal mass of
either a $D^{**0}(2420)$ or a $D^{**0}(2460)$, it is
combined with a  $\pi^-$ or $\rho^-$ to form a
$B^-$ candidate.
CLEO~II has also looked for $D^{**}$ production in the channels
$B^-\to D^+ \pi^- \pi^-$ and $\bar{B^0}\to D^0 \pi^- \pi^+$.
Since $D^{**0}(2420)\to D \pi$ is forbidden, only the
$D^{**0}(2460)$ is searched for in the $D \pi \pi$ final state.

Fig.~\ref{dsspi} shows candidate $B$ mass distributions
obtained by  CLEO~II for the four combinations of
$D^{**0}(2460)$ or $D^{**0}(2420)$, and $\pi^-$ or $\rho^-$. In the
$D^{**0}(2420) \pi^-$ mode, there is
a significant signal of 8.5 events on a background of 1.5 events.
In this channel CLEO~II quotes the
branching ratio given in Table~\ref{kh1}, while
for the other three channels, they give upper limits.
ARGUS has also found evidence for $B \to D^{**}(2420) \pi^-$ using
a partial reconstruction technique in which they observe a fast and slow pion
from the $D^{**}$ decay but
do not reconstruct the $D^0$ meson\cite{Krieger}.

\begin{table}[htb]
\caption{$D_s$ decay channels used to reconstruct $B\to DD_s$ decays.}
\label{TDinf}
\hfill{
\begin{tabular}{ll}
ARGUS \cite{ARGUSDDs} & CLEO 1.5 \cite{DDcleo}\\ \hline
$D_s^+ \to \phi \pi^+ $& $D_s^+ \to \phi \pi^+ $\\
$D_s^+ \to \phi \pi^+ \pi^0 $&  \\
$D_s^+ \to \phi \pi^+ \pi^+ \pi^- $&  \\
$D_s^+ \to K_s K^+ $& $D_s^+ \to K_s K^+ $\\
$D_s^+ \to K_s K^{*+}$ & \\
$D_s^+ \to \bar{K}^{*0} K^+$ & $D_s^+ \to \bar{K}^{*0} K^+ $\\
$D_s^+ \to K^{*0} \bar{K}^{*}+ $& $D_s^+ \to \bar{K}^{*0} K^{*+}$ \\
$D_s^+ \to \eta ' \pi^+$ &
\end{tabular}
}
\hfill
\end{table}

\subsection{Exclusive Decays to $D$ and $D_s$ Mesons}
\label{doubledees}

Another important class of modes are decays to two charmed mesons.
As shown in Fig. ~\ref{Fcsdiag} (a)
the production of an isolated pair of charmed mesons
($D_s^{(*)}$ and $D^{(*)}$) proceeds through a Cabibbo favored
spectator diagram in which
the $s\overline{c}$ pair from the virtual $W^-$ hadronizes into a
$D_s^-$ or a $D_s^{*-}$ meson and the remaining spectator quark and the
$c$ quark form a $D^{(*)}$ meson.
These modes have been observed by the CLEO~1.5\cite{DDcleo}
and ARGUS\cite{ARGUSDDs} experiments.
The decay channels listed in Table~\ref{TDinf} are used to form
$D_s$ meson candidates.
B mesons are then reconstructed in eight decay modes:
$D_s^-D^+$, $D_s^-D^0$,
$D_s^{*-}D^+$, $D_s^{*-}D^0$,
$D_s^-D^{*+}$, $D_s^-D^{*0}$,
$D_s^{*-}D^{*+}$, and $D_s^{*-}D^{*0}$.
The results of the ARGUS experiment are shown in Fig. \ref{Fargusdd}.
\begin{figure}[htb]
\unitlength 1.0in
\vskip 10 mm
\begin{center}
\begin{picture}(3.0,3.0)(0.0,0.0)
\end{picture}
\vskip 15 mm
\caption[]{
Distributions of the efficiency corrected
$\psi$ and $K^*$ helicity angles in
$B \to \psi K^*$ decays from CLEO~II.
The overlaid smooth curves are projections of the
unbinned maximum likelihood fit described in the text.}
\label{expol}
\end{center}
\end{figure}

\begin{figure}[htb]
\unitlength 1.0in
\begin{center}
\begin{picture}(3.0,3.0)(0.0,0.0)
\end{picture}
\bigskip
\bigskip
\caption{Beam constrained mass distributions from CLEO~II
for (a) $B^-$ events and (b) $\bar{B^0}$ events.}
\label{FBM}
\end{center}
\end{figure}

\begin{figure}[htb]
\unitlength 1.0in
\vskip 10 mm
\begin{center}
\begin{picture}(2.5,2.2)(0.0,0.0)
\end{picture}
\vskip 15 mm
\caption[]{(a) The $\psi K^+$ mass distribution
from the CDF experiment
 (b)  The $\psi K^{*0}$ mass
distribution from the CDF experiment. The solid
line indicates the fitted region.}\label{cdfbd}
\end{center}
\end{figure}

\begin{figure}[htb]
\unitlength 1.0in
\vskip 10 mm
\begin{center}
\begin{picture}(2.5,2.2)(0.0,0.0)
\end{picture}
\vskip 15 mm
\caption[]{(a) The $\psi K^+ K^-$ mass distribution
from the CDF experiment for
 $K^+ K^-$ mass within 10 MeV$/c^2$ of the $\phi$ mass (solid) and
for the normalized $\phi$ sideband region (b)  The $K^+ K^-$ mass
distribution for $\psi K^+ K^-$ combinations within 20 MeV$/c^2$
of 5380 MeV$/c^2$.}\label{cdfbs}
\end{center}
\end{figure}

\begin{table}[htb]
\caption{Upper limits (90\% C.L) on color suppressed $B$ decays.}\label{Tbrcol}
\begin{tabular}{lcc}
Decay Mode &  Events &   U. L. (\%)  \\ \hline
$\bar{B^0} \to D^{0} \pi^0$    & $<20.7  $   & $<0.048$   \\
$\bar{B^0} \to D^{0} \rho^0$   & $<19.0$     & $<0.055$   \\
$\bar{B^0} \to D^{0} \eta$     & $<9.5$      & $<0.068$    \\
$\bar{B^0} \to D^{0} \eta^{'}$ & $<3.5 $     & $<0.086$   \\
$\bar{B^0} \to D^{0} \omega $  & $<12.7 $     & $<0.063$    \\
$\bar{B^0} \to D^{*0} \pi^0$   & $<11.0 $     & $<0.097$   \\
$\bar{B^0} \to D^{*0} \rho^0$  & $<8.1$      & $<0.117$    \\
$\bar{B^0} \to D^{*0} \eta$    & $<2.3 $     & $<0.069$   \\
$\bar{B^0} \to D^{*0} \eta^{'}$  & $<2.3 $   & $<0.27$    \\
$\bar{B^0} \to D^{*0} \omega$    & $<9.0 $     & $<0.21$
\end{tabular}
\end{table}

\begin{table}[htb]
\caption{Upper limits on ratios of branching fractions
for color suppressed to normalization modes.}\label{Tratcol}
\begin{tabular}{cc}
Ratio of Branching Ratios & CLEO~II (90\% C.L.)  \\ \hline
${\cal B}(\bar{B^0} \to D^0 \pi^0)/{\cal B}(B^- \to D^0 \pi^-)$
                                                    & $< 0.09$   \\
${\cal B}(\bar{B^0} \to D^0 \rho^0)/{\cal B}(B^- \to D^0 \rho^-)$
                                                    & $< 0.05 $   \\
${\cal B}(\bar{B^0} \to D^0 \eta)/{\cal B}(B^- \to D^0 \pi^-)$
                                                    & $< 0.12 $   \\
${\cal B}(\bar{B^0} \to D^0 \eta^{'})/{\cal B}(B^- \to D^0 \pi^-)$
                                                    & $< 0.16 $   \\
${\cal B}(\bar{B^0} \to D^0 \omega)/{\cal B}(B^- \to D^0 \rho^-)$
                                                    & $< 0.05 $   \\
${\cal B}(\bar{B^0} \to D^{*0}\pi^0)/{\cal B}(B^- \to D^{*0} \pi^-)$
                                                    & $< 0.20 $   \\
${\cal B}(\bar{B^0} \to D^{*0} \rho^0)/{\cal B}(B^- \to D^{*0} \rho^-)$
                                                    & $< 0.07 $   \\
${\cal B}(\bar{B^0} \to D^{*0} \eta)/{\cal B}(B^- \to D^{*0} \pi^-)$
                                                    & $< 0.14 $   \\
${\cal B}(\bar{B^0} \to D^{*0} \eta^{'})/{\cal B}(B^-\to D^{*0}\pi^-)$
                                                    & $< 0.54 $   \\
${\cal B}(\bar{B^0} \to D^{*0} \omega)/{\cal B}(B^- \to D^{*0} \rho^-)$
                                                    & $< 0.09 $
\end{tabular}
\end{table}

\begin{table}[htb]
\caption{Measurements of the $\bar{B^0}$ and $B^-$  Masses [MeV].}
\label{Tmabs}
\begin{tabular}{lccc}
Experiment& $M_{\bar{B^0}}$ & $M_{B^-}$ & $M_{\bar{B^0}}-M_{B^-}$ \\ \hline
ARGUS   & $5279.6\pm 0.7 \pm 2.0 $  & $ 5280.5\pm 1.0 \pm 2.0$ &
$-0.9\pm 1.2 \pm 0.5$  \\
CLEO 87 & $5278.0 \pm 0.4 \pm 2.0$  & $5278.3 \pm 0.4 \pm 2.0$ &
$-0.4\pm 0.6 \pm 0.5$  \\
CLEO 93 & $5279.2 \pm 0.2 \pm 2.0$  & $5278.8 \pm 0.2 \pm 2.0$ &
$0.41\pm 0.25 \pm 0.19$  \\ \hline
Average & $5278.9 \pm 0.2 \pm 2.0   $ & $5278.7 \pm 0.2 \pm 2.0$ & $  0.2\pm
0.3$
\end{tabular}
\end{table}

\begin{table}[htb]
\caption{$B^-$ Branching fractions [\%]}
\label{kh1}
 \begin{tabular}{llll}
Mode & ARGUS & CLEO 1.5 & CLEO II \\
\hline
 $B^- \rightarrow D^0 \pi ^-$ & $ 0.22 \pm 0.09  \pm 0.06  \pm 0.01 $ &
$ 0.56 \pm 0.08  \pm 0.05  \pm 0.02 $ &
$ 0.55 \pm 0.04  \pm 0.05  \pm 0.02  $\\
 $B^- \rightarrow D^0 \rho ^-$ & $ 1.45 \pm 0.45  \pm 0.41  \pm 0.06 $ &
 &
$ 1.35 \pm 0.12  \pm 0.14  \pm 0.04  $\\
 $B^- \rightarrow D^{0} \pi ^+ \pi ^- \pi ^-$ &  &
$ 1.24 \pm 0.31  \pm 0.14  \pm 0.05 $ &
 \\
 $B^- \rightarrow D^{*0} \pi ^-$ & $ 0.39 \pm 0.14  \pm 0.10  \pm 0.02 $ &
$ 0.99 \pm 0.25  \pm 0.17  \pm 0.04 $ &
$ 0.52 \pm 0.07  \pm 0.07  \pm 0.03  $\\
 $B^- \rightarrow D^{*0} \rho ^-$ & $ 0.96 \pm 0.58  \pm 0.36  \pm 0.04 $ &
 &
$ 1.68 \pm 0.21  \pm 0.27  \pm 0.07  $\\
 $B^- \rightarrow D_J^{(*)0} \pi ^-$ & $ 0.13 \pm 0.07  \pm 0.03  \pm 0.01 $ &
$ 0.13 \pm 0.07  \pm 0.01  \pm 0.01 $ &
 \\
 $B^- \rightarrow D^{*+} \pi ^- \pi ^- \pi ^0$ & $ 1.68 \pm 0.65  \pm 0.38  \pm
0.07 $ &
 &
 \\
 $B^- \rightarrow D_J^{(*)0} \rho ^-$ & $ 0.33 \pm 0.20  \pm 0.08  \pm 0.01 $ &
 &
 \\
 $B^- \rightarrow D^{*0} \pi ^- \pi ^- \pi ^+$ &  &
 &
$ 0.94 \pm 0.20  \pm 0.17  \pm 0.02  $\\
 $B^- \rightarrow D^{*0} a_1 ^-$ &  &
 &
$ 1.88 \pm 0.40  \pm 0.34  \pm 0.04  $\\
 $B^- \rightarrow D^+ \pi^- \pi ^- $ &  &
 &
$ <0.14 $  (90\% C.L.)\\
  $B^- \rightarrow D^{*+} \pi ^- \pi ^-$ & $ 0.24 \pm 0.13  \pm 0.05  \pm 0.01
$ &
$ <0.37$  (90\% C.L.)&
 $ 0.19 \pm 0.07  \pm 0.03  \pm 0.01  $\\
 $B^- \rightarrow D^{**0}(2420) \pi^- $ & $ 0.30 \pm 0.08  \pm 0.06  \pm 0.01 $
&
 &
$ 0.11 \pm 0.05  \pm 0.02  \pm 0.01  $\\
 $B^- \rightarrow D^{**0}(2420) \rho^- $ &  &
 &
$ <0.14 $  (90\% C.L.)\\
  $B^- \rightarrow D^{**0}(2460) \pi^- $ &  &
 &
$ <0.13 $  (90\% C.L.)\\
  $B^- \rightarrow D^{**0}(2460) \rho^- $ &  &
 &
$ <0.47 $  (90\% C.L.)\\
  $B^- \rightarrow D^0 D_s^-$ & $ 1.65 \pm 0.82  \pm 0.42  \pm 0.07 $ &
$ 1.66 \pm 0.70  \pm 0.71  \pm 0.07 $ &
 \\
 $B^- \rightarrow D^0 D_s^{*-}$ & $ 1.10 \pm 0.82  \pm 0.30  \pm 0.04 $ &
 &
 \\
 $B^- \rightarrow D^{*0} D_s^-$ & $ 0.77 \pm 0.53  \pm 0.18  \pm 0.03 $ &
 &
 \\
 $B^- \rightarrow D^{*0} D_s^{*-}$ & $ 1.84 \pm 0.95  \pm 0.44  \pm 0.07 $ &
 &
 \\
 $B^- \rightarrow \psi K^-$ & $ 0.08 \pm 0.04  \pm 0.01 $ &
$ 0.09 \pm 0.02  \pm 0.02 $ &
$ 0.11 \pm 0.01  \pm 0.01  $\\
 $B^- \rightarrow \psi ' K^-$ & $ 0.20 \pm 0.09  \pm 0.04 $ &
$ <0.05$  (90\% C.L.)&
 $ 0.06 \pm 0.02  \pm 0.01  $\\
 $B^- \rightarrow \psi K^{*-}$ & $ 0.19 \pm 0.13  \pm 0.03 $ &
$ 0.15 \pm 0.11  \pm 0.03 $ &
$ 0.18 \pm 0.05  \pm 0.02  $\\
 $B^- \rightarrow \psi ' K^{*-}$ & $ <0.53 $ (90\% C.L.)&
 $ <0.38$  (90\% C.L.)&
 $ <0.30 $  (90\% C.L.)\\
  $B^- \rightarrow \psi K^- \pi ^+ \pi ^-$ & $ <0.19 $ (90\% C.L.)&
 $ 0.14 \pm 0.07  \pm 0.03 $ &
 \\
 $B^- \rightarrow \psi ' K^- \pi ^+ \pi ^-$ & $ 0.21 \pm 0.12  \pm 0.04 $ &
 &
 \\
 $B^- \rightarrow \chi_{c1} K^-$ & $ 0.22 \pm 0.15  \pm 0.07 $ &
 &
$ 0.10 \pm 0.04  \pm 0.01  $\\
 $B^- \rightarrow \chi_{c1} K^{*-}$ &  &
 &
$ <0.21 $  (90\% C.L.)\\
 \end{tabular}
\end{table}
\begin{table}[htb]
\caption{$\bar{B}^0$ Branching fractions in [\%]}
\label{kh2}
 \begin{tabular}{llll}
Mode & ARGUS & CLEO 1.5 & CLEO II \\
\hline
 $\bar{B}^0 \rightarrow D^+ \pi ^-$ & $ 0.48 \pm 0.11  \pm 0.08  \pm 0.07 $ &
$ 0.27 \pm 0.06  \pm 0.03  \pm 0.04 $ &
$ 0.29 \pm 0.04  \pm 0.03  \pm 0.05  $\\
 $\bar{B}^0 \rightarrow D^+ \rho ^-$ & $ 0.90 \pm 0.50  \pm 0.27  \pm 0.14 $ &
 &
$ 0.81 \pm 0.11  \pm 0.12  \pm 0.13  $\\
 $\bar{B}^0 \rightarrow D^+ \pi ^- \pi ^- \pi ^+$ &  &
$ 0.80 \pm 0.21  \pm 0.09  \pm 0.12 $ &
 \\
 $\bar{B}^0 \rightarrow D^{*+} \pi ^-$ & $ 0.26 \pm 0.08  \pm 0.04  \pm 0.01 $
&
$ 0.44 \pm 0.11  \pm 0.05  \pm 0.02 $ &
$ 0.26 \pm 0.03  \pm 0.04  \pm 0.01  $\\
 $\bar{B}^0 \rightarrow D^{*+} \rho ^-$ & $ 0.65 \pm 0.28  \pm 0.26  \pm 0.03 $
&
$ 2.11 \pm 0.89  \pm 1.23  \pm 0.08 $ &
$ 0.74 \pm 0.10  \pm 0.14  \pm 0.02  $\\
 $\bar{B}^0 \rightarrow D^{*+} \pi ^- \pi ^- \pi ^+$ & $ 1.12 \pm 0.28  \pm
0.33  \pm 0.04 $ &
$ 1.76 \pm 0.31  \pm 0.29  \pm 0.07 $ &
$ 0.63 \pm 0.10  \pm 0.11  \pm 0.02  $\\
 $\bar{B}^0 \rightarrow D^{*+} a_1^-$ &  &
 &
$ 1.26 \pm 0.20  \pm 0.22  \pm 0.03  $\\
 $\bar{B}^0 \rightarrow D^{0} \pi ^+ \pi^- $ &  &
 &
$ <0.16 $  (90\% C.L.)\\
  $\bar{B}^0 \rightarrow D^{**+}(2460) \pi^- $ &  &
 &
$ <0.22 $  (90\% C.L.)\\
  $\bar{B}^0 \rightarrow D^{**+}(2460) \rho^- $ &  &
 &
$ <0.49 $  (90\% C.L.)\\
  $\bar{B}^0 \rightarrow D^+ D_s^-$ & $ 1.09 \pm 0.83  \pm 0.44  \pm 0.16 $ &
$ 0.58 \pm 0.33  \pm 0.24  \pm 0.09 $ &
 \\
 $\bar{B}^0 \rightarrow D^+ D_s^{*-}$ & $ 1.73 \pm 1.09  \pm 0.67  \pm 0.26 $ &
 &
 \\
 $\bar{B}^0 \rightarrow D^{*+} D_s^-$ & $ 0.80 \pm 0.57  \pm 0.22  \pm 0.03 $ &
$ 1.17 \pm 0.66  \pm 0.52  \pm 0.05 $ &
 \\
 $\bar{B}^0 \rightarrow D^{*+} D_s^{*-}$ & $ 1.49 \pm 0.80  \pm 0.43  \pm 0.06
$ &
 &
 \\
 $\bar{B}^0 \rightarrow \psi K^0$ & $ 0.09 \pm 0.07  \pm 0.02 $ &
$ 0.07 \pm 0.04  \pm 0.02 $ &
$ 0.08 \pm 0.02  \pm 0.01  $\\
 $\bar{B}^0 \rightarrow \psi ' K^0$ & $ <0.31 $ (90\% C.L.)&
 $ <0.16$  (90\% C.L.)&
 $ <0.08 $  (90\% C.L.)\\
  $\bar{B}^0 \rightarrow \psi \bar{K}^{*0}$ & $ 0.13 \pm 0.06  \pm 0.02 $ &
$ 0.13 \pm 0.06  \pm 0.03 $ &
$ 0.17 \pm 0.03  \pm 0.02  $\\
 $\bar{B}^0 \rightarrow \psi ' \bar{K}^{*0}$ & $ <0.25 $ (90\% C.L.)&
 $ 0.15 \pm 0.09  \pm 0.03 $ &
$ <0.19 $  (90\% C.L.)\\
  $\bar{B}^0 \rightarrow \psi K^{-} \pi ^+$ &  &
$ 0.12 \pm 0.05  \pm 0.03 $ &
 \\
 $\bar{B}^0 \rightarrow \psi ' K^- \pi ^+$ & $ <0.11 $ (90\% C.L.)&
  &
 \\
 $\bar{B}^0 \rightarrow \chi_{c1} K^0$ &  &
 &
$ <0.27 $  (90\% C.L.)\\
  $\bar{B}^0 \rightarrow \chi_{c1} \bar{K}^{*0}$ &  &
 &
$ <0.21 $  (90\% C.L.)\\
 \end{tabular}
\end{table}
\begin{table}[htb]
\caption{World average $B^-$ branching fractions [\%]}
\label{kh3}
\begin{tabular}{ll}
Mode & Branching Fraction \\
\hline
$B^- \rightarrow D^0 \pi ^-$ & $0.49 \pm 0.05 \pm 0.02 $ \\
$B^- \rightarrow D^0 \rho ^-$ & $1.36 \pm 0.18 \pm 0.05 $ \\
$B^- \rightarrow D^{0} \pi ^+ \pi ^- \pi ^-$ & $1.24 \pm 0.34 \pm 0.05 $ \\
$B^- \rightarrow D^{*0} \pi ^-$ & $0.52 \pm 0.08 \pm 0.02 $ \\
$B^- \rightarrow D^{*0} \rho ^-$ & $1.54 \pm 0.31 \pm 0.06 $ \\
$B^- \rightarrow D_J^{(*)0} \pi ^-$ & $0.13 \pm 0.05 \pm 0.01 $ \\
$B^- \rightarrow D^{*+} \pi ^- \pi ^- \pi ^0$ & $1.68 \pm 0.76 \pm 0.07 $ \\
$B^- \rightarrow D_J^{(*)0} \rho ^-$ & $0.33 \pm 0.21 \pm 0.01 $ \\
$B^- \rightarrow D^{*0} \pi ^- \pi ^- \pi ^+$ & $0.94 \pm 0.26 \pm 0.04 $ \\
$B^- \rightarrow D^{*0} a_1 ^-$ & $1.88 \pm 0.52 \pm 0.08 $ \\
$B^- \rightarrow D^+ \pi^- \pi ^- $ & $<0.14 $  (90\% C.L.)\\
$B^- \rightarrow D^{*+} \pi ^- \pi ^-$ & $0.20 \pm 0.07 \pm 0.01 $ \\
$B^- \rightarrow D^{**0}(2420) \pi^- $ & $0.16 \pm 0.05 \pm 0.01 $ \\
$B^- \rightarrow D^{**0}(2420) \rho^- $ & $<0.14 $  (90\% C.L.)\\
$B^- \rightarrow D^{**0}(2460) \pi^- $ & $<0.13 $  (90\% C.L.)\\
$B^- \rightarrow D^{**0}(2460) \rho^- $ & $<0.47 $  (90\% C.L.)\\
$B^- \rightarrow D^0 D_s^-$ & $1.65 \pm 0.68 \pm 0.07 $ \\
$B^- \rightarrow D^0 D_s^{*-}$ & $1.10 \pm 0.88 \pm 0.04 $ \\
$B^- \rightarrow D^{*0} D_s^-$ & $0.77 \pm 0.56 \pm 0.03 $ \\
$B^- \rightarrow D^{*0} D_s^{*-}$ &$ 1.84 \pm 1.05 $ \\
$B^- \rightarrow \psi K^-$ &$ 0.10 \pm 0.01 $ \\
$B^- \rightarrow \psi ' K^-$ &$ 0.07 \pm 0.02 $ \\
$B^- \rightarrow \psi K^{*-}$ &$ 0.17 \pm 0.05 $ \\
$B^- \rightarrow \psi ' K^{*-}$ & $<0.53 $  (90\% C.L.)\\
$B^- \rightarrow \psi K^- \pi ^+ \pi ^-$ &$ 0.14 \pm 0.08 $ \\
$B^- \rightarrow \psi ' K^- \pi ^+ \pi ^-$ &$ 0.21 \pm 0.13 $ \\
$B^- \rightarrow \chi_{c1} K^-$ &$ 0.10 \pm 0.04 $ \\
$B^- \rightarrow \chi_{c1} K^{*-}$ & $<0.21 $  (90\% C.L.)\\
\end{tabular}
\end{table}
\begin{table}[htb]
\caption{World average $\bar{B}^0$ branching fractions [\%]}
\label{kh4}
 \begin{tabular}{ll}
Mode & Branching Fraction \\
\hline
$\bar{B}^0 \rightarrow D^+ \pi ^-$ & $0.30 \pm 0.04 \pm 0.04 $ \\
$\bar{B}^0 \rightarrow D^+ \rho ^-$ & $0.82 \pm 0.16 \pm 0.12 $ \\
$\bar{B}^0 \rightarrow D^+ \pi ^- \pi ^- \pi ^+$ & $0.80 \pm 0.23 \pm 0.12 $ \\
$\bar{B}^0 \rightarrow D^{*+} \pi ^-$ & $0.28 \pm 0.04 \pm 0.01 $ \\
$\bar{B}^0 \rightarrow D^{*+} \rho ^-$ & $0.74 \pm 0.16 \pm 0.03 $ \\
$\bar{B}^0 \rightarrow D^{*+} \pi ^- \pi ^- \pi ^+$ & $0.79 \pm 0.13 \pm 0.03 $
\\
$\bar{B}^0 \rightarrow D^{*+} a_1^-$ & $1.26 \pm 0.30 \pm 0.05 $ \\
$\bar{B}^0 \rightarrow D^{0} \pi ^+ \pi^- $ & $<0.16 $  (90\% C.L.)\\
$\bar{B}^0 \rightarrow D^{**+}(2460) \pi^- $ & $<0.22 $  (90\% C.L.)\\
$\bar{B}^0 \rightarrow D^{**+}(2460) \rho^- $ & $<0.49 $  (90\% C.L.)\\
$\bar{B}^0 \rightarrow D^+ D_s^-$ & $0.66 \pm 0.38 \pm 0.10 $ \\
$\bar{B}^0 \rightarrow D^+ D_s^{*-}$ & $1.73 \pm 1.28 \pm 0.26 $ \\
$\bar{B}^0 \rightarrow D^{*+} D_s^-$ & $0.93 \pm 0.50 \pm 0.04 $ \\
$\bar{B}^0 \rightarrow D^{*+} D_s^{*-}$ & $1.49 \pm 0.91 \pm 0.06 $ \\
$\bar{B}^0 \rightarrow \psi K^0$ &$ 0.08 \pm 0.02 $ \\
$\bar{B}^0 \rightarrow \psi ' K^0$ & $<0.31 $  (90\% C.L.)\\
$\bar{B}^0 \rightarrow \psi \bar{K}^{*0}$ &$ 0.15 \pm 0.03 $ \\
$\bar{B}^0 \rightarrow \psi ' \bar{K}^{*0}$ &$ 0.15 \pm 0.09 $ \\
$\bar{B}^0 \rightarrow \psi K^{-} \pi ^+$ &$ 0.12 \pm 0.06 $ \\
$\bar{B}^0 \rightarrow \psi ' K^- \pi ^+$ & $<0.11 $  (90\% C.L.)\\
$\bar{B}^0 \rightarrow \chi_{c1} K^0$ & $<0.27 $  (90\% C.L.)\\
$\bar{B}^0 \rightarrow \chi_{c1} \bar{K}^{*0}$ & $<0.21 $  (90\% C.L.)\\
\end{tabular}
\end{table}

Improvements in the size of the signals for these modes are expected from
CLEO~II
which has measured additional $D_s$ modes with neutrals (e.g. $D_s\to \eta \pi,
\eta\rho$)\cite{CLEOds}.
 Partial reconstruction techniques
are also being
investigated for $B\to D^{*} D_s^{*}$ .

\section{COLOR SUPPRESSED B DECAY}
\label{B->psi-K(*)}

\subsection{Exclusive $B$ Decays to Charmonium}
\label{intro-B->psi-K(*)}

In $B$ decays to charmonium the $c$ quark from the
$b$ combines with a $\bar{c}$ quark from the virtual $W^-$ to
form a charmonium state. This process is described by the color suppressed
diagram shown in Fig.~\ref{Fcsdiag}(b).
By comparing $B$ meson decays to different final states with
charmonium mesons the dynamics of this decay mechanism can be investigated.

The decay modes $\bar{B^0} \to \psi K^0$ and  $\bar{B^0} \to \psi' K^0$ are
of special interest since the final states are
CP eigenstates. These decays are of great importance
for the investigation of
one of the three CP violating angles accessible to study in $B$ decays.
It is also possible to use the decay
$\bar{B^0} \to \psi K^{*0}$, $K^{*0} \to K^0 \pi^0$ which has a
somewhat higher branching ratio, but this final state
consists of a mixture of CP eigenstates.
It has even CP if the
orbital angular momentum L is 0 or 2 and odd CP for L=1.
If both CP states are present the CP asymmetry will be diluted.
A measurement of CP violation in this channel is only possible if one of the
CP states dominates, or if a detailed moments analysis of the various decay
components is performed \cite{Idunit}.
Recent measurements of the polarization in the decay $\bar{B^0}
\to\psi \bar{K^{*0}}$  allow us to determine the
fractions of the two CP states.

B meson candidates are formed by combining a charmonium
and a strange meson candidate. CLEO~1.5 and ARGUS have observed
signals for some of these modes.
Using the procedures outlined in Sec.~\ref{B-recon} the
beam constrained mass distributions shown
in Fig.~\ref{bpsik} and Fig.~\ref{bpsipk} are obtained by CLEO~II.
The branching ratios are listed in Tables~\ref{kh1}
and \ref{kh2} . Recently, CDF has reported signals
for $B\to \psi K^{*0}$ and $B\to \psi K^-$ (see Fig.~\ref{cdfbd}).

The ratio of vector to pseudoscalar meson production
\begin{equation}
{{\cal B}(B \to \psi K^*)\over{{\cal B}
(B \to \psi K)}} = 1.64 \pm 0.34 .
\end{equation}
can be calculated using factorization and the ratio of the
$B\to K^*$ and $B\to K$ form factors. However, it
is not certain that factorization is a good approximation for the
color suppressed diagram. The revised BSW model\cite{Neubie}
predicts a value of 1.61 for this ratio, which is close to the
experimental value. Another test is the ratio
\begin{equation}
{{\cal B}(B \to \psi' K^*)\over{{\cal B}
(B \to \psi' K)}} = 1.9 \pm 1.1.
\end{equation}
This can be compared to the revised BSW model which predicts 1.85
for this ratio.
Evidence for the decay mode $B\to \chi_{c} K$ has been reported by CLEO~II
and  ARGUS. The average branching fraction is
${\cal B}(B^-\to \chi_c K^-) = (0.133\pm 0.065) \%$.

\subsection{Polarization in $B\to\psi K^*$ }

The polarization in $B\to\psi K^*$ is studied using the methods described
for the
$\bar{B^0}\to D^{*+}\rho^-$ polarization measurement
in Section \ref{pol-D*-rho}.
After integration over the azimuthal angle between the $\psi$ and the
$K^*$ decay planes, the angular distribution in $B \to \psi K^*$ decays
can be written as
\begin{equation}
 {d^2\Gamma\over{d\cos\Theta_{\psi}d\cos\Theta_{K^*}}}
\propto {1\over{4}}\sin^2\Theta_{K^*}
(1+\cos^2\Theta_{\psi})(|H_{+1}|^2+|H_{-1}|^2)
+\cos^2\Theta_{K^*}\sin^2\Theta_{\psi}|H_{0}|^2 , \label{psipolar}
\end{equation}
where the $K^*$ helicity angle $\Theta_{K^*}$ is the angle between
the kaon direction in the $K^*$ rest frame and the $K^*$ direction in the
$B$ rest frame and $\Theta_{\psi}$ is the corresponding $\psi$ helicity angle,
and $H_{\pm1,0}$ are the helicity amplitudes.
The fraction of longitudinal polarization in $B \to \psi K^*$
is determined by an unbinned fit to the $\psi$ and $K^*$ helicity angle
distributions. CLEO~II finds
\begin{equation}
{\left({\Gamma_L\over{\Gamma}}\right)}_{B \to \psi K^*} = 0.84
\pm 0.06 \pm 0.08
\end{equation}
The efficiency corrected distributions in
each of the helicity angles $\cos\Theta_{\psi}$ and $\cos\Theta_{K^*}$
are shown in Fig.~\ref{expol}.

ARGUS has also determined the polarization in
$B \to \psi K^*$ decays and found that their data are consistent
with 100\% longitudinal polarization\cite{argpol}. They find
$\Gamma_L/\Gamma \:>\: 78$\% (90\% C.L.).
These results can be compared to the theoretical predictions of
Kr\"amer and Palmer\cite{Krampalm}
which again depend on the assumption of factorization and on the
unmeasured $B\to K^*$ form factor. Using the BSW model to estimate the form
factor, they find $\Gamma_{L}/\Gamma= 0.57$. Using HQET to extrapolate
from the E691 measurements of the $D\to K^*$ form factor, they obtain
$\Gamma_{L}/\Gamma=0.73$.

Although the decay mode $B \to \psi K^*$  may not be completely
polarized,  it is still dominated by a single CP eigenstate.
This mode will therefore be useful for measurements of CP violation.

\subsection{Exclusive Decays to a $D^{0 (*)}$ and a Neutral Meson.}
\label{color-supress}

We now discuss searches for $B$ decays which can occur
via an internal W-emission graph but which
do not yield charmonium mesons in the final
state. Naively, one expects that
these decays will be suppressed relative to decays which
occur via the external W-emission graph.
For the internal graph, in the absence
of gluons, the colors of the quarks from the virtual $W$ must
match the colors of the $c$ quark
and the accompanying spectator antiquark.
In this simple picture, one expects that the suppression
factor should be  $1/18$ for decays involving $\pi^0$, $\rho^0$
and $\omega$ mesons\cite{Dpi}.
In heavy quark decays the effects
of gluons cannot be neglected, and QCD based calculations
\cite{Neubie} predict suppression factors of order $1/50$.
If color suppressed $B$ decay modes are not greatly suppressed
then these modes could also be useful for CP violation studies \cite{Dunietz}.

CLEO~II has searched for color suppressed decay modes of $B$ mesons which
contain a single $D^0$ or $D^{*0}$ meson in the final state\cite{wex}.
The relevant color suppressed modes are listed in Table~\ref{Tbrcol}.
The decay channels used are $\eta \to \gamma \gamma$,
$\omega \to \pi^+ \pi^- \pi^0$ and $\eta^{'} \to
\eta \pi^+ \pi^-$, followed by $\eta \to \gamma \gamma$\cite{BReta}.
For decays of a pseudoscalar meson into a final state containing a
pseudoscalar and a vector meson (V), a helicity angle cut of
$|\cos \Theta_{V}| \; > \; 0.4$ is used\cite{omeg}.
No signals were observed.
Upper limits \cite{PDGul} on the
branching ratios for color suppressed modes are given in Table~\ref{Tbrcol}.
Upper limits on
the ratios of color suppressed modes to normalization modes are given in
Table~\ref{Tratcol}.
These limits show that there is color suppression of these $B$ decay modes.

\section{B MESON MASSES}
\label{mass-diff}

\subsection{Masses of the $\bar{B^0}$ and $B^-$ Mesons.}

We now discuss measurements of the $\bar{B^0}$ and $B^-$ masses
and the mass difference between them.
For these analyses only fully reconstructed $B$ decays in modes with good
signal to background are used. As an example,
CLEO~II uses the modes $B^-\to\psi K^-$,
$\bar{B^0}\to\psi K^{*0}$,
$B^-\to D^0\pi^-$, $B^-\to D^0\rho^-$,
$B^-\to D^{*0}\pi^-$, $B^-\to D^{*0}\rho^-$,
$\bar{B^0}\to D^+\pi^-$, $\bar{B^0}\to D^+\rho^-$,
$\bar{B^0}\to D^{*+}\pi^-$, and $\bar{B^0}\to D^{*+}\rho^-$.
With tight cuts,
there are 362 $B^-$ and 340 $B^0$ candidates reconstructed.
The $M_B$ distributions for the sum of these modes are
shown in Fig.~\ref{FBM}.

The absolute values of the $B^-$ and $\bar{B^0}$ masses are limited in
accuracy by the knowledge of the beam energy.
A correction of (-1.1$\pm$0.5) MeV is made for initial state radiation
as described in Ref.~\cite{Cdsr}.
The systematic error from the uncertainty in the absolute value of the
CESR/DORIS energy scale is determined by calibrating to the known
$\Upsilon (1S)$ mass.
The mass difference is determined more accurately than the masses themselves,
because the beam energy uncertainty cancels, as do many systematic errors
associated with the measurement errors on the charged tracks and neutral pions.
There are several models which predict the isospin mass
difference\cite{MDone}, which give values between 1.2 and 2.3 MeV
which are larger than the experimental results given in Table \ref{Tmabs}.
However, papers by Goity and Hou and by Lebed \cite{quatsch}
discuss models that can lead to small values of the mass difference.
That the $\bar{B^0} - B^-$ mass difference is much smaller than the
corresponding mass differences
 in the $K$ and $D$ mesons is somewhat surprising.

\subsection{Measurement of the $B_s$ Mass}
\label{Bs-mass}

First evidence for exclusive $B_s$ decays has recently been reported
by the CDF\cite{cdfbs}, OPAL\cite{opalbs}
and ALEPH collaborations\cite{alephbs} .
CDF observes a signal of $14.0\pm 4.7$ events
in the $B_s\to \psi \phi$ mode (see Fig.~\ref{cdfbs})
and determines
the $B_s$ mass to be $5383.3 \pm 4.5 \pm 5.0$~MeV.
ALEPH finds two unambiguous $B_s$ events
in the $B_s\to D_s^- \pi^+$ and  $B_s \to \psi' \phi$
modes and obtains
a mass of $5368.6\pm 5.6 \pm 1.5$~MeV.
OPAL finds one $B_s$ candidate in the $\psi\phi$ mode with a mass of
$5360 \pm 70$ MeV.
By reconstructing
 exclusive $B^-$ and $\bar{B}^0$ decays (see  Fig.~\ref{cdfbd}),
the high energy experiments calibrate their $B_s$ measurements
relative to the known $B^-$ and $\bar{B}^0$ masses.
The three $B_s$ mass measurements are consistent
with each other and with a non-relativistic quark model
prediction of a mass in the range $5345-5388$~MeV\cite{Kwros}.
In the near future, the $B_s$ mass will be measured more precisely,
and it is expected that exclusive $\Lambda_b$ decays will also be
reconstructed.

\section{THEORETICAL INTERPRETATION OF HADRONIC B DECAY}

\subsection{Introduction}

The simple spectator diagram for two-body hadronic $B$ meson decays
that occur through the Cabibbo favored $b\to c$ transition
is described by the Hamiltonian\cite{vud}:
\begin{equation}
H ={G_F\over \sqrt 2}V_{cb}
\left\{\left[(\bar d u)+(\bar s c)\right]
(\bar c b)\right\} \label{Eraw}
\end{equation}
where $(\bar q_i q_j)=\bar q_i\gamma_{\mu}(1-\gamma_5)q_j$,
$G_F$ is the Fermi coupling constant, and $V_{cb}$ is the CKM matrix element.

The spectator diagram is modified by hard gluon exchange
between the initial and final quark lines. The effect of these exchanges can
be taken into account by use of the renormalization group, with the
result that an additional term is added to the Hamiltonian, which now
contains two pieces, the original term multiplied by
a coefficient $c_1(\mu)$, and an additional term multiplied by $c_2(\mu)$:
\begin{equation}
H_{eff}={G_F\over \sqrt 2}V_{cb} \left\{c_1(\mu)\left[(\bar d
u)+(\bar s c)\right] (\bar c b)+ c_2(\mu)\left[(\bar c u)(\bar d b)+(\bar c
c)(\bar s b)\right] \right\} \label{Eheff}
\end{equation}
The $c_i$ are Wilson coefficients that can be calculated from QCD.
However, the calculation is inherently uncertain  because it is unclear
at what mass scale, $\mu$,  these coefficients should be evaluated.
The usual scale is taken to be $\mu \sim m_b^{2}$.
Defining
\begin{equation}
c_{\pm}(\mu)=c_1(\mu)\pm c_2(\mu) \label{Ecees}
\end{equation}
the leading-log approximation gives\cite{Neubie}
\begin{equation}
c_{\pm}(\mu)=\left({\alpha_s(M_{W}^{2})\over\alpha_s(\mu)}\right)
^{\displaystyle {-6\gamma_{\pm}\over (33-2n_f)}} \label{Ecpmcal}
\end{equation}
where $\gamma_-=-2\gamma_+=2$, and $n_f$ is the
number of active flavors, which is usually taken to be five in this case.

The additional term in the Hamiltonian in Eq.\ (\ref{Eheff}) corresponds to the
``color suppressed'' diagram.
The quark pairings in this diagram are different from those in the spectator
diagram, and lead to the decay modes discussed in section \ref{B->psi-K(*)}.
{}From the observation of the $B \to \psi X_s$ decays, where $X_s$ is a strange
meson, the magnitude of the color-suppressed term can be deduced.
In $B^-$ decays, both spectator and color-suppressed
diagrams are present and can interfere.
By comparing the rates for $B^-$ and $\bar{B^0}$ decays, both the
size and the relative sign of the color suppressed term can be determined
(see Sec.~\ref{a1-a2}).

For comparisons between theoretical models and data we will use a standard set
of values for the couplings, $V_{cb} = 0.041$ and $V_{ub} /V_{cb} = 0.075$, and
for the $B$ meson lifetime, $\tau_{B} = 1.44\pm 0.04$ ps \cite{Micha}.

\subsection{Factorization}

Factorization is the assumption that two body hadronic decays of $B$ mesons
can be expressed as the product of two independent hadronic currents, one
describing the formation of a meson from the converted $b$ quark and the
light spectator quark, and the other describing the production of a meson by
the hadronization of the virtual $W^-$. This description is expected to be
valid for the external spectator decays where the large energy carried by the
$W^-$
causes the products of the $W^-$ to be well separated from the spectator
quark system \cite{Bjorken},\cite{DG}. It has also been used to calculate
color-suppressed and
penguin diagrams, although it is not known whether factorization is a correct
assumption for these diagrams.

There are number of tests of the factorization hypothesis that can be
made by comparing rates and polarizations for semileptonic and hadronic
$B$ decays. These will be discussed in section \ref{test-factor}.
If factorization holds, then measurements of hadronic $B$ decays
can be compared to the theoretical models, and
used to extract fundamental parameters of the Standard Model.
For instance the CKM matrix element $V_{ub}$ could be obtained from
$\bar{B^0} \to \pi^+\pi^-$ or $ \bar{B^0} \to D_s^- \pi^+$,  and the decay
constant $f_{D_s}$ could be determined from $\bar{B^0} \to D_s^- D^{*+}$.

\subsection{Phenomenological Models of Hadronic $B$ Decay}

Several groups have developed models of hadronic $B$ decays
based on the factorization approach.
To compute rates
for all hadronic $B$ decays the magnitude and sign of the color
amplitude must also be known. It is difficult to calculate this
amplitude from first principles in QCD.
Instead a
phenomenological approach was adopted by Bauer, Stech and Wirbel
\cite{Stech}, in which two undetermined coefficients were assigned to
the effective charged current,  $a_1$,
and the effective neutral current, $a_2$, parts of the $B$ decay
Hamiltonian. In reference \cite{Stech} these coefficients were determined from
a fit to a subset of the experimental data on charm decays.
The values of $a_1$ and $a_2$ can be related to the QCD coefficients
$c_1$ and $c_2$ by
$a_{1} = c_{1} + \xi c_{2}$ and $a_{2} = c_{2} + \xi c_{1}$
where $\xi= 1 /N_{\rm color}$.
The values $a_1 = 1.26$ and $a_2 = -0.51$
that give the best fit to the experimental data on charm decay
correspond to $1/N_{\rm color} \sim 0$ \cite{Neubie}.
However, there is no rigorous
theoretical justification for this choice of $N_{\rm{color}}$
\cite{Shifman}.
In section \ref{a1-a2} we will discuss the determination of the values of $a_1$
and $a_2$ from a fit to the $B$ meson decay data.

\subsection{Heavy Quark Effective Theory}

It has recently been appreciated that there is a symmetry of QCD
that is useful in understanding systems containing one heavy quark.
This symmetry arises when the quark becomes sufficiently heavy
to make its mass irrelevant to the nonperturbative dynamics of the
light quarks. This allows the heavy quark degrees of freedom to
be treated in isolation from the the light quark degrees of freedom.
Heavy quark effective theory (HQET) has been developed by
Isgur and Wise \cite{ISGW} who define a single universal form factor,
$\xi(v.v^{'})$, known as the Isgur-Wise function. In this function $v$ and
$v^{'}$ are the four velocities of the initial and final state heavy quarks.
In the heavy quark limit all the
form factors for hadronic matrix elements such as $B\to D^*$ and
$B\to D$ can be related to this single
function. The value of this function can then be determined from a measurement
of the $D^* l \nu$ rate as a function of $q^2$ \cite{ISGW}.

The evaluation of amplitudes for hadronic decays requires not only the
assumption of factorization, but also the input of hadronic form factors and
meson decay constants. As a result of the development of HQET it is now
believed that many of the hadronic form factors for $b \to c $ transitions
can be calculated quite
well in an essentially model-independent way. This has been done by
several groups \cite{Neubie},\cite{Bari}. The comparison of these theoretical
predictions with the experimental results can be used to test the range of
validity of HQET, and the extent to which $1/M_Q$ corrections to the heavy
quark symmetry are needed.

\section{TESTS OF THE FACTORIZATION HYPOTHESIS }
\label{test-factor}

\subsection{Branching Ratio Tests}

The large samples of reconstructed hadronic $B$ decays
have made possible the precise measurements of branching ratios
discussed in section \ref{BDpiDrho}. As an example of the use of these results
to test the factorization hypothesis we will
consider the specific case of $\bar{B^0}\to D^{*+}\pi^-$.
The amplitude for this reaction is
\begin{equation}
A ={G_F\over \sqrt 2}V_{cb}V_{ud}^*
\langle \pi^- | (\bar{d} u) | 0 \rangle
\langle D^{*+} | (\bar c b) | \bar{B^0} \rangle.\label{EHeffDP}
\end{equation}
The $V_{ud}$ CKM factor arises from the $W^-\to\bar u d$ vertex. The first
hadron current that creates the $\pi^-$ from the vacuum is related to
the pion decay constant, $f_{\pi}$, by:
\begin{equation}
\langle \pi^-(p) | (\bar d u) | 0 \rangle = -if_{\pi}p_{\mu}.\label{Efpi}
\end{equation}
The other hadron current can be found from the
semileptonic decay $\bar{B^0}\to D^{*+}\ell^- \bar{\nu_{\ell}}$.
Here the amplitude is the product of a lepton current and the hadron
current that we seek to insert in Eq.~(\ref{EHeffDP}).
Factorization can be tested experimentally by
verifying whether the relation
\begin{equation} {\Gamma\left(\bar{B^0}\to
D^{*+}\pi^-\right)\over\displaystyle{d\Gamma\over
\displaystyle dq^2}
\left(\bar{B^0}\to D^{*+} \ell ^- \bar{\nu_{l}}\right)\biggr|_{q^2=m^2_{\pi}}}
=
6\pi^2{ c_1^2}
f_{\pi}^2|V_{ud}|^2 ,\label{Efact}
\end{equation}
is satisfied. Here
$q^2$ is the four momentum transfer from the $B$ meson to the $D^*$
meson. Since $q^2$
is also the mass of the lepton-neutrino system, by
setting $q^2 = m_{\pi}^2=0.019 ~ GeV^2$
we are simply requiring that the lepton-neutrino system has
the same kinematic properties as does the
pion in the hadronic decay. $V_{ud}$ and $f_{\pi}$ have well
measured values of 0.975 and 131.7~MeV respectively.
For the coefficient $c_1$ we will use the value $1.12\pm 0.1$ deduced from
perturbative QCD \cite{qcd}. The error in $c_1$ reflects the uncertainty
in the mass scale at which the coefficient $c_1$ should be evaluated.
In the original test of equation~(\ref{Efact}),  Bortoletto and Stone
\cite{Bort} found that the equation was satisfied for $c_1$=1.
In the following discussion we will denote the left hand side of
Eq.~(\ref{Efact}) by
$R_{Exp}$ and the right hand side by $R_{Theo}$.

This type of factorization  test
can be extended to larger $q^2$ values by using
other $\bar{B^0}\to D^{*+} X^-$ decays, e.g. $ X^- =\rho^-$ or
$a_1^-$. For the $\rho^-$ case Eq.~(\ref{Efact}) becomes:
\begin{equation}
 R = { {\Gamma(\bar{B}^0 \to D^{*+} \rho^-)}\over{
{{d\Gamma}\over{dq^2}} {(B\to D^{*} l ~\nu)|}_{q^2=m_{\rho}^2}} }
 = {6 \pi^2  c_1^2 f_{\rho}^2  |V_{ud}|^2}  \label{Efctr}
\end{equation}
where the semileptonic decay
is evaluated at $ q^2 = m_{\rho}^2=0.60$  GeV$^2$.
The decay constant on the right hand side of this equation can be
determined from $ e^+ e^- \to \rho^0$ which gives  $ f_\rho=215 \pm 4$ MeV.
\footnote{A second method uses the relation
$\Gamma(\tau^- \to \nu \rho^-)=~
0.804 {G_F^2 \over{16 \pi}} |V_{ud}^2| M_{\tau}^3 f_\rho^2$, where the
$\rho$ width has been taken into account \cite{Pham}.
This gives f$_{\rho} = 212.0 \pm 5.3$ MeV \cite{narrow}.}
For the factorization test with $\bar{B^0} \to D^{*+} a_1^-$
we use $f_{a_1} = 205 \pm 16$ MeV \cite{ir} determined from $\tau$ decay.
To derive numerical predictions
for branching ratios, we must interpolate the observed differential
$q^2$ distribution \cite{width} for
$B \to D^* \ell ~\nu$ to $q^2=m_\pi^2$, $m_\rho^2$, and $m_{a_1}^2$,
respectively. Until this distribution is measured more precisely we use
theoretical models to perform this interpolation.
The results from three models are given in Table~\ref{TFactst}. The
differences between the models for $B \to D^* \ell ~ \nu$ is small
(See Fig.~\ref{dlnu}).
\begin{figure}[htb]
\unitlength 1.0in
\vskip 10 mm
\begin{center}
\begin{picture}(3.0,2.8)(0.0,0.0)
\end{picture}
\vskip 15 mm
\caption[]{The $q^2$ distribution for the decay
$\bar{B}^0 \to D^{*+} \ell^-\bar{\nu}_{\ell}$. This is a weighted average from
CLEO and ARGUS data. The curves are fits using various models of semileptonic
decays (from Ref.~\protect\cite{Bort}).}
\label{dlnu}
\end{center}
\end{figure}
\begin{table}[htb]
\caption{Ingredients for Factorization Tests.}\label{TFactst}
\begin{tabular}{cc}
$ \vert c_1 \vert $ & $1.12 \pm 0.1$  \\
$ f_{\pi}$ & $131.74 \pm 0.15 $ MeV   \\
$ f_{\rho}$ & $215\pm 4$ MeV \\
$ f_{a_1}$ & $205\pm 16$ MeV   \\
$ V_{ud}$ & $0.975 \pm 0.001 $   \\
%
%
$ {{d {\cal B}}\over{dq^2}}(B \to D^* l ~\nu)\vert_{q^2=m_{\pi}^2} (WSB)$ &
0.0023 GeV$^{-2}$   \\
$ {{d {\cal B}}\over{dq^2}}(B \to D^* l ~\nu)\vert_{q^2=m_{\pi}^2} (ISGW)$ &
0.0020 GeV$^{-2}$   \\
$ {{d {\cal B}}\over{dq^2}}(B \to D^* l ~\nu)\vert_{q^2=m_{\pi}^2} (KS)$ &
0.0024 GeV$^{-2}$   \\
$ {{d {\cal B}}\over{dq^2}}(B \to D^* l ~\nu)\vert_{q^2=m_{\rho}^2} (WSB)$ &
0.0025 GeV$^{-2}$   \\
$ {{d {\cal B}}\over{dq^2}}(B \to D^* l ~\nu)\vert_{q^2=m_{\rho}^2} (ISGW)$ &
0.0024 GeV$^{-2}$   \\
$ {{d {\cal B}}\over{dq^2}}(B \to D^* l ~\nu)\vert_{q^2=m_{\rho}^2} (KS)$ &
0.0027 GeV$^{-2}$   \\
$ {{d {\cal B}}\over{dq^2}}(B \to D^* l ~\nu)\vert_{q^2=m_{a_1}^2} (WSB)$ &
0.0032 GeV$^{-2}$   \\
$ {{d {\cal B}}\over{dq^2}}(B \to D^* l ~\nu)\vert_{q^2=m_{a_1}^2} (ISGW)$ &
0.0030 GeV$^{-2}$   \\
$ {{d {\cal B}}\over{dq^2}}(B \to D^* l ~\nu)\vert_{q^2=m_{a_1}^2} (KS)$ &
0.0033 GeV$^{-2}$
\end{tabular}
\end{table}
Using the extrapolation of the $q^2$ spectrum\cite{Dstbr}
from the WSB model as the central value, we obtain from Eqs.~(\ref{Efact}) and
(\ref{Efctr}) the results given in Table ~\ref{Tfactc}.
\begin{table}[htb]
\caption{Comparison of $R_{Exp}$ and $R_{Theo}$}\label{Tfactc}
\begin{tabular}{lcc}
 & $R_{Exp}$ (GeV$^2$) & $R_{Theo}$ (GeV$^2$)  \\ \hline
$\bar{B}^0 \to D^{*+}\pi^- $
& $1.22\pm 0.18$  & $1.22 \pm 0.17$   \\
$\bar{B}^0 \to D^{*+}\rho^- $
& $2.96\pm 0.65$ & $3.26 \pm 0.46$   \\
$ \bar{B}^0 \to D^{*+} a_1^- $
& $3.9\pm 0.9$  & $3.0 \pm 0.50$
\end{tabular}
\end{table}
Some of the systematic uncertainties in $R_{Exp}$
cancel if we form ratios of branching fractions, as does the
QCD coefficient $c_1$ in $R_{Theo}$. Thus in the case of
$D^{*+}\rho^-$/$D^{*+}\pi^-$,
the expectation from factorization is given by
$R_{Theo}(\rho)$/$R_{Theo}(\pi)$ times the ratio of the
semileptonic branching ratios evaluated at the appropriate $q^2$ values.
In Table~\ref{Tfacthh} we show the comparison between the measured ratios
and two theoretical predictions by Reader and Isgur \cite{ir}, and
the revised BSW model \cite{Neubie}.
\begin{table}[htb]
\caption{Ratios of $B$ decay widths.}\label{Tfacthh}
\begin{tabular}{lcccc}
 & Exp. & Factorization & RI Model & BSW Model \\ \hline
${\cal B}(\bar{B}^0 \to D^{*+}\rho^-) /
{\cal B}(\bar{B}^0 \to D^{*+}\pi^-)$
& $2.64 \pm 0.68$ & $2.90 \pm 0.26$ & 2.2 -- 2.3 & 2.8  \\
${\cal B}(\bar{B}^0 \to D^{*+} a_1^-) /
{\cal B}(\bar{B}^0 \to D^{*+}\pi^-)$
& $4.5 \pm 1.2$ & $3.4 \pm 0.27$ & 2.0 -- 2.1 &3.4
\end{tabular}
\end{table}
At the present level of precision, there is good
agreement between the experimental results and the expectation from
factorization for the $ q^2$ range $ 0 < q^2 < m_{a_1}^2$.
Note that it is  possible that factorization will be a poorer
approximation for decays will smaller energy release or larger $q^2$.
Factorization tests can be extended to higher $q^2$  using
$B\to D^{*} D_s^{(*)}$ decays as
 will be discussed in section \ref{facapply} .

\subsection{Factorization and Angular Correlations}
\label{fac-ang-cor}

More subtle tests of the factorization hypothesis can be performed by examining
the polarization in $B$ meson decays into two vector mesons, as
suggested by K\"orner and Goldstein\cite{Kg}.
Again, the underlying principle is to compare the hadronic decays to the
appropriate semileptonic decays evaluated at a fixed value in $q^2$.
For instance, the ratio of longitudinal to transverse polarization
($\Gamma_{L}/\Gamma_{T}$)
in $\bar{B^0} \to D^{*+} \rho^{-}$ should be equal to the corresponding ratio
for $B\to D^{*}\ell \nu$
 evaluated at $ q^2={m_\rho}^2=0.6~ \rm{GeV}^2$.
\begin{equation}
 {{\Gamma_{L}}\over{\Gamma_{T}}} ({\bar{B^0} \to D^{*+} \rho^{-}})
= {{\Gamma_{L}}\over{\Gamma_{T}}}
{(B\to D^*\ell\nu)|}_{q^2=m_{\rho}^2}
\end{equation}
The advantage of this method is that it is not affected by QCD
corrections \cite{lepage}.

For $B \to D^*\ell\nu$ decay, longitudinal polarization
dominates at low $q^2$, whereas near
$ q^2= q^2_{\rm max}$ transverse polarization dominates. There is a simple
physical argument for the behaviour of the form factors
near these two kinematic limits. Near $ q^2=q^2_{\rm max}$,
the $D^*$ is almost at rest and its small velocity is uncorrelated with the
$D^*$ spin, so all three $D^*$ helicities
are equally likely and we expect $\Gamma_T / \Gamma_L$ = 2.
At $q^2=0$, the $D^*$ has the maximum possible momentum, while the lepton and
neutrino are collinear and travel in the direction opposite to the $D^*$.
The lepton and neutrino helicities are aligned to give
$S_z= 0$, so near $q^2=0$ longitudinal polarization is dominant.

For $\bar{B^0} \to D^{*+} \rho^-$,
we expect $88\%$ longitudinal polarization from the argument
described above \cite{Rosfac}. Similar results have been obtained by
Neubert\cite{neub} and Kramer \etal \cite{Kramfac} .

\begin{figure}[htb]
\unitlength 1.0in
\vskip 15 mm
\begin{center}
\begin{picture}(2.5,2.5)(0.0,0.0)
\end{picture}
\vskip 15 mm
\caption[]{
The differential branching ratio for
$\bar{B^0} \to D^{*+} \ell \bar{\nu}_{\ell}$. The curves show the theoretical
prediction for producing transversely (dashed) and longitudinally (dash-dotted)
polarized $D^*$ mesons, as well as the total decay rate (solid) (from
Ref.~\protect\cite{neub}).}\label{neuba}
\end{center}
\end{figure}

\noindent
Fig.~\ref{neuba} shows the prediction of Neubert for transverse and
longitudinal polarization in $B \to D^*\ell\nu$ decays.
Using this figure we find $\Gamma_L /\Gamma$ to be
85\% at $q^2={m_\rho}^2=0.6$.
The agreement  between these predictions and the
experimental result (Sec.~\ref{pol-D*-rho})
\begin{equation}
\Gamma_L /\Gamma \; = \;  90 \pm 7 \pm 5 \%
\end{equation}
supports the factorization hypothesis in hadronic $B$ meson decay
for $q^2$ values up to $m_{\rho}^2$.

\subsection{Tests of Spin Symmetry in HQET}
\label{spin-sym}

In HQET the effect of the heavy quark magnetic moment does not enter to
lowest order \cite{Mannel}, and the assumption of factorization leads to the
following predictions based on the spin symmetry of HQET:
\begin{equation}
 \Gamma (\bar{B^0} \to D^+ \pi^-) = \Gamma (\bar{B^0} \to D^{*+}\pi^-)
\end{equation}
and
\begin{equation}
 \Gamma (\bar{B^0} \to D^+ \rho^-) =
 \Gamma (\bar{B^0} \to D^{*+}\rho^-).
\end{equation}
After correcting for phase space and deviations from
heavy quark symmetry it is predicted that
${\cal B}(\bar{B^0} \to D^+ \pi^-) = 1.03~ {\cal B}(\bar{B^0} \to D^{*+}\pi^-)$
and ${\cal B}(\bar{B^0} \to D^+ \rho^-) = 0.89 ~
{\cal B}(\bar{B^0} \to D^{*+} \rho^-)$.
A separate calculation by Blok and Shifman using a QCD sum rule approach
predicts that
${\cal B}(\bar{B^0} \to D^+ \pi^-) = 1.2 {\cal B}(\bar{B^0} \to D^{*+} \pi^-)$.
This differs from the HQET prediction due to the presence of non-factorizable
contributions \cite{BS}.

{}From the experimental data we find
\begin{equation}
{{{\cal B}(\bar{B^0} \to D^+ \pi^-)}\over{{\cal B}(\bar{B^0} \to D^{*+}
\pi^-)}}
\; = \; 1.07 \pm  0.21 \pm 0.14
\end{equation}
and
\begin{equation}
{{{\cal B}(\bar{B^0} \to D^+ \rho^-)}\over{ {\cal B}(\bar{B^0} \to D^{*+}
\rho^-)}}\; = \; 1.11 \pm 0.32 \pm 0.16
\end{equation}
The second error bar is due to the uncertainty in the D branching
fractions.
The two ratios of branching fractions are consistent with the
expectations from HQET spin symmetry and with the prediction
from Blok and Shifman that includes nonfactorizable contributions.
Similar tests will be possible using the modes $B\to D^{(*)} D_s^{(*)}$ once
more precise measurements of the branching ratios are available.

Mannel \etal \cite{Mannel} observe that by using a combination of HQET,
factorization, and data on $B\to D^*\ell\nu$,
they can obtain model dependent predictions for
${\cal B} (\bar{B^0}\to D^+ \rho^-)/ {\cal B}(\bar{B^0} \to D^+ \pi^-)$.
Using three parameterizations of the universal Isgur-Wise form factor
\cite{param}, they predict this ratio to be 3.05, 2.52, or 2.61.
{}From the measurements of the branching ratios we obtain
\begin{equation}
{{{\cal B}(\bar{B^0}\to D^+ \rho^-)}\over{ {\cal B}(\bar{B^0} \to D^{+}
\pi^-)}}\; = \; 2.7 \pm 0.6
\end{equation}
The systematic errors from the $D$ branching fractions
cancel in this ratio.
Again we find good agreement with the prediction from HQET
combined with factorization.

\subsection{Applications of Factorization}
\label{facapply}

If factorization holds, hadronic $B$ decays can be used to extract information
about semileptonic
decays. For example, we can determine the so far unmeasured
rate $B\to D^{**}(2420)\ell\nu$ from
the branching ratio of $B\to D^{**}(2420)\pi$.
By  assuming that
the rate for $B\to D^{**}(2420)\pi$ is related to
$d\Gamma/dq^2 (B \to D^{**}(2420) \ell \nu)$ evaluated at $q^2 = m_{\pi}^2$.
Using the model of Colangelo \etal \cite{Bari}
to determine the shape of
the form factors we obtain the ratio
$$
\frac{\Gamma(B \to D^{**}(2420) \ell \nu)}{\Gamma(B \to D^{**}(2420)\pi)}
= 3.2
$$
Combining this result
with the experimental value in Table~\ref{kh3} we predict
${\cal B} (D^{**}(2420) l \nu ) = 0.51 \pm 0.16 \%$

A second application of factorization is the determination of $f_{D_s}$
using the decays $B \to D^*D_s$.
The rate for $\bar{B^0}\to D^{*+}D_s$ is related
to the differential rate for
$\bar{B^0}\to D^{*+}\ell^-\nu$ at $q^2 = m_{D_s}^2$ if factorization continues
to be valid at larger values of $q^2$:
\begin{equation} {\Gamma\left(\bar{B^0}\to
D^{*+} D_{s}^{-}\right)\over\displaystyle{d\Gamma\over
\displaystyle dq^2}
\left(\bar{B^0}\to D^{*+}\ell^-\nu\right)\biggr|_{q^2=m^2_{D_s}}} =
6\pi^2 \delta { c_1^2}
f_{D_s}^2|V_{cs}|^2 ,\label{Efacts}
\end{equation}
The factor $\delta =0.37$ is required because the $D^*$ in the $B \to D^*D_s$
final state is fully polarized, and should be compared with the rate of
$B\to D^*\ell\nu$ at $q^2=m^2_{D_s}$ in that polarization state only.
Using the above result and the average branching ratio
for ${\cal B}(B\to D^{*+} D_{s}^-)=0.9\pm 0.5 \%$, we obtain
$$ f_{D_s} = (288 \pm 64) \sqrt{3.7\%/B(D_s \to \phi \pi^+)} ~\rm{MeV}$$
This result can be compared to the value
$$ f_{D_s} = (344 \pm 37 \pm 52) \sqrt{B(D_s \to \phi \pi^+)/3.7\%} ~\rm{MeV}$$
that has recently been obtained from a
direct measurement of $D_s\to \mu \nu$ decays in continuum charm events
\cite{CLNS9314}. Both these values of $f_{D_s}$ are consistent with the
theoretical predictions which are in the range $f_{D_s}=200-290$~MeV
\cite{Lattices}, \cite{Potentials}, \cite{QCDsum}.
If both the $D_s \to \phi \pi^+$ branching ratio and $f_{D_s}$
are measured more precisely, then measurements of the
branching ratios of $B\to D^* D_s$ decays can be used
to test factorization in $B$ decay at $q^2 = m_{D_s}^{2}$. In the near future,
it will also be possible to test factorization in this $q^2$ range
by measuring $\Gamma_{L}/\Gamma$ in $B \to D^* D_{s}^*$ decays.

\section{DETERMINATION OF THE COLOR SUPPRESSED AMPLITUDE}
\label{eff-color-supp}

\subsection{Color Suppression in $B$ Decay}

In the decays of charmed mesons the effect of color suppression
is obscured by the effects of final state interactions (FSI), and
soft gluon effects which enhance $W$ exchange diagrams.
Table~\ref{Tcolsuprat} gives
ratios of several charmed meson decay modes with approximately
equal phase space factors where the mode in the numerator is color suppressed
while the mode in the denominator is  an external spectator decay.
With the exception of the
 decay $D^0\to \bar{K}^0\rho^0$ it is clear that the color
suppressed decays do not have significantly smaller branching ratios.
\begin{table}[htb]
\caption{Measured Ratios of color suppressed to external spectator branching
fractions.}\label{Tcolsuprat}
\begin{tabular}{cc}
Mode & Branching fraction \cite{PDG} \\ \hline
${\cal B}(D^0 \to \bar{K^0}\rho^0) / {\cal B}(D^0 \to K^- \rho^+)$
         & $0.08 \pm 0.04$   \\
${\cal B}(D^0 \to K^0 \pi^0) / {\cal B}(D^0 \to K^- \pi^+)$
         & $0.57 \pm 0.13$   \\
${\cal B}(D^0\to \bar{K^{*0}} \pi^0) / {\cal B}(D^0 \to K^{*-} \pi^+)$
         & $0.47\pm 0.23$  \\
${\cal B}(D^0 \to \pi^0 \pi^0) / {\cal B}(D^0 \to \pi^- \pi^+) \cite{CLpipi}$
         & $0.77 \pm 0.25$   \\
${\cal B}(D_s^{+} \to \bar{K^{*0}} K^+) / {\cal B}(D_s \to \phi \pi^+)$
         & $0.95\pm 0.10$   \\
${\cal B}(D_s^{+} \to \bar{K^0} K^+) / {\cal B}(D_s \to \phi \pi^+)$
         & $1.01 \pm 0.16$ \\
\end{tabular}
\end{table}

When the BSW model is used to fit the data on charm decays
it gives values of $a_1=1.26$ and $a_2 = -0.51$.
The BSW model assumes that the values of the coefficients can be
extrapolated from $\mu = m_{c}^2$ to $\mu = m_{b}^2$
taking into account the evolution of the strong coupling constant
$\alpha_s$. This extrapolation gives the predictions $a_1=1.1$ and
$a_2=-0.24$ for $B$ decays.
The smaller magnitude of $a_2$ means that in contrast to the charm sector
one expects to find a more consistent pattern of color suppression in $B$ meson
decays.
Another approach uses the factorization hypothesis, HQET
and model dependent form factors (RI model)\cite{ir}. In this approach,
$a_1$ and $a_2$ are determined from QCD (with
$ 1 /N_{\rm color} =1 /3$), and color suppressed $B$ decays are
expected to occur at about $1/1000$ the rate of unsuppressed decays.
The observation of color suppressed $B$ decays at a much greater
level would indicate the breakdown of the factorization hypothesis.
In Section ~\ref{color-supress} we obtained upper limits for color suppressed
$B$ decays with a $D^0$ or $D^{*0}$ meson in the final state.
In Table~\ref{Tbrcolcomp} these results are compared to the predictions
of the BSW and the RI models.

\begin{table}[htb]
\caption{Branching fractions of color suppressed $B$ decays
and comparisons with models.}\label{Tbrcolcomp}
\begin{tabular}{lcccc}
Decay Mode & U. L. (\%) & BSW (\%) &
 $\cal{B}$ (BSW) & RI~model(\%)  \\ \hline
$\bar{B^0} \to D^{0} \pi^0$     &$<0.048$ & $0.012$
 & $0.20 a_2^{2} (f_{D}/220 \rm{MeV})^2$ & $0.0013 - 0.0018$   \\
$\bar{B^0} \to D^{0} \rho^0$    &$<0.055$ & $0.008$
 & $0.14 a_2^{2} (f_{D}/220 \rm{MeV})^2$ & $0.00044$   \\
$\bar{B^0} \to D^{0} \eta$      &$<0.068$ & $0.006$
& $0.11 a_2^{2} (f_{D}/220 \rm{MeV})^2 $             &              \\
$\bar{B^0} \to D^{0} \eta^{'}$  &$<0.086$ & $0.002$
& $ 0.03 a_2^{2}(f_{D}/220 \rm{MeV})^2$  &              \\
$\bar{B^0} \to D^{0} \omega $   &$<0.063$ & $0.008$
& $0.14 a_2^{2}(f_{D}/220 \rm{MeV})^2$   &              \\
$\bar{B^0} \to D^{*0} \pi^0$    &$<0.097$ & $0.012$
& $ 0.21 a_2^{2}(f_{D*}/220 \rm{MeV})^2$ & $0.0013-0.0018$   \\
$\bar{B^0} \to D^{*0} \rho^0$   &$<0.117$  & $0.013$
& $ 0.22 a_2^{2}(f_{D*}/220 \rm{MeV})^2$ & $0.0013 -0.0018$   \\
$\bar{B^0} \to D^{*0} \eta$     &$<0.069$ & $0.007$
& $0.12 a_2^{2}(f_{D*}/220 \rm{MeV})^2$   &   \\
$\bar{B^0} \to D^{*0} \eta^{'}$ &$<0.27$  & $ 0.002$
& $0.03 a_2^{2}(f_{D*}/220 \rm{MeV})^2$   &   \\
$\bar{B^0} \to D^{*0} \omega$   &$<0.21$  & $0.013$
& $ 0.22 a_2^{2}(f_{D*}/220 \rm{MeV})^2$  &
\end{tabular}
\end{table}

In contrast to charm decays, color suppression seems to be operative
in hadronic decays of $B$ mesons. The limits on the color suppressed
modes with $D^{0(*)}$ and neutral mesons are still above the level
expected by the two models, but we can already exclude a prediction
by Terasaki \cite{tera} that
${\cal{B}}(\bar{B^0} \to D^0 \pi^0) \approx 1.8 {\cal{B}}(\bar{B^0} \to
D^+\pi^-)$.
To date, the only color suppressed $B$ meson decay modes
that have been observed are final states
which contain charmonium mesons e.g. $B\to \psi K$ and $B\to \psi K^*$
\cite{psicomment}.

\subsection{Determination of $|a_1|$, $|a_2|$ and
the Relative Sign of ($a_2/a_1$)}
\label{a1-a2}

In the BSW model \cite{Neubie} , the branching
fractions of the $\bar{B}^0$ normalization
modes are proportional to $a_1^2$ while
the branching fractions of the
$B\to\psi$ decay modes depend only on $a_2^2$.
 A fit to the
branching ratios for the modes
$\bar{B^0}\to D^+\pi^-$, $D^+\rho^-$, $D^{*+}\pi^-$ and $D^{*+}\rho^-$
using the model of Neubert \etal\ yields
\begin{equation}
|a_1| = 1.07 \pm 0.04  \pm  0.06
\end{equation}
and a fit to
the modes with $\psi$ mesons in the final state gives
\begin{equation}
|a_2| = 0.23 \pm 0.01 \pm 0.01
\end{equation}
The first error
on $|a_1|$ and $|a_2|$ includes the
uncertainties from the charm or charmonium branching ratios,
 the experimental systematics associated with detection
efficiencies and background subtractions as well as the statistical
errors from the branching ratios.
The second  error quoted is the uncertainty due to
the $B$ meson production fractions and lifetimes.
 We have assumed that the ratio of $B^+ B^-$ and $B^0 \bar{B^0}$ production
at the $\Upsilon(4S)$ is one
\cite{Micha},\cite{mgs},
and assigned an uncertainty of 10\% to it.

\begin{table}[htb]
\label{Tbswcol}
\caption{Predicted branching fractions in terms of BSW parameters $a_1$, $a_2$
}
\begin{tabular}{lcc}
Mode & Neubert \etal \cite{Neubie} & Deandrea \etal \cite{DBGN} \\ \hline

$\bar{B^0} \to D^+ \pi^- $        &$ 0.264 a_1^2 $ & $0.276 a_1^2$ \\
$\bar{B^0} \to D^+ \rho^-$        &$ 0.621 a_1^2 $ & $0.713 a_1^2$ \\
$\bar{B^0} \to D^{*+} \pi^-$      &$ 0.254 a_1^2 $ & $0.276 a_1^2$ \\
$\bar{B^0} \to D^{*+} \rho^-$     &$ 0.702 a_1^2 $ & $0.943 a_1^2$ \\
$ B^- \to D^0 \pi^- $             &
$ 0.265 [a_1 +1.230 a_2 ( f_D/220)]^2 $ &
$ 0.276 [a_1 +1.155 a_2 ( f_D/220)]^2 $  \\
$ B^- \to D^0 \rho^-$             &
$  0.622 [a_1 + 0.662 a_2 ~( f_D/220)]^2 $ &
$ 0.713 [a_1 +0.458 a_2 ( f_D/220)]^2 $  \\
$ B^- \to D^{*0} \pi^-$           &
$  0.255 [a_1 +1.292 a_2 ~( f_{D^*}/220)]^2$ &
$ 0.276 [a_1 + 1.524 a_2 ( f_{D^*}/220)]^2 $  \\

$ B^-  \to D^{*0} \rho^-$         &
 $  0.703 [a_1^2 + 1.487 a_1 a_2 ~( f_{D^*}/220) $   &
$ 0.943 [a_1^2 + 1.31 a_1 a_2 ~( f_{D^*}/220) $  \\
 & $+0.635 a_2^2 (f_{D^*}/220)^2]$ &
  $+ 0.53 a_2^2 ( f_{D^*}/220)^2]$ \\
$ B^- \to \psi K^- $               &$ 1.819 a_2^2 $ & $ 1.634 a_2^2$ \\
$ B^- \to  \psi K^{*-}$            &$ 2.932 a_2^2 $ & $ 2.393 a_2^2$ \\
$ \bar{B^0} \to \psi \bar{K}^0$    &$ 1.817 a_2^2 $ & $ 1.634 a_2^2$ \\
$ \bar{B^0} \to \psi \bar{K}^{*0} $&$ 2.927 a_2^2 $ & $ 2.393 a_2^2$
\end{tabular}
\end{table}

\begin{table}[htb]
\caption{Ratios of normalization modes to determine the sign of
$a_2/a_1$. The magnitude of $a_2/a_1$ is the value in the
BSW model which agrees with our result for $B\to \psi$ modes.}\label{Tbswexpc}
\begin{tabular}{ccccc}
Ratio &$a_2/a_1 =-0.22 $ & $a_2/a_1 =0.22 $ & Experiment & RI~ model \\ \hline
$R_1 $& 0.53  & 1.61 & $1.63 \pm 0.36$ &$1.20-1.28$  \\
$R_2 $& 0.73  & 1.31 & $1.66 \pm 0.46$ &$1.09-1.12$  \\
$R_3 $& 0.51  & 1.65 & $1.86 \pm 0.39$ &$1.19-1.27$  \\
$R_4 $& 0.70  & 1.36 & $2.08 \pm 0.61$ &$1.10-1.36$
\end{tabular}
\end{table}

By comparing branching ratios of $B^-$ and $\bar{B^0}$ decay modes it is
possible to determine the the sign of $a_2$ relative to $a_1$.
The BSW model, Ref.~\cite{Neubie} predicts the following ratios:
\begin{equation}
R_1 = {{\cal B}(B^- \to D^0 \pi^-) \over {\cal B}(\bar{B^0}\to D^+ \pi^-)}
                = (1 + 1.23 a_2/a_1)^2  \label{colrate1}
\end{equation}
\begin{equation}
R_2 = {{\cal B}(B^- \to D^0 \rho^-)
\over {\cal B}(\bar{B^0} \to D^+ \rho^-)}
                = (1 + 0.66 a_2 /a_1)^2  \label{colrate2}
\end{equation}
\begin{equation}
R_3 = {{\cal B}(B^- \to D^{*0} \pi^-)
         \over {\cal B}(\bar{B^0} \to D^{*+} \pi^-)}
                     =(1 + 1.29 a_2/a_1)^2  \label{colrate3}
\end{equation}
\begin{equation}
R_4 = {{\cal B}(B^- \to D^{*0} \rho^-)
          \over{\cal B}(\bar{B^0} \to D^{*+} \rho^-)}
                     \approx (1 + 0.75 a_2/a_1)^2   \label{colrate4}
\end{equation}

Table~\ref{Tbswexpc} shows a comparison between the
experimental results and
the two allowed solutions in the BSW model.
In the experimental ratios the systematic errors due to
detection efficiencies partly cancel.
In the ratios $R_3$ and $R_4$ the $D$ meson branching ratio uncertainties
do not contribute to the systematic error.

A least squares fit to the ratios $R_1$ - $R_3$ gives
$a_2/a_1 = 0.27 \pm 0.08 \pm 0.06$
where we have ignored uncertainties in the
theoretical predictions.
$R_4$ is not included in the fit since
the model prediction in this case is not thought to be reliable \cite{volkie}.
The second error is due to the uncertainty in
the $B$ meson production fractions and lifetimes
which enter into the determination of $a_1/a_2$ in the combination
$(f_+  \tau_{+}/ f_{0} \tau_{0})$.
 As this ratio increases,
the value of $a_2/a_1$ decreases.
The allowed range of $(f_+  \tau_{+}/ f_{0} \tau_{0})$
excludes a negative value of $a_2/a_1$.

Other uncertainties in the magnitude\cite{fdvari}
 of $f_D$, $f_{D^*}$ and in the hadronic form
factors can change the magnitude of $a_2/a_1$ but not its sign.
The numerical factors which
multiply $a_2/a_1$ include the ratios of $B \to \pi$($B\to\rho$)
to $B\to D$ ($B\to D^*$) form
factors, as well as the ratios of the meson decay constants. We
assume values of 220~MeV for $f_D$ and $f_{D^*}$ \cite{rosfd}.
To investigate the model dependence of the result we have recalculated
$|a_1|$, $|a_2|$, and $a_2/a_1$ in the model of Deandrea \etal\ We find
$|a_1| = 0.99 \pm 0.04 \pm 0.06$,
$|a_2| = 0.25 \pm 0.01 \pm 0.01$, and
$a_2/a_1 = 0.25 \pm 0.07 \pm 0.05$, consistent with the results discussed
above. A different set of $B \to \pi$ form factors can be calculated using
QCD sum rules. Using the form factors determined by Belyaev, Khodjamirian
and R\"uckl \cite{brueckl} and
by Ball \cite{pballff}, $a_2/a_1$ changes by 0.04. Kamal and Pham
have also considered the effect of uncertainties in form factors,
the effects of final state interactions, and annihilation terms. They
conclude that these may change the magnitude of $a_2/a_1$ but
not its sign \cite{KPham}.

\begin{table}[htb]
\caption{Predicted (BSW) and measured ratios of widths of
$D^+$ and $D^0$ modes in charm decay.}\label{Tbcharm}
\begin{tabular}{cccc}
Mode &$a_2/a_1 =-0.40 $ & $a_2/a_1 =0.40 $ & Ratio of widths (exp)\cite{PDG} \\
\hline
$D^+ \to \bar{K}^0\pi^+ /D^0\to K^- \pi^+$ &
   0.26           &  2.2   &  $0.28  \pm 0.05 $   \\
$D^+\to \bar{K}^0\rho^+ /D^0\to K^-\rho^+$ &
   0.58        &  1.5  & $ 0.36 \pm 0.10   $   \\
$D^+\to \bar{K}^{*0}\pi^+ /D^0\to K^{*-}\pi^+$ &
   0.05       &   3.2   & $0.17\pm  0.07 $   \\
$D^+\to \bar{K}^{*0}\rho^+ /D^0\to K^{*-}\rho^+ $ &
   0.34       &   2.0   & $0.25 \pm 0.12 $
\end{tabular}
\end{table}

The magnitude of $a_2$ determined from this fit is consistent with the value
of $a_2$ determined from the fit to the $B\to\psi$ decay modes.
The sign of $a_2$ disagrees with the theoretical
extrapolation from the fit to charmed meson decays using
the BSW model\cite{oldfit}.
Table~\ref{Tbcharm} compares the corresponding charm decay ratios
to the theoretical expectations for positive and negative values of $a_2/a_1$.

\section{RARE HADRONIC DECAYS}

\subsection{Introduction}

There are hadronic $B$ meson decays that cannot be produced
by the usual $b\to c$ transition. The results of the
experimental search for these rare
decay modes provides important information on the mechanisms of $B$ meson
decay and significant progress is being made
with the collection of large samples of $B$ mesons by the CLEO II experiment.
As an indication of this we will discuss the
first observation of radiative penguin decay as well as new
experimental results on the decays $\bar{B^0}\to\pi^+\pi^-$ and
$\bar{B^0}\to K^-\pi^+$ where a statistically significant signal has been
observed in the sum of the two modes.

Decays of the kind $B\to D_s X_u$, where the $X_u$ system hadronizes as
pions, can occur via a $b\to u$ spectator diagram where the $W$ forms a
$c\bar{s}$ pair. Since other contributing diagrams are expected to be
negligible these decays may provide a clean environment in which to measure
$V_{ub}$ in hadronic decays. Decays of the kind $\bar{B^0}\to D_s^+ X_s^-$,
where $X_s$ is a strange meson, are also interesting since they are
associated with a $W$ exchange diagram.

\begin{figure}[htb]
\begin{center}
\vskip 15 mm
\unitlength 1.0in
\begin{picture}(2.,1.2)(0,0)
\end{picture}
\caption{Rare $B$ meson decay diagrams: (a)
$b \to u$ spectator and (b) gluonic penguin.}
\label{rarefeyn}
\end{center}
\end{figure}

Charmless hadronic decays
 such as $\bar{B^0}\to\pi^+ \pi^-$, $B^-\to\pi^-\pi^0$,
$\bar{B^0}\to\pi^{\pm}\rho^{\mp}$ and $B^-\to\pi^0 \rho^-$, are expected to be
produced by the $b\to u$ spectator diagram (Fig. \ref{rarefeyn}(a)), although
there is a possible small
contribution from a $b\to d$ penguin diagram (Fig. \ref{rarefeyn}(b)). The
decay
$\bar{B^0}\to \pi^+\pi^-$ has been discussed as a possible place to observe
CP violation in the $B$ meson system \cite{CPpipi}. The final state is
a CP eigenstate, and CP violation can arise from interference between the
amplitude for the direct decay via the $b\to u$ spectator diagram, and the
amplitude for the decay following $B^0\bar{B^0}$ mixing. In this decay the
CP violating angle is different from the one accessible in
$\bar{B^0}\to \psi K_s$, so the measurement is complementary.
There is a possible complication if the $b\to d$ penguin contribution
to the amplitude is significant. This could be resolved if measurements are
made on other rare hadronic decay modes to determine the role of the penguin
amplitude in any observed CP violating effect \cite{CPpipi}.

Decays to charmless hadronic final states containing an $s$ quark are expected
to have a significant contribution from a $b\to s$ penguin diagram, although
they can also occur through a Cabibbo suppressed $b\to u$ spectator diagram.
The inclusive rates for the hadronic penguin diagrams $b\to sg$
and $b\to sq\bar{q}$ are estimated
to be about 1\% from the parton model, but predictions for the
hadronization into exclusive final states are uncertain because the simple
assumptions about factorization of the amplitude used for the spectator
diagram may not be valid for loop diagrams.

\subsection{Decays to $D_s$ Mesons}

These decays have recently been searched for by ARGUS \cite{ARGUSDspi}
and CLEO~II \cite{CLEODspi}. The upper limits
are given in Table~\ref{TABDspi}
along with theoretical predictions by Choudury\etal \cite{CISS},
and Deandrea \etal \cite{DBGN}.

\begin{table} [hbt]
\caption{Theoretical predictions and experimental upper limits (90\% C.L.)
for $B$ decays to $D_s$.
All numbers quoted are branching fractions $\times 10^{-5}$}
\label{TABDspi}

\begin{center}
\begin{tabular}{lcccc}
$B$ Decay          & Choudury & Deandrea & ARGUS    & CLEO II  \\ \hline
$D_s^+\pi^-$       &   1.9    &    8.1   & $<$170.0 & $<$27.0  \\
$D_s^{*+}\pi^-$    &   2.7    &    6.1   & $<$120.0 & $<$44.0  \\
$D_s^+\rho^-$      &   1.0    &    1.2   & $<$220.0 & $<$66.0  \\
$D_s^{*+}\rho^-$   &   5.4    &    4.5   & $<$250.0 & $<$74.0  \\
$D_s^+\pi^0$       &   1.8    &    3.9   & $<$90.0  & $<$20.0  \\
$D_s^{*+}\pi^0$    &   1.3    &    3.0   & $<$90.0  & $<$32.0  \\
$D_s^+\eta$        &          &    1.1   &          & $<$46.0  \\
$D_s^{*+}\eta$     &          &    0.8   &          & $<$75.0  \\
$D_s^+\rho^0$      &    0.5   &    0.6   & $<$340.0 & $<$37.0  \\
$D_s^{*+}\rho^0$   &    2.8   &    2.4   & $<$200.0 & $<$48.0  \\
$D_s^+\omega$      &          &    0.6   & $<$340.0 & $<$48.0  \\
$D_s^{*+}\omega$   &          &    2.4   & $<$190.0 & $<$68.0  \\
$D_s^+ K^-$        &          &          & $<$170.0 & $<$23.0  \\
$D_s^{*+} K^-$     &          &          & $<$120.0 & $<$17.0  \\
$D_s^+ K^{*-}$     &          &          & $<$460.0 & $<$97.0  \\
$D_s^{*+} K^{*-}$  &          &          & $<$580.0 & $<$110.0 \\
\end{tabular}
\end{center}
\end{table}
The experimental limits are still at least a factor of three above the
theoretical predictions. If we compare the limits to the predictions of
Deandrea \etal ~we note that the best constraint on $|V_{ub}/V_{cb}|$ will
come from the CLEO~II limit on $\bar{B^0}\to D_s^+\pi^-$, but that this model
dependent limit is still above the range
$0.06<|V_{ub}/V_{cb}|<0.10$ allowed by the recent semileptonic
 data \cite{btoulnu}.
Combining several $D_s X_u$ modes the sensitivity to $V_{ub}$ can be slightly
improved. For example, using the BSW model CLEO obtains an upper limit
of $|V_{ub}/V_{cb}|< 0.15$ (90\% C.L.) \cite{CLEODspi}.

\subsection{Charmless Hadronic $B$ Decay}

Predictions of branching ratios for charmless hadronic decays were made by
Bauer, Stech and Wirbel \cite{Stech} using the $b\to u$ spectator diagram and
the assumption of factorization. The possible contributions from penguin
diagrams were neglected. These predictions have recently been updated
by Deandrea \etal \cite{DBGN} using new estimates of the hadronic form factors.
We compare their results to the experimental upper limits in Table
\ref{TABbsw}.
\begin{table} [hbt]

\caption{Theoretical predictions and experimental upper limits (90\% C.L.)
for charmless hadronic $B$ decays. All numbers quoted are branching fractions
$\times 10^{-5}$.}
\label{TABbsw}
\begin{center}
\begin{tabular}{lcccc}
$B$ Decay     &Deandrea &  ARGUS  &  CLEO 1.5 & CLEO II \\ \hline
$\pi^+\pi^-$  &  1.8  & $<$13.0 & $<$7.7 & $<$2.9  \\
$\pi^{\pm}\rho^{\mp}$ & 5.2 & $<$52.0 &  &$<$29.0 \\
$\rho^+\rho^-$&  1.3  &         &        &         \\
$\pi^{\pm}a_1^{\mp}$ &   & $<$90.0 & $<$49.0 &     \\
$\pi^0\pi^0$  &  0.06 &         &        &         \\
$\pi^0\rho^0$ &  0.14 & $<$40.0 &        &         \\
$\rho^0\rho^0$&  0.05 & $<$28.0 & $<$29.0 &        \\
$\pi^-\pi^0$  &  1.4  & $<$24.0 &        &         \\
$\pi^-\rho^0$ &  0.7  & $<$15.0 & $<$17.0&         \\
$\pi^0\rho^-$ &  2.7  & $<$55.0 &        &         \\
$\rho^-\rho^0$&  0.7  & $<$100.0 &       &         \\
$p\bar{p}\pi^-$&      & ($52\pm 14\pm 19$)& $<$14.0 &
\end{tabular}
\end{center}
\end{table}
We have included in Table \ref{TABbsw} the reported observation of the decay
$B^-\to p\bar{p}\pi^-$ by ARGUS \cite{ppbarpi}, even though this has not
been confirmed by later data from either ARGUS or CLEO \cite{ppbarpi2}.

There are two recent sets of theoretical predictions by
Deshpande \etal \cite{Desh} and Chau \etal \cite{Chau} that
take into account both penguin and spectator contributions and
make predictions for a large number of charmless hadronic $B$ decays.
A selection of these predictions are shown in table~\ref{TABbsg}. Large
contributions from the penguin amplitude are expected in decays such as
$B\to K^{(*)}\phi$ and $B\to K^{(*)}\pi$.
However, the decays $B\to K\rho$ are predicted to
have very small penguin amplitudes due to cancellations in the contributions
to the amplitude \cite{Desh}.

\begin{table} [hbt]
\caption{Theoretical predictions and experimental upper limits (90\% C.L.)
for $b\to s$ decays. All numbers quoted are branching fractions
$\times 10^{-5}$}
\label{TABbsg}
\begin{center}
\begin{tabular}{lccccc}
$B$ Decay & Deshpande & Chau & ARGUS & CLEO 1.5 & CLEO II \\ \hline
$K^-\pi^+$   & 1.1 &  1.7  & $<$18.0 & $<$7.7  & $<$2.6 \\
$K^-\rho^+$  &  0  &  0.2  &         &         & $<$11.0\\
$K^0\pi^0$   & 0.5 &  0.6  &         &         &        \\
$K^0\rho^0$  & 0.01&  0.04 & $<$16.0 & $<$50.0 &        \\
$K^{*-}\pi^+$& 0.6 &  1.9  & $<$62.0 & $<$38.0 &        \\
$K^{*0}\pi^0$& 0.3 &  0.5  &         &         &        \\
$K^-\pi^0$   & 0.6 &  0.8  &         &         &        \\
$K^-\rho^0$  & 0.01&  0.06 & $<$18.0 & $<$8.0  &        \\
$K^0\pi^-$   & 1.1 &  1.2  & $<$9.6  & $<$10.0 &        \\
$K^0\rho^-$  &  0  &  0.03 &         &         &        \\
$K^{*0}\pi^-$& 0.6 &  0.9  & $<$17.0 & $<$15.0 &        \\
$K^{*-}\pi^0$& 0.3 &  0.9  &         &         &        \\
$K^0\phi$    & 1.1 &  0.9  & $<$36.0 & $<$42.0 & $<$9.4 \\
$K^{*0}\phi$ & 3.1 &  0.9  & $<$32.0 & $<$38.0 & $<$19.0\\
$K^-\phi$    & 1.1 &  1.4  & $<$18.0 & $<$9.0  & $<$1.7 \\
$K^{*-}\phi$ & 3.1 &  0.8  & $<$130.0&         & $<$12.0\\
\end{tabular}
\end{center}
\end{table}

New upper limits have been presented for
$\bar{B}^0\to\pi^+ \pi^-$ \cite{PRLkpi} and $\bar{B^0}\to\pi^{\pm}\rho^{\mp}$
\cite{DPFkpi}. The CLEO~II search for $\bar{B}^0\to\pi^+ \pi^-$ is
discussed in detail in the next section.
CLEO~II also has a new limit on $\bar{B^0}\to K^-\pi^+$ \cite{PRLkpi}, and
preliminary results on $\bar{B^0}\to K^-\rho^+$ \cite{DPFkpi} as well as the
$B\to K^{(*)}\phi$ modes \cite{CLEOphi}. The CLEO~II limits on
$\bar{B^0}\to K^-\pi^+$ and $B^-\to K^-\phi$,
which are expected to have a large penguin amplitude, are close
to the theoretical predictions.

The experimental sensitivities to branching ratios
have now reached the $10^{-5}$ range.
Since the theoretical predictions for several $B$ decay modes
are in this range, it is possible that some signals will be observed soon.
By measuring a sufficient number of charmless $B$ decay modes
(e.g. $\bar{B}^0 \to \pi^- \pi^+$, $B^- \to \pi^- \pi^0$,
$\bar{B}^0 \to \pi^0 \pi^0$) it may be possible to isolate the spectator
and penguin contributions.

\subsection{New Experimental Results on $\bar{B^0}\to \pi^+\pi^-$ and
$\bar{B^0}\to K^- \pi^+$}
\label{newpipi}

The decay modes
$\bar{B^0}\to\pi^+\pi^-$, $\bar{B^0}\to K^-\pi^+$, and $\bar{B^0}\to K^+ K^-$
\cite{bkk}, have been searched for by CLEO~II using a
data sample of 1.37~fb$^{-1}$ taken on the $\Upsilon$(4S)\cite{PRLkpi}. A
sample of
0.64~fb$^{-1}$ taken just below the resonance is used to study the continuum
background. Since $B$ mesons are produced nearly at rest on the $\Upsilon$(4S),
the final state has two nearly back-to-back tracks with momenta
about 2.6~GeV/c.
We distinguish candidates for $B$ meson decays from continuum background using
the difference, $\Delta E$, between the total energy of the two tracks and the
beam
energy, and the beam-constrained mass,
$M_B$. The r.m.s. resolutions on $\Delta E$ and $M_B$ are 25~MeV and 2.5~MeV
respectively.

Separation between $\pi^-\pi^+$, $K^-\pi^+$ and $K^-K^+$ events is provided by
the
$\Delta E$ variable, and by $dE/dx$ information from the 51-layer main drift
chamber. The $\Delta E$ shift between $K\pi$ and $\pi\pi$ events is 42~MeV
if $E_1$ and $E_2$ are determined using the pion mass. This is
1.7$\sigma_{\Delta E}$. The $dE/dx$ separation between kaons and pions at
2.6~GeV/c is found to be $(1.8\pm 0.1)\sigma$ from a study of a sample of
$D^{*+}$-tagged $D^0\to K^-\pi^+$ decays. Thus, in the CLEO II experiment
the total separation between $K\pi$ and $\pi\pi$ events is $2.5\sigma$.

The background arises almost entirely from the continuum where the two-jet
structure of the events can produce high momentum, back-to-back tracks.
These events can be discriminated against by calculating the angle, $\theta_T$,
between the thrust axis of the candidate tracks, and the thrust axis of the
rest of the event. The distribution of $\cos\theta_T$ is peaked at $\pm$1 for
continuum events, and is nearly flat for $B\bar{B}$ events. A cut is made at
$|\cos\theta_T|<0.7$. Additional discrimination is provided by a Fisher
discriminant
 \cite{CLNSKpi},\cite{Fisher}, $\cal{F}$ = $\sum_{i=1}^{n}\alpha_i y_i$.
The inputs
$y_i$  are the direction of the candidate thrust axis, the $B$ meson flight
direction, and nine variables measuring the energy flow of the rest of the
event. The coefficients $\alpha_i$ are chosen to maximize the separation
between $B\bar{B}$ signal events and continuum background events.
The optimal cut on the Fischer discriminant is
84\% efficient for signal and 40\% efficient for background.

Two approaches are used to evaluate the amount of signal in the data sample.
In the first approach a cut is made on ${\cal {F}}$ and
events are classified as
$\pi\pi$, $K\pi$ or $KK$ according to the
most probable hypothesis from the $dE/dx$ information.
The signal and background numbers are given in
Table~\ref{Kpi}. The efficiency for the correct identification of a signal
event in this analysis is 19\%.
The background is estimated using sidebands in the continuum and on-resonance
data and scaling factors
from Monte Carlo studies. There is no $B\bar{B}$ background
in the signal region.

\begin{table} [hbt]

\caption{Event yields, fitted branching fractions and 90\% C.L. upper limits
for
$B^0\to\pi^+\pi^-$, $B^0\to K^+\pi^-$ and $B^0\to K^+ K^-$}
\label{Kpi}

\begin{center}
\begin{tabular}{lccccc}
Decay mode& Signal Events & Background & $N_{fit}$ & B.R.$\times 10^{-5}$ &
U.L.$\times 10^{-5}$\\ \hline
$\pi^+\pi^-$ & 6 & 1.4$\pm$0.2 & $7.2_{-3.5}^{+4.3}$ &
 $1.3_{-0.6}^{+0.8}\pm0.2$ &           $<2.9$  \\
$K^+\pi^-$   & 6 & 1.5$\pm$0.2 & $6.4_{-3.1}^{+3.9}$ &
 $1.1_{-0.6}^{+0.7}\pm0.2$ &           $<2.6$  \\
$K^+K^-$     & 0 & 1.1$\pm$0.1 & $0.0_{-0.0}^{+0.8}$ &
 $0.0_{-0.0}^{+0.2}$       &           $<0.7$  \\
$\pi^+\pi^-$ or $K^+\pi^-$ & 12 & 2.9$\pm$0.3 & $13.6_{-3.9}^{+4.7}$ &
 $2.4_{-0.7}^{+0.8}\pm0.2$ &\\
\end{tabular}
\end{center}
\end{table}

\begin{figure}[hbt]

\vspace{-1.0cm}
\vspace{-3.5cm}
\vskip 4mm
\caption{Likelihood contours in the CLEO~II analysis
for the fit to $N_{\pi\pi}$ and $N_{K\pi}$.
The best fit is indicated by the cross, the 1, 2, 3, and 4$\sigma$ contours by
solid lines,
and the $1.28\sigma$ contour by the dotted line.}
\label{contour}
\end{figure}

To increase the efficiency of the search and to exploit the information
contained in the distributions of the $\Delta E$, $M_B$, $\cal{F}$ and $dE/dx$
variables a second analysis is performed.
The cuts described in the previous paragraph are removed, and
an unbinned maximum-likelihood fit is made. In this fit the signal and
background distributions are defined by probability density functions
derived from Monte Carlo studies. The fit determines the relative contributions
of $\pi^-\pi^+$, $K^-\pi^+$ and $K^-K^+$ to the signal and background. The best
fit values
for the signal yields $N_{\pi\pi}$, $N_{K\pi}$ and $N_{KK}$, are given
in Table~\ref{Kpi}. Fig.~\ref{contour} shows the $n\sigma$ contours in the
plane $N_{\pi\pi}\; vs.\; N_{K\pi}$, and Fig.~\ref{Kpiproj} shows the
projections
of the likelihood fit onto the $M_B$ and $\Delta E$ axes compared to the
events observed. The efficiency for
a signal event to be included in the likelihood analysis is 38\%.

\begin{figure}[hbt]

\vspace{-1.5cm}
\vspace{-2.6cm}
\vskip 4mm
\caption{CLEO~II results on $B\to \pi^+ \pi^-$ and $K^+ \pi^-$.
Comparison of on-resonance data (histogram) with projections of
the likelihood fit (solid curve). (a) Projection onto $M_B$ after cuts on
$\Delta E$ and $\cal{F}$ (b) Projection onto $\Delta E$ after cuts on $M_B$
and $\cal{F}$. The shaded portions of the histogram are $\pi\pi$ events, the
unshaded are $K\pi$ events. The dotted and dot-dashed lines in (b) indicate the
fit projections for $K\pi$ and $\pi\pi$ separately.}
\label{Kpiproj}
\end{figure}

The best fit value shown in Fig. \ref{contour} is more than 4$\sigma$ away
from the point $N_{\pi\pi} = N_{K\pi} = 0$. After including the effect
of systematic errors on the sum of $N_{\pi\pi}$ and $N_{K\pi}$ \cite{CLNSKpi},
it has been concluded that the significance of the sum is sufficient to claim
the observation of a signal for charmless hadronic $B$ decays.
It should be emphasized that
the present data do not have sufficient statistical precision
to allow any conclusion to be reached about the relative importance of the two
decays. While the
 CLEO~II experiment does not measure signals
for the
individual decays $\bar{B^0}\to\pi^+\pi^-$ and $\bar{B^0}\to K^-\pi^+$,
 it does set stringent upper limits (Table \ref{Kpi}).

Studies have been made of the amount of additional data that might be required
to measure signals in the individual modes, and it is estimated that a sample
of about 3 fb$^{-1}$ is sufficient, assuming that the best fit continues to
give
the same yields for $N_{\pi\pi}$ and $N_{K\pi}$. Note that the
separation between $\pi\pi$ and $K\pi$ provided by $\Delta E$ and $dE/dx$
is adequate for this analysis, as can be seen from the nearly circular
form of the contours in Fig. \ref{contour}.

\section{ELECTROMAGNETIC PENGUIN DECAYS}

\subsection{Observation of $B\to K^* (892)\gamma$}

The first observation of the electromagnetic decay $B\to K^*\gamma$
has been reported by CLEO~II\cite{PRLbsg}.
A data sample of 1.38 pb$^{-1}$ taken on the $\Upsilon$(4S) resonance
was searched for both
 $\bar{B}^0\to \bar{K}^{*0}\gamma$ and $B^-\to K^{*-}\gamma$,
where the $\bar{K}^{*0}$ was detected
in its $K^-\pi^+$ decay mode, and the $K^{*-}$
in both the $K^-\pi^0$ and $K_s\pi^-$ decay modes.
If a $K^*$ candidate is within 75~MeV of the known $K^*$ mass then
it is combined with an isolated photon with an energy between 2.1 and 2.9~GeV.
The photon candidate must not be matched to a charged track, and
must have a shower shape consistent with an isolated
photon.
If the photon candidate forms a $\pi^0$($\eta$) meson when combined with
any another photon with energy greater than 30(200)~MeV it is rejected.

Candidates for $B$ meson decays are identified using the variables
$\Delta E = E_{K^*} + E_{\gamma} - E_{beam}$ and $M_B$.
The r.m.s.
resolutions on $\Delta E$ and $M_B$ are 40~MeV and 2.8~MeV respectively.
\begin{table} [hbt]

\caption{Summary of results for $B\to K^*\gamma$}
\label{TABksg}

\begin{center}
\begin{tabular}{lccc}
&$\bar{B}^0\to K^{*0}\gamma$&  \multicolumn{2}{c}{$B^-\to K^{*-}\gamma$}\\
&$\bar{K}^{*0}\to K^-\pi^+$&$K^{*-}\to K_s\pi^-$&$K^{*-}\to K^-\pi^0$\\ \hline
Signal Events        &  8  &  2  &  3  \\
Continuum Background & 1.1$\pm$0.2 & 0.05$\pm$0.03 & 0.8$\pm$0.3 \\
$B\bar{B}$ Background& 0.30$\pm$0.15 & 0.01$\pm$0.01 & 0.10$\pm$0.05 \\
Detection Efficiency & (11.9$\pm$1.8)\% & (2.0$\pm$0.3)\% & (3.1$\pm$0.5)\%\\
Branching Ratio & (4.0$\pm$1.7$\pm$0.8)$\times 10^{-5}$&
\multicolumn{2}{c}{(5.7$\pm$3.1$\pm$1.1)$\times 10^{-5}$}\\
\end{tabular}
\end{center}
\end{table}

There are two main sources of background from the continuum, $q\bar{q}$ jets
and initial state radiation (ISR). These backgrounds are suppressed by
applying cuts on the shape variables $R_2<0.5$, $|\cos\theta_T|<0.7$,
and $0.25<s_{\perp}<0.60$. The upper restriction on $s_{\perp}$ is useful
for rejecting ISR background.
By transforming the event into the frame where the photon is at
rest, and defining new shape variables $R_2^{'}$ and $\cos\theta_T^{'}$ in this
frame,
the ISR background can be further suppressed.
There is a small amount of background
to $B\to K^*\gamma$ from other $B\bar{B}$ events.
The size of this background was determined from a high statistics
Monte Carlo study. This study includes a feeddown
from other $b\to s\gamma$ decays, which was
estimated using the theoretical models for $b\to s\gamma$ discussed in the
next section.
The remaining background is mainly due to continuum $e^+e^-$ annihilation.
This contribution has been determined using $\Delta E, \; M_B$ sidebands in
both the $\Upsilon(4S)$ and continuum data and
 scaling factors determined from Monte
Carlo studies.

Supporting evidence that the events in the signal region are due to
the decay $B\to K^*\gamma$ comes from a likelihood analysis similar to
the one described in section \ref{newpipi}.
In this analysis the distributions of the events in the variables
$M_B$, $\Delta E$, $M_{K^*}$, $\cos\Theta_{K^*}$ (the $K^*$
helicity angle), $\cos\theta_B$,
$R_2$, $R_2'$, $s_{\perp}$, and $\cos\theta_T$
are compared to the distributions expected from Monte Carlo
samples of signal and continuum background events \cite{CLNSksg}.
This analysis gives results completely consistent with the signal
and background yields given in Table~\ref{TABksg}.

\begin{figure}[hbt]
\vspace{-1.5cm}
\vspace{-2.5cm}
\vskip 4mm
\caption{The beam-constrained mass distribution
from CLEO~II for $B\to K^*\gamma$
candidates: $K^-\pi^+\gamma$ solid, $K^-\pi^0\gamma$ shaded, $K_s\pi^-\gamma$
unshaded}
\label{FIGksg}
\end{figure}

The eight $\bar{K}^{*0}\gamma$ and five $K^{*-}\gamma$
events in the signal region, $|\Delta E|<90$~MeV and $5.274<M_B<5.286$~GeV, are
a clear signal for the decay $B\to K^*\gamma$ (Fig.~\ref{FIGksg}).
The yields in the observed modes are consistent.
Assuming that $\bar{B}^0\to \bar{K}^{*0}\gamma$ and $B^-\to K^{*-}\gamma$ are
equal, the average branching ratio is $(4.5\pm 1.5\pm 0.9)\times 10^{-5}$.
This is in agreement with theoretical predictions from the
electromagnetic penguin diagram \cite{Ali}.

\subsection{Experimental Constraints on the $ b \to s \gamma$ Inclusive Rate}

At present,
due to the uncertainties in the hadronization, only the inclusive $b \to s
\gamma$
rate can be reliably compared with theoretical calculations.
This rate can be measured from the endpoint of the
inclusive photon spectrum in $B$ decay.
The signal for $b\to s\gamma$ is expected to peak in the region
$2.2<E_{\gamma}<2.7$~GeV, with only 15\% of the rate expected to be
outside this range \cite{Ali}.

\begin{figure}[hbt]
\vspace{-1.5cm}
\vspace{-3.0cm}
\vskip 4mm
\caption{Inclusive gamma spectrum
from CLEO~II for (a) On resonance (solid line) and scaled
off resonance (dashed line) (b) excess of gammas from $B$ decays}
\label{FIGbsg}
\end{figure}

The experimental situation is shown in Fig.~\ref{FIGbsg}.
In the CLEO~II data there is an excess of $69\pm 43$ events from $B$ decays in
the region
$2.2<E_{\gamma}<2.7$~GeV.
The contribution in the signal region due to photons coming from
$b\to c$ decays is thought to be small compared to the expected signal
from $b\to s\gamma$. A conservative upper limit is being quoted
by not making a $b\to c$ subtraction,
and by assuming that only 70\% of the $b\to s\gamma$ rate is in the energy
window \cite{PCK}.

An alternative approach to measuring the inclusive rate is to use the
observed exclusive rate for $B\to K^*\gamma$.
However, the fraction of the inclusive rate that hadronizes to a particular
exclusive final state is not very well understood. Ali \etal \cite{Ali}
predict the mass distribution of the $X_s$ system using an estimate of the
Fermi momentum of the
spectator quark ($p_{F}=300$ MeV)\cite{CLEOpf} .
By integrating this spectrum up to
1~GeV and assuming this region is dominated by the $K^*$ resonance,
the fraction of $K^*(892)\gamma$ is estimated to be (13$\pm$3)\%.
Other authors have made predictions between 5\% and 40\%
for the fraction of $K^*(892)\gamma$ \cite{ksg}. A reasonable estimate that
covers most of the theoretical predictions is (13$\pm$6)\%.
Combining this number with the measured branching ratio
for $B\to K^*\gamma$ a lower limit can be obtained for the inclusive rate.
This approach would become more useful if additional exclusive channels with
higher mass $X_s$ systems were to be observed.

Using the inclusive measurement for the upper limit, and the observation
of $B\to K^*\gamma$ for the lower limit, the present 95\% C.L. limits on the
inclusive rate are \cite{PCK}:
\begin{equation}
0.8\times 10^{-4} < {\cal{B}}(b\to s\gamma) < 5.4\times 10^{-4}
\end{equation}
It is anticipated that CLEO~II will improve both of these limits, and will
eventually observe a signal in the inclusive photon spectrum.

Searches have also been made for $b\to s\gamma$ processes
at LEP. The L3 experiment has set an upper limit of $1.2\times 10^{-3}$
(90\% C.L.) on the inclusive $b\to s\gamma$ rate \cite{L3shit}.
The exclusive decays $\bar{B}^0\to \bar{K}^{*0}\gamma$ and $B_s\to \phi\gamma$
have been searched for by the DELPHI experiment using
the particle identification capabilities of the RICH detector. Upper limits
of $3.6\times 10^{-4}$ and $19.0\times 10^{-4}$ are obtained for
these two decays \cite{Battaglia}.
The results of the searches performed at the $Z^0$ are
inferior to the results from CLEO~II due to the smaller number
of $B$ mesons produced, and the large backgrounds from non $b\bar{b}$
processes. In the L3 analysis this large background is subtracted
according to the predictions of the JETSET Monte Carlo. Another problem
at LEP is that the photons have higher energies
and are more difficult to distinguish from $\pi^0$ and $\eta$ mesons.

\subsection{Theoretical Implications of $b \to s \gamma$}

There are many calculations of the inclusive rate for $b\to s\gamma$
\cite{Ali}, \cite{bsg}.
The rate has a logarithmic dependence on the top quark mass, $m_t$, and is
proportional to the product of CKM matrix elements $|V_{ts}V_{tb}|^2$.
Large leading order QCD corrections increase the rate by a factor of about 3.5.
Using the limits $100 < m_t < 200$~GeV, and allowing the
range of mass scales, $\mu$, at which the QCD corrections are evaluated,
to vary between $m_b/2$ and $2m_b$, Ali and Greub calculate the inclusive
rate to be $(3.0\pm 1.2)\times 10^{-4}$.
This prediction is completely consistent with the experimental results
discussed in the previous section.
Ali and Greub have used the
experimental bounds to determine the range of possible values
for the ratio CKM matrix element $|V_{ts}/V_{cb}|$:
$$0.50 < |V_{ts}/V_{cb}| < 1.67$$
which is expected from unitarity to be close to 1.

The ratio
$|V_{td}/V_{ts}|$ can be determined
 from a comparison of the decay rates
for $B\to\rho\gamma$ (or $B\to \omega \gamma$)
and $B\to K^*\gamma$. In this ratio many of the theoretical uncertainties are
expected to cancel.

There has been recent interest in $b\to s\gamma$ as a probe of physics beyond
the standard model \cite{Hewett,SUSY}. There are possible additional
contributions to the loop from a charged Higgs boson and from supersymmetric
particles. Hewett \cite{Hewett} has considered two Higgs doublet models and
shown that contributions comparable to the standard model are expected for a
charged Higgs mass of order 100~GeV. In supersymmetric models there
are also contributions from loops containing charginos, neutralinos and
squarks that tend to cancel the
charged Higgs and standard model contributions (in unbroken supersymmetry all
contributions to the loop diagram cancel exactly)\cite{cancel}.
Several recent papers
\cite{SUSY} investigate the parameter space allowed by $b\to s\gamma$
for particular models of the breaking of the supersymmetry. For most of the
parameter space the charged Higgs contribution is the dominant one, and
the present CLEO~II upper limit on $b\to s\gamma$ constrains the charged Higgs
mass to be greater than 200~GeV.
This is more restrictive than constraints from direct searches at existing
high energy colliders.
An important consequence of this limit
would be to forbid the decay $t\to b H^+$.
The limit on the charged Higgs mass can be avoided in supersymmetric models
if the stop mass is small
since this leads to a large
negative contribution from the chargino-stop loop. For this case the rate for
$b\to s\gamma$ could even become smaller than the standard model prediction.

Other constraints on new physics have been  derived from the bounds
on $b \to s \gamma$.
If there are anomalous $W- W- \gamma$ couplings, these
can significantly modify the rate for $b\to s\gamma$.
The CLEO measurements exclude
 certain regions of the parameter space of
anomalous dipole and quadrupole couplings of the W boson that cannot
be explored by direct studies of $W^+ -\gamma$ production at hadron colliders
\cite{Chiawwg}.


\subsection{$b\to s \ell ^+\ell ^-$ Decays}

The $b\to s\gamma$ diagram can be modified by replacing the real photon by a
virtual photon or by a virtual $Z^0$ or  other neutral boson
that produces a lepton pair (see Fig.~\ref{kstgll}).
 This penguin diagram leads to both
$B\to K \ell ^+ \ell ^-$ and $B\to K^* \ell ^+\ell ^-$ decays, since the $B\to
K$ transition is
no longer forbidden by angular momentum conservation as it was for
$b\to s\gamma$. Although the penguin amplitude for $b\to s \ell ^+\ell ^-$ is
smaller
than $b\to s\gamma$ the final states can be identified easily,
and are particularly favorable for study at hadron colliders.
As in the radiative penguin decay discussed previously, the
process $b\to s \ell ^+ \ell ^-$
is sensitive to high mass physics including charged Higgs bosons
 and non-standard neutral particles.

\begin{figure}[htb]
\begin{center}
\unitlength 1.0in
\begin{picture}(3.,0.8)(0,0)
\end{picture}
\vskip 15 mm
\caption{Diagrams for the decays $B\to K^{(*)} \ell ^+ \ell ^-$.}
\label{kstgll}
\end{center}
\end{figure}

The penguin amplitude has been calculated by a number of authors
\cite{PengKll}, with results for the inclusive $b\to s e^+ e^-$ rate
of $ (1-2)\times 10^{-5}$ and for the $b\to s \mu^+ \mu^-$ rate of
$ (4-8) \times 10^{-6}$.
 The exclusive channels $K^* \ell ^+\ell ^-$
and $K \ell ^+\ell ^-$ are expected to comprise $5-30\%$
of the inclusive rate.  However, the theoretical
description of $b\to s\ell ^+\ell ^-$ is more complicated
than $b\to s\gamma$, since the
final states $K^{(*)}\ell ^+\ell ^-$ can also be produced
via ``long distance'' contributions from the hadronic
decay $B\to K^{(*)}\Psi$ followed by $\Psi\to \ell ^+\ell ^-$ where
$\Psi$ stands for a real or virtual charmonium state \cite{LongD}.
Ali, Mannel and Morozumi \cite{AliKll} have performed
 a full analysis of $b\to s \ell^+\ell^-$
including both the penguin and the long distance contributions.
Their predictions for the inclusive $b\to s \ell^+ \ell^-$ rate are in the
range $(2-6)\times 10^{-6}$ excluding the regions
close to the $\psi$ and $\psi'$ mass where the long distance contributions
dominate. There is interference between the penguin and long distance
amplitudes over a wide range of dilepton masses. Ali \etal\  point out that the
sign of the interference is controversial, and that information about the
interference can be obtained both from the dilepton mass distribution, and
from the forward-backward asymmetry of the lepton pair.

Experimental searches have
been made by CLEO and ARGUS at the $\Upsilon (4S)$, and by UA1 in $p\bar{p}$
collisions. The CLEO and ARGUS analyses \cite{CLEOKll,ARGUSKll} make a simple
veto on dilepton masses consistent with a real $\psi$ or $\psi'$, and see
almost no background in their beam-constrained mass plots. The UA1 analysis
\cite{UA1Kll} selects a range $3.9 < M(\ell ^+\ell ^-) < 4.4$~GeV which is
believed
to have small long distance contributions and no radiative tail from the
$\Psi$.
 UA1 performs both an exclusive
search for $\bar{B}^0\to \bar{K}^{*0}\mu^+\mu^-$ and an inclusive search for
$B\to X_s\mu^+\mu^-$. The upper limits from all the experimental measurements
are summarized in Table~\ref{Kll}. These upper limits are all well above
the theoretical predictions, but a detailed comparison of the various limits
is complicated by the different dilepton mass ranges considered by the
different analyses. The limits from UA1 suggest that $b\to s \ell ^+\ell ^-$
decays
might eventually be observed at hadron colliders, as well as by
$\Upsilon (4S)$ experiments that have collected large enough
data samples to be sensitive to these decay modes.
\begin{table} [hbt]

\caption{Experimental upper limits (90\% C.L.)
for $b\to s\ell ^+\ell ^-$ decays. All numbers quoted are branching fractions
$\times 10^{-5}$}
\label{Kll}

\begin{center}
\begin{tabular}{lcccc}
$B$ Decay       &   ARGUS  &  CLEO I  & CLEO 1.5 & UA1    \\ \hline
$K^0e^+e^-$     &  $<$15.0 & $<$56.0  &          &        \\
$K^- e^+e^-$     &  $<$9.0  & $<$24.0  & $<$5.7   &        \\
$K^0\mu^+\mu^-$ &  $<$26.0 & $<$39.0  &          &        \\
$K^- \mu^+\mu^-$ &  $<$22.0 & $<$36.0  & $<$17.0  &        \\
$\bar{K}^{*0}e^+e^-$  &  $<$29.0 &          & $<$6.9   &        \\
$K^{*-}e^+e^-$  &  $<$63.0 &          &          &        \\
$\bar{K}^{*0}\mu^+\mu^-$&$<$34.0 &          & $<$16.0  & $<$2.3 \\
$K^{*-}\mu^+\mu^-$&$<$110.0&          &          &        \\
$X_s\mu^+\mu^-$ &          &          &          & $<$5.0 \\
\end{tabular}
\end{center}
\end{table}

\section{PURELY LEPTONIC B DECAY}

\subsection{$B$ Decays to Two Leptons}

The Standard Model allows $B^0$ and $B_s$ mesons to decay to $e^+e^-$
$\mu^+\mu^-$ or $\tau^+\tau^-$ via box diagrams or loop diagrams involving
both $W$ and $Z$ propagators (see Fig.~\ref{dilepfig})
 \cite{RareAli}. The largest branching fraction
is predicted to be $4\times 10^{-7}$ for $B_s\to\tau^+\tau^-$, and the
smallest $2\times 10^{-15}$ for $B^0\to e^+e^-$. The decays to the lighter
leptons are suppressed by a helicity factor which is proportional to
$m_{\ell} ^2$, and the $B^0$ decays are suppressed
relative to the $B_s$ decays by the factor $|V_{td}/V_{ts}|^2$.
Decays to the final states $e^{\pm}\mu^{\mp}$, $e^{\pm}\tau^{\mp}$ and
$\mu^{\pm}\tau^{\mp}$ are all forbidden in the Standard Model by lepton
family number conservation.

A search for $B^0$ decays to two leptons has been made by CLEO~II
\cite{CLEOll}, and there are also searches for
$B^0\to\mu^+\mu^-$ by the UA1 and CDF collaborations at hadron colliders
\cite{UA1Kll,CDFll}. The 90\% C.L. upper limits on the allowed processes are
$5.9\times 10^{-6}$ for $B^0\to e^+e^-$ (CLEO~II), and
$3.2\times 10^{-6}$ (CDF), $5.9\times 10^{-6}$ (CLEO~II) and
$8.3\times 10^{-6}$ (UA1) for $B^0\to\mu^+\mu^-$. The hadron collider
experiments can undoubtedly set similar limits on $B_s\to\mu^+\mu^-$, and
presumably have not done so because the $B_s$ mass
was unknown until recently (see section \ref{Bs-mass} ) .

\begin{figure}[htb]
\begin{center}
\unitlength 1.0in
\begin{picture}(3.,0.5)(0,0)
\end{picture}
\bigskip
\vskip 5 mm
\caption{Diagrams for the dilepton decays
 of $B$ mesons.}
\label{dilepfig}
\end{center}
\end{figure}

CLEO~II also sets limits on the lepton-flavor changing decays of
$5.9\times 10^{-6}$ for $B^0\to e^{\pm}\mu^{\mp}$,
$7.9\times 10^{-4}$ for $B^0\to e^{\pm}\tau^{\mp}$  and
$1.2\times 10^{-3}$ for $B^0\to\mu^{\pm}\tau^{\mp}$.

Several recent papers consider the relative sensitivity of
various lepton-flavor changing decays to non-Standard Model couplings
\cite{SherYuan},\cite{Campbell} .
Sher and Yuan
argue that larger Yukawa couplings
are expected for third generation quarks,
and that these larger couplings not only enhance the sensitivity of the decays,
but also make them less dependent on the detailed parameterization of the new
couplings\cite{SherYuan}.
They make a comparison of $B$ and $K$ decays which suggests that
$B_s\to\tau\mu$ has the best sensitivity, although it is unclear how to
search for this channel experimentally. The more accessible channel
$B_s\to\mu e$ could also have better sensitivity than the
equivalent decay $K_{L}\to\mu e$, even
though the upper limit on the latter is now
in the $10^{-11}$ range.
$B^0$ decays are less sensitive than $B_s$ decays but are still of
interest because they can be
searched for in experiments at the $\Upsilon (4S)$.

\subsection{The Decays $B\to\tau\nu$, $B\to\mu\nu$ and $B\to e\nu$.}

The decay $B^+\to\tau^+\nu$ proceeds through the annihilation of the
constituent quarks in analogy to the $\pi^+\to\mu^+\nu$ decay.
The branching fraction is given by:
$$ {\cal{B}}(B^+\to\tau^+\nu) = \frac{G_F^2m_Bm_\tau^2}{8\pi}
\left(1-\frac{m_\tau^2}{m_B^2}\right)
f_B^2|V_{ub}|^2\tau_B $$
All the parameters in this equation are well known except the
decay constant $f_B$ and the CKM matrix element $V_{ub}$.
Given a more accurate knowledge of $V_{ub}$ from other measurements
and the experimental observation of the decay $B^+\to\tau^+\nu$, it
would be possible to determine a value for $f_B$. The measurement of
this decay constant is of fundamental importance for $B$ physics
since it enters into many other $B$ decay measurements, including
most notably $B\bar{B}$ mixing \cite{mixing}.

The present theoretical estimates of $f_B$ from lattice QCD and QCD sum rules
are in the range $f_B = (180\pm 50)$~MeV \cite{fB}.
Using this value of $f_B$ and our standard values of $V_{ub}$ and $\tau_B$,
we obtain a prediction of ${\cal{B}}(B^+\to\tau^+\nu) = 4.0\times 10^{-5}$.
The decays $B^+\to\mu^+\nu$ and $B^+\to e^+\nu$ have smaller branching ratios
of $1.4\times 10^{-7}$ and $3.3\times 10^{-12}$ respectively. The decays to the
muon and electron are suppressed  relative to the tau decay by a helicity
factor proportional to $m_{\ell} ^2$.

CLEO~II is investigating ways of observing these decay modes \cite{DPFTaunu}.
For $B^+\to\tau^+\nu$ the method that has been tried is to start with a sample
of 716 fully reconstructed $B^-$ mesons, including many of the decay channels
that have been discussed earlier in this review. The signature for the decay
of the $B^+$ meson to $\tau^+\nu$ , $\tau^+ \to \ell ^+
 \nu \bar{\nu}$ or $\tau^+ \to \pi^+ \nu$ is the presence of
one and only one charged track and no neutral showers with energies greater
than 200~MeV in addition to those tracks and showers that were used to
reconstruct the $B^-$ meson. No candidates are observed
among the 716 reconstructed $B^-$ events. The efficiency for finding a
$B^+\to\tau^+\nu$ event among these fully reconstructed $B^-$ events,
including the $\tau \to l \nu \bar{\nu}$ and $ \tau \to \pi \nu$
 branching fractions, is
 $(26.0\pm 3.3)$\%. This leads to a 90\% C.L. upper limit of
${\cal{B}}(B^+\to\tau^+\nu) < 1.3$\%.

The technique that CLEO~II is using for $B^+\to\mu^+\nu$
is somewhat different.
For this mode it is not necessary to fully reconstruct the other $B$ meson.
 The $B$ meson decays almost at rest into a $\mu^+$
and a neutrino which are back-to-back and have energies of about 2.65~GeV.
The muon is well identified and has little background. The neutrino is
``detected'' by calculating the missing momentum $p_{miss}$ of the whole event.
If all the decay products of the other $B^-$ have been measured by the CLEO~II
detector $p_{miss}$ will be a good estimator of the neutrino momentum.
Then the analysis proceeds as if this were a fully reconstructed $B$ decay,
with the calculation of the energy difference, $\Delta E$, and the
beam-constrained mass, $M_B$. Since there is often a small amount of
undetected energy from the decay of the other $B$, a cut is made
at $-200 < \Delta E < +400$~MeV. This analysis is almost
background free, and gives a 90\% C.L. upper limit of
${\cal{B}}(B^+\to\mu^+\nu) < 2.0\times 10^{-5}$. A similar limit is
obtained for $B^+\to\ e^+\nu$.

The CLEO~II limits of $B^+\to\tau^+\nu$ and $B^+\to\mu^+\nu$ are both
two orders of magnitude above the theoretical predictions, giving rather
uninteresting limits on $f_B$ of 3.2 and 2.1 GeV for $|V_{ub}/V_{cb}|=0.075$.
However, it should be noted
that the experimental searches are not yet background limited, so the
upper limits on the branching ratios
will improve linearly with future increases in statistics.
Efforts are being made to improve the $B\to \tau\nu$
analysis either
by relaxing the requirements for full reconstruction, or by adding
more $\tau$ decay modes.

\section{CONCLUSIONS}

In the past two years significant progress in
the area of hadronic $B$ decay has been made.
The data are now of sufficient quality to perform non-trivial
tests of the factorization hypothesis including comparisons
of rates for $D^{*+} X^-$
(where $X^-=\pi^- ,\rho^-$, or $a_1^-$)
with rates for  $D^{*+} \ell ^- \bar{\nu}$
at $q^2=M_X^2$, as well as comparisons of  the polarizations in
$D^{*+}\rho^-$ with $D^{*+} \ell^-\bar{\nu}_\ell$. In all cases the
factorization hypothesis is consistent with the data at the present
level of experimental precision and for $q^2 < m_{a_1}^2$.

Improved measurements of branching ratios of two-body
decays with a final state $\psi$ meson
have been reported from CLEO~II and ARGUS.
The decay $B \to \psi K^*$ is
strongly polarized with
$\Gamma_L / \Gamma = (84 \pm 6 \pm 8)$ \%; therefore this mode
will be  useful for measuring CP violation.

There is no evidence for
color suppressed decays to a charmed meson and light
neutral
hadron in the final state.
The most stringent limit,
${\cal B}(\bar{B^0}\to D^0\pi^0) / {\cal B}(\bar{B^0}\to D^+\pi^-) < 0.07$
from CLEO~II, is
still above the level where these
color suppressed $B$ decays are expected in most
models.
The observation of $B \to \psi$ modes
shows that color suppressed decays are present. Using results on exclusive
$B \to \psi$ decays from CLEO~1.5, CLEO~II and ARGUS,
 we find a value of the BSW parameter
$|a_2|\; = \; 0.23 \pm 0.01 \pm  0.01$. We also report a new value for
the BSW parameter
$|a_1|\; = \; 1.07 \pm 0.04 \pm 0.06$.
By comparing rates for $B^-$ and $\bar{B}^0$ modes, it has
 been shown that the sign of
$a_2/a_1$ is positive, in contrast to what is found in charm decays.

There has  been dramatic progress in the study of rare decays.
CLEO~II has reported evidence for charmless
hadronic $B$ decay in the sum of $B\to K^+ \pi^-$ and $B\to \pi^+ \pi^-$
and has observed the first direct evidence
for the radiative penguin decay $B \to K^{*} \gamma$
with a branching ratio of $(4.5\pm 1.5 \pm 0.9) \times 10^{-5}$.
CLEO~II also sets
 limits on the inclusive process
$0.8\times 10^{-4} < {\cal{B}}(b\to s\gamma) < 5.4\times 10^{-4}$ .
These results restrict the allowed range for $V_{ts}$ and constrain
physics beyond the Standard Model.

Large samples of reconstructed hadronic decays will be obtained
in the next few years by the CLEO~II collaboration as a result
of further improvements in the luminosity of CESR, and in the
performance of the CLEO~II detector.
This will permit accurate tests of the factorization hypothesis
over the full $q^2$ range. The large tagged sample can be used
to study inclusive properties of $B^+$ and $B^0$ decays and
constrain $f_{B}$ via $B^+\to \tau^+ \nu$.
Measurements of  additional decays to final states with charmonium
mesons will be performed and
other color suppressed decays will be observed.

A larger data sample
should allow further results to be obtained on rare $B$ decays
including the observation of $B^0\to \pi^+ \pi^-$, $B^0\to K^+ \pi^-$
and a measurement of the inclusive process $b\to s \gamma$.
The measurement of several rare hadronic decays would
provide information on the relative importance of the
penguin and spectator amplitudes. Additional electromagnetic
penguin decays such as $B\to K^{**} \gamma$,
 $B\to K^* \ell ^+ \ell ^-$ and $B\to \rho \gamma$
may be observed. These provide further constraints on
Standard model parameters and extensions of the Standard Model.

A goal of the study of
 hadronic and rare $B$ decays is to eventually
 measure CP asymmetries in $B$ decays such as
$B\to \psi K_s$ and $B\to \pi^+ \pi^-$.
In order to test
the consistency of the Standard Model description of
CP violation in these decays, the mechanisms
of $B$ decay must be well understood. This review shows that
rapid progress is being made in understanding $B$ decay.

\acknowledgements

We thank  H. Albrecht, A. Ali, C. Bebek, P. Colangelo, J. Hewett, N. Isgur,
G. Kane, S. Menary,
M. Neubert, V. Rieckert, J. Rodriguez, and M. Zoeller
for useful discussions. We thank our colleagues from the CLEO
and ARGUS experiments for useful comments and for their contributions
to the work discussed in this review.
We thank the National Science Foundation,
the Department of Energy, the University of Hawaii,
Ohio State University and Syracuse University for their support.

\clearpage
\newpage

\centerline{\Large{\bf APPENDIX}}
\addcontentsline{toc}{section}{APPENDIX}
Tables XVI and XVII
contain the $B$ meson branching fraction
as measured by the ARGUS, CLEO 1.5 and CLEO II experiments. In this appendix
we list more technical information
 found in the ARGUS \cite{ThirdB} -\cite{FifthB},
\cite{ARGUSDDs}, \cite{arguspsi} and CLEO \cite{SecondB}, \cite{DDcleo},
\cite{fastpsi}, \cite{SixthB}
publications. This includes the number of signal events and the reconstruction
efficiencies.
Note that different experiments used different procedures to obtain branching
ratios in modes where several $D$ or $\psi$ decay channels were used (see Sec.
\ref{thatsit}.
The  information provided here will be useful
to estimate the signal yields for future $B$ experiments and also
to rescale the $B$ meson branching ratios when more precise measurements
of the charmed meson branching fraction become available.
\begin{table}[htb]
\caption{Detailed $B^-$ branching ratios. Experiment: ARGUS}
\begin{tabular}{lll}
\label{argus_bm}
$ B^- $ decay & Signal events & Branching ratio [\%]\\
\hline
$B^- \rightarrow D^0 \pi ^-$ & $  12 \pm  5 $& $ 0.22 \pm 0.09 \pm 0.06 \pm
0.01 $ \\
$B^- \rightarrow D^0 \rho ^-$ & $  19 \pm  6 $& $ 1.45 \pm 0.45 \pm 0.41 \pm
0.06 $ \\
$B^- \rightarrow D^{*0} \pi ^-$ & $   9 \pm  3 $& $ 0.39 \pm 0.14 \pm 0.10 \pm
0.02 $ \\
$B^- \rightarrow D^{*0} \rho ^-$ & $   7 \pm  4 $& $ 0.96 \pm 0.58 \pm 0.36 \pm
0.04 $ \\
$B^- \rightarrow D_J^{(*)0} \pi ^-$ & $   6 \pm  3 $& $ 0.13 \pm 0.07 \pm 0.03
\pm 0.01 $ \\
$B^- \rightarrow D^{*+} \pi ^- \pi ^- \pi ^0$ & $  26 \pm 10 $& $ 1.68 \pm 0.65
\pm 0.38 \pm 0.07 $ \\
$B^- \rightarrow D_J^{(*)0} \rho ^-$ & $   5 \pm  3 $& $ 0.33 \pm 0.20 \pm 0.08
\pm 0.01 $ \\
$B^- \rightarrow D^{*+} \pi ^- \pi ^-$ & $  11 \pm  6 $& $ 0.24 \pm 0.13 \pm
0.05 \pm 0.01 $ \\
$B^- \rightarrow D^0 D_s^-$ & $ 4.4 \pm 2.2 $& $ 1.65 \pm 0.82 \pm 0.42 \pm
0.07 $ \\
$B^- \rightarrow D^0 D_s^{*-}$ & $ 2.3 \pm 1.8 $& $ 1.10 \pm 0.82 \pm 0.30 \pm
0.04 $ \\
$B^- \rightarrow D^{*0} D_s^-$ & $ 2.0 \pm 1.4 $& $ 0.77 \pm 0.53 \pm 0.18 \pm
0.03 $ \\
$B^- \rightarrow D^{*0} D_s^{*-}$ & $ 4.8 \pm 2.5 $& $ 1.84 \pm 0.95 \pm 0.44
\pm 0.07 $ \\
$B^- \rightarrow \psi K^-$ & $   6 $& $ 0.08 \pm 0.04 \pm 0.01 $\\
$B^- \rightarrow \psi ' K^-$ & $   5 $& $ 0.20 \pm 0.09 \pm 0.04 $\\
$B^- \rightarrow \psi K^{*-}$ & $   2 $& $ 0.19 \pm 0.13 \pm 0.03 $\\
$B^- \rightarrow \psi ' K^{*-}$ & $ < 3.9  $& $ < 0.53 $ at $90 $\% C.L. \\
$B^- \rightarrow \psi K^- \pi ^+ \pi ^-$ & $ <   8  $& $ < 0.19 $ at $90 $\%
C.L. \\
$B^- \rightarrow \psi ' K^- \pi ^+ \pi ^-$ & $   3 $& $ 0.21 \pm 0.12 \pm 0.04
$\\
$B^- \rightarrow \chi_{c1} K^-$ & $   4 \pm 2.0 $& $ 0.22 \pm 0.15 \pm 0.07 $\\
\end{tabular}
\end{table}
\begin{table}[htb]
\caption{Detailed $\bar{B}^0$ branching ratios. Experiment: ARGUS}
\begin{tabular}{lll}
\label{argus_b0}
$ \bar{B}^0 $ decay & Signal events & Branching ratio [\%]\\
\hline
$\bar{B}^0 \rightarrow D^+ \pi ^-$ & $  22 \pm  5 $& $ 0.48 \pm 0.11 \pm 0.08
\pm 0.07 $ \\
$\bar{B}^0 \rightarrow D^+ \rho ^-$ & $   9 \pm  5 $& $ 0.90 \pm 0.50 \pm 0.27
\pm 0.14 $ \\
$\bar{B}^0 \rightarrow D^{*+} \pi ^-$ & $  12 \pm  4 $& $ 0.26 \pm 0.08 \pm
0.04 \pm 0.01 $ \\
$\bar{B}^0 \rightarrow D^{*+} \rho ^-$ & $  19 \pm  9 $& $ 0.65 \pm 0.28 \pm
0.26 \pm 0.03 $ \\
$\bar{B}^0 \rightarrow D^{*+} \pi ^- \pi ^- \pi ^+$ & $  26 \pm  7 $& $ 1.12
\pm 0.28 \pm 0.33 \pm 0.04 $ \\
$\bar{B}^0 \rightarrow D^+ D_s^-$ & $ 2.4 \pm 1.8 $& $ 1.09 \pm 0.83 \pm 0.44
\pm 0.16 $ \\
$\bar{B}^0 \rightarrow D^+ D_s^{*-}$ & $ 3.2 \pm 2.0 $& $ 1.73 \pm 1.09 \pm
0.67 \pm 0.26 $ \\
$\bar{B}^0 \rightarrow D^{*+} D_s^-$ & $ 2.6 \pm 1.8 $& $ 0.80 \pm 0.57 \pm
0.22 \pm 0.03 $ \\
$\bar{B}^0 \rightarrow D^{*+} D_s^{*-}$ & $ 3.9 \pm 2.0 $& $ 1.49 \pm 0.80 \pm
0.43 \pm 0.06 $ \\
$\bar{B}^0 \rightarrow \psi K^0$ & $   2 $& $ 0.09 \pm 0.07 \pm 0.02 $\\
$\bar{B}^0 \rightarrow \psi ' K^0$ & $ < 2.3  $& $ < 0.31 $ at $90 $\% C.L. \\
$\bar{B}^0 \rightarrow \psi \bar{K}^{*0}$ & $   6 $& $ 0.13 \pm 0.06 \pm 0.02
$\\
$\bar{B}^0 \rightarrow \psi ' \bar{K}^{*0}$ & $ < 3.9  $& $ < 0.25 $ at $90 $\%
C.L. \\
$\bar{B}^0 \rightarrow \psi ' K^- \pi ^+$ & $ < 2.3  $& $ < 0.11 $ at $90 $\%
C.L. \\
\end{tabular}
\end{table}
\begin{table}[htb]
\caption{Detailed ${B}^-$ branching ratios. Experiment: CLEO 1.5}
\begin{tabular}{llllll}
\label{cleo15_bm}
$B^- $ decay & Signature & Signal & Eff. & BR [\%] & Branching ratio [\%]\\
\hline
$B^- \rightarrow D^0 \pi ^-$ & & & & & $ 0.56 \pm 0.08 \pm 0.05 \pm 0.02 $ \\
 & $D^0 \rightarrow K^- \pi ^+$ &$  19 \pm  5 $ & 0.42 &$ 0.50 \pm 0.12 $ & \\
 & $D^0 \rightarrow K^- \pi ^+ \pi ^+\pi ^-$ &$  25 \pm  6 $ & 0.27 &$ 0.49 \pm
0.10 $ & \\
 & $D^0 \rightarrow \bar{K}^0 \pi ^+ \pi ^-$ &$  10 \pm  4 $ & 0.05 &$ 1.32 \pm
0.55 $ & \\
$B^- \rightarrow D^{0} \pi ^+ \pi ^- \pi ^-$ & & & & & $ 1.24 \pm 0.31 \pm 0.14
\pm 0.05 $ \\
 & $D^0 \rightarrow K^- \pi ^+$ &$  34 \pm  8 $ & 0.32 &$ 1.24 \pm 0.31 $ & \\
$B^- \rightarrow D^{*0} \pi ^-$ & & & & & $ 0.99 \pm 0.25 \pm 0.17 \pm 0.04 $
\\
 & $D^0 \rightarrow K^- \pi ^+$ &$   9 \pm  3 $ & 0.13 &$ 0.97 \pm 0.35 $ & \\
 & $D^0 \rightarrow K^- \pi ^+ \pi ^+\pi ^-$ &$  12 \pm  4 $ & 0.08 &$ 0.94 \pm
0.35 $ & \\
$B^- \rightarrow D_J^{(*)0} \pi ^-$ & & & & & $ 0.13 \pm 0.07 \pm 0.01 \pm 0.01
$ \\
 & $D^0 \rightarrow K^- \pi ^+$ &$ 2.2 \pm 1.5 $ & 0.22 &$ 0.15 \pm 0.10 $ & \\
 & $D^0 \rightarrow K^- \pi ^+ \pi ^+\pi ^-$ &$ 1.8 \pm 1.5 $ & 0.13 &$ 0.11
\pm 0.09 $ & \\
$B^- \rightarrow D^{*+} \pi ^- \pi ^-$ & & & & & $<0.37 $ \\
 & $D^0 \rightarrow K^- \pi ^+$ &$ <  8 $ & 0.22 &$ < 0.54 $ & \\
 & $D^0 \rightarrow K^- \pi ^+ \pi ^+\pi ^-$ &$ <3.5 $ & 0.11 &$ < 0.24 $ & \\
$B^- \rightarrow D^0 D_s^-$ & & & & & $ 1.66 \pm 0.70 \pm 0.71 \pm 0.07 $ \\
 &   &$ 5.0 \pm 2.2 $ & 0.07 &$ 1.66 \pm 0.70 $ & \\
$B^- \rightarrow \psi K^-$ & & & & & $ 0.09 \pm 0.02 \pm 0.02 $ \\
 & $\psi  \rightarrow \mu ^+ \mu ^- , e^+ e^-$ &$  11 \pm  3 $ & 0.41 &$ 0.09
\pm 0.02 $ & \\
$B^- \rightarrow \psi ' K^-$ & & & & & $<0.05 $ \\
 & $\psi ' \rightarrow \mu ^+ \mu ^- , e^+ e^-$ &$ <2.3 $ & 0.53 &$ < 0.10 $ &
\\
 & $\psi ' \rightarrow \psi \pi ^+ \pi ^-$ &$ <2.3 $ & 0.23 &$ < 0.11 $ & \\
$B^- \rightarrow \psi K^{*-}$ & & & & & $ 0.15 \pm 0.11 \pm 0.03 $ \\
 & $\psi  \rightarrow \mu ^+ \mu ^- , e^+ e^-$ &$   2 \pm  1 $ & 0.05 &$ 0.15
\pm 0.11 $ & \\
$B^- \rightarrow \psi ' K^{*-}$ & & & & & $<0.38 $ \\
 & $\psi ' \rightarrow \mu ^+ \mu ^- , e^+ e^-$ &$ <2.3 $ & 0.08 &$ < 0.70 $ &
\\
 & $\psi ' \rightarrow \psi \pi ^+ \pi ^-$ &$ <2.3 $ & 0.03 &$ < 0.82 $ & \\
$B^- \rightarrow \psi K^- \pi ^+ \pi ^-$ & & & & & $ 0.14 \pm 0.07 \pm 0.03 $
\\
 & $\psi  \rightarrow \mu ^+ \mu ^- , e^+ e^-$ &$   6 \pm  3 $ & 0.14 &$ 0.14
\pm 0.07 $ & \\
\end{tabular}
\end{table}
\begin{table}[htb]
\caption{Detailed $\bar{B}^0$ branching ratios. Experiment: CLEO 1.5}
\begin{tabular}{llllll}
\label{cleo15_b0}
$\bar{B}^0 $ decay & Signature & Signal & Eff. & BR [\%] & Branching ratio
[\%]\\
\hline
$\bar{B}^0 \rightarrow D^+ \pi ^-$ & & & & & $ 0.27 \pm 0.06 \pm 0.03 \pm 0.04
$ \\
 & $D^+ \rightarrow K^- \pi ^+ \pi ^+$ &$  17 \pm  4 $ & 0.33 &$ 0.23 \pm 0.06
$ & \\
 & $D^+ \rightarrow \bar{K}^0 \pi ^+$ &$   4 \pm  2 $ & 0.09 &$ 0.56 \pm 0.30 $
& \\
$\bar{B}^0 \rightarrow D^+ \pi ^- \pi ^- \pi ^+$ & & & & & $ 0.80 \pm 0.21 \pm
0.09 \pm 0.12 $ \\
 & $D^+ \rightarrow K^- \pi ^+ \pi ^+$ &$  27 \pm  9 $ & 0.22 &$ 0.40 \pm 0.19
$ & \\
 & $D^+ \rightarrow \bar{K}^0 \pi ^+$ &$  11 \pm  4 $ & 0.06 &$ 2.70 \pm 1.00 $
& \\
$\bar{B}^0 \rightarrow D^{*+} \pi ^-$ & & & & & $ 0.44 \pm 0.11 \pm 0.05 \pm
0.02 $ \\
 & $D^0 \rightarrow K^- \pi ^+$ &$   8 \pm  3 $ & 0.34 &$ 0.36 \pm 0.14 $ & \\
 & $D^0 \rightarrow K^- \pi ^+ \pi ^+\pi ^-$ &$   9 \pm  3 $ & 0.19 &$ 0.38 \pm
0.12 $ & \\
$\bar{B}^0 \rightarrow D^{*+} \rho ^-$ & & & & & $ 2.11 \pm 0.89 \pm 1.23 \pm
0.08 $ \\
 & $D^0 \rightarrow K^- \pi ^+$ &$   2 \pm  1 $ & 0.02 &$ 1.35 \pm 0.90 $ & \\
 & $D^0 \rightarrow K^- \pi ^+ \pi ^+\pi ^-$ &$   4 \pm  2 $ & 0.02 &$ 1.90 \pm
0.95 $ & \\
$\bar{B}^0 \rightarrow D^{*+} \pi ^- \pi ^- \pi ^+$ & & & & & $ 1.76 \pm 0.31
\pm 0.29 \pm 0.07 $ \\
 & $D^0 \rightarrow K^- \pi ^+$ &$  18 \pm  4 $ & 0.15 &$ 0.17 \pm 0.05 $ & \\
 & $D^0 \rightarrow K^- \pi ^+ \pi ^+\pi ^-$ &$  18 \pm  5 $ & 0.08 &$ 1.81 \pm
0.57 $ & \\
$\bar{B}^0 \rightarrow D^+ D_s^-$ & & & & & $ 0.58 \pm 0.33 \pm 0.24 \pm 0.09 $
\\
 &   &$ 3.0 \pm 1.7 $ & 0.10 &$ 0.65 \pm 0.31 $ & \\
$\bar{B}^0 \rightarrow D^{*+} D_s^-$ & & & & & $ 1.17 \pm 0.66 \pm 0.52 \pm
0.05 $ \\
 &   &$ 3.0 \pm 1.7 $ & 0.05 &$ 1.17 \pm 0.57 $ & \\
$\bar{B}^0 \rightarrow \psi K^0$ & & & & & $ 0.07 \pm 0.04 \pm 0.02 $ \\
 & $\psi  \rightarrow \mu ^+ \mu ^- , e^+ e^-$ &$   3 \pm  2 $ & 0.15 &$ 0.07
\pm 0.04 $ & \\
$\bar{B}^0 \rightarrow \psi ' K^0$ & & & & & $<0.16 $ \\
 & $\psi ' \rightarrow \mu ^+ \mu ^- , e^+ e^-$ &$ <2.3 $ & 0.18 &$ < 0.30 $ &
\\
 & $\psi ' \rightarrow \psi \pi ^+ \pi ^-$ &$ <2.3 $ & 0.07 &$ < 0.35 $ & \\
$\bar{B}^0 \rightarrow \psi \bar{K}^{*0}$ & & & & & $ 0.13 \pm 0.06 \pm 0.03 $
\\
 & $\psi  \rightarrow \mu ^+ \mu ^- , e^+ e^-$ &$   7 \pm  3 $ & 0.21 &$ 0.13
\pm 0.06 $ & \\
$\bar{B}^0 \rightarrow \psi ' \bar{K}^{*0}$ & & & & & $ 0.15 \pm 0.09 \pm 0.03
$ \\
 & $\psi ' \rightarrow \mu ^+ \mu ^- , e^+ e^-$ &$   2 \pm  1 $ & 0.25 &$ 0.19
\pm 0.13 $ & \\
 & $\psi ' \rightarrow \psi \pi ^+ \pi ^-$ &$   1 \pm  1 $ & 0.10 &$ 0.11 \pm
0.11 $ & \\
$\bar{B}^0 \rightarrow \psi K^{-} \pi ^+$ & & & & & $ 0.12 \pm 0.05 \pm 0.03 $
\\
 & $\psi  \rightarrow \mu ^+ \mu ^- , e^+ e^-$ &$   7 \pm  3 $ & 0.19 &$ 0.12
\pm 0.05 $ & \\
\end{tabular}
\end{table}
\begin{table}[htb]
\caption{Detailed ${B}^-$ branching ratios. Experiment: CLEO II}
\begin{tabular}{llllll}
\label{cleoii_bm}
$B^- $ decay & Signature & Signal & Eff. & BR [\%] & Branching ratio [\%]\\
\hline
$B^- \rightarrow D^0 \pi ^-$ & & & & & $ 0.55 \pm 0.04 \pm 0.05 \pm 0.02 $ \\
 & $D^0 \rightarrow K^- \pi ^+$ &  $ 76.3 \pm 9.1 $ &  0.43 & $ 0.48 \pm 0.06
$& \\
  & $D^0 \rightarrow K^- \pi ^+ \pi ^0$ &  $ 134 \pm 15 $ &  0.19 & $ 0.62 \pm
0.07 $& \\
  & $D^0 \rightarrow K^- \pi ^+ \pi ^+\pi ^-$ &  $  94 \pm 11 $ &  0.22 & $
0.57 \pm 0.07 $& \\
 $B^- \rightarrow D^0 \rho ^-$ & & & & & $ 1.35 \pm 0.12 \pm 0.14 \pm 0.04 $ \\
 & $D^0 \rightarrow K^- \pi ^+$ &  $  80 \pm  9 $ &  0.16 & $ 1.40 \pm 0.18 $&
\\
  & $D^0 \rightarrow K^- \pi ^+ \pi ^0$ &  $  42 \pm  9 $ &  0.04 & $ 1.04 \pm
0.23 $& \\
  & $D^0 \rightarrow K^- \pi ^+ \pi ^+\pi ^-$ &  $ 90.4 \pm 12.1 $ &  0.08 & $
1.53 \pm 0.20 $& \\
 $B^- \rightarrow D^{*0} \pi ^-$ & & & & & $ 0.52 \pm 0.07 \pm 0.07 \pm 0.03 $
\\
 & $D^0 \rightarrow K^- \pi ^+$ &  $ 13.3 \pm 3.8 $ &  0.16 & $ 0.36 \pm 0.13
$& \\
  & $D^0 \rightarrow K^- \pi ^+ \pi ^0$ &  $ 37.7 \pm 6.9 $ &  0.08 & $ 0.63
\pm 0.12 $& \\
  & $D^0 \rightarrow K^- \pi ^+ \pi ^+\pi ^-$ &  $ 20.0 \pm 4.9 $ &  0.08 & $
0.52 \pm 0.13 $& \\
 $B^- \rightarrow D^{*0} \rho ^-$ & & & & & $ 1.68 \pm 0.21 \pm 0.27 \pm 0.07 $
\\
 & $D^0 \rightarrow K^- \pi ^+$ &  $ 25.7 \pm 5.4 $ &  0.06 & $ 1.74 \pm 0.37
$& \\
  & $D^0 \rightarrow K^- \pi ^+ \pi ^0$ &  $ 43.8 \pm 7.8 $ &  0.03 & $ 2.24
\pm 0.40 $& \\
  & $D^0 \rightarrow K^- \pi ^+ \pi ^+\pi ^-$ &  $ 16.9 \pm 4.6 $ &  0.03 & $
1.19 \pm 0.35 $& \\
 $B^- \rightarrow D^{*0} \pi ^- \pi ^- \pi ^+$ & & & & & $ 0.94 \pm 0.20 \pm
0.17 \pm 0.02 $ \\
 & $D^0 \rightarrow K^- \pi ^+$ &  $ 5.5 \pm 2.9 $ &  0.05 & $ 0.51 \pm 0.26 $&
\\
  & $D^0 \rightarrow K^- \pi ^+ \pi ^0$ &  $ 27.7 \pm 7.2 $ &  0.02 & $ 1.74
\pm 0.45 $& \\
  & $D^0 \rightarrow K^- \pi ^+ \pi ^+\pi ^-$ &  $  15 \pm  4 $ &  0.03 & $
1.26 \pm 0.37 $& \\
 $B^- \rightarrow D^{*0} a_1 ^-$ & & & & & $ 1.88 \pm 0.40 \pm 0.34 \pm 0.04 $
\\
 & $D^0 \rightarrow K^- \pi ^+$ &  $ 5.5 \pm 2.9 $ &  0.05 & $ 1.02 \pm 0.52 $&
\\
  & $D^0 \rightarrow K^- \pi ^+ \pi ^0$ &  $ 27.7 \pm 7.2 $ &  0.02 & $ 3.42
\pm 0.90 $& \\
  & $D^0 \rightarrow K^- \pi ^+ \pi ^+\pi ^-$ &  $ 15.0 \pm 4.5 $ &  0.03 & $
2.52 \pm 0.74 $& \\
 $B^- \rightarrow D^+ \pi^- \pi ^- $ & & & & & $<0.14 $ \\
 & $D^+ \rightarrow K^- \pi ^+ \pi ^+$ & $ < 10.3$ &  0.11 & $ <0.14 $& \\
 $B^- \rightarrow D^{*+} \pi ^- \pi ^-$ & &$14.1 \pm 5.4$ & & & $ 0.19 \pm 0.07
\pm 0.03 \pm 0.01 $ \\
$B^- \rightarrow D^{**0}(2420) \pi^- $ & & & & & $ 0.11 \pm 0.05 \pm 0.02 \pm
0.01 $ \\
 & $D^{**0} \rightarrow D^{*+} \pi^- $ &  $ 8.5 \pm 3.8 $ &  & $ 0.11 \pm 0.05
$& \\
 $B^- \rightarrow D^{**0}(2420) \rho^- $ & & & & & $<0.14 $ \\
 & $D^{**0} \rightarrow D^{*+} \pi^- $ &  $ 3.4 \pm 2.1 $ &  & $ <0.14 $& \\
 $B^- \rightarrow D^{**0}(2460) \pi^- $ & & & & & $<0.13 $ \\
 & $D^{**0} \rightarrow D^{*+} \pi^- $ &  $ 3.5 \pm 2.3 $ &  & $ <0.28 $& \\
  & $D^{**0} \rightarrow D^+ \pi^- $ & $ < 5.6$ &  0.21 & $ <0.13 $& \\
 $B^- \rightarrow D^{**0}(2460) \rho^- $ & & & & & $<0.47 $ \\
 & $D^{**0} \rightarrow D^{*+} \pi^- $ &  $ 3.2 \pm 2.4 $ &  & $ <0.50 $& \\
  & $D^{**0} \rightarrow D^+ \pi^- $ & $ < 6.1$ &  0.08 & $ <0.47 $& \\
\end{tabular}
\end{table}
\begin{table}[htb]
\caption{Detailed ${B}^-$ branching ratios. Experiment: CLEO II}
\begin{tabular}{llllll}
\label{cleoii_bm2}
$B^- $ decay & Signature & Signal & Eff. & BR [\%] & Branching ratio [\%]\\
\hline
 $B^- \rightarrow \psi K^-$ & & & & & $ 0.110 \pm 0.015 \pm 0.009 $ \\
 & $\psi  \rightarrow \mu ^+ \mu ^- , e^+ e^-$ &  $ 58.7 \pm 7.9 $ &  0.47 & $
0.11 \pm 0.01 $& \\
 $B^- \rightarrow \psi ' K^-$ & &$7.0 \pm 2.6$ & & & $ 0.061 \pm 0.023 \pm
0.009 $ \\
$B^- \rightarrow \psi K^{*-}$ & & & & & $ 0.178 \pm 0.051 \pm 0.023 $ \\
 & $K^{*-} \rightarrow K^- \pi ^0$ &  $ 6.0 \pm 2.4 $ &  0.07 & $ 0.22 \pm 0.09
$& \\
  & $K^{*-} \rightarrow K_s \pi ^-$ &  $ 6.6 \pm 2.7 $ &  0.17 & $ 0.13 \pm
0.06 $& \\
 $B^- \rightarrow \psi ' K^{*-}$ & & & & & $<0.30 $ \\
 & $K^{*-} \rightarrow K^- \pi ^0$ &  $   1 \pm  1 $ &  & $ <0.56 $& \\
  & $K^{*-} \rightarrow K_s \pi ^-$ &  $   1 \pm  1 $ &  & $ <0.36 $& \\
 $B^- \rightarrow \chi_{c1} K^-$ & & & & & $ 0.097 \pm 0.040 \pm 0.009 $ \\
 & $\chi_{c1} \rightarrow \gamma \psi $ &  $ 6.0 \pm 2.4 $ &  0.20 & $ 0.10 \pm
0.04 $& \\
 $B^- \rightarrow \chi_{c1} K^{*-}$ & & & & & $<0.21 $ \\
 & $K^{*-} \rightarrow K^- \pi ^0$ &  0 &  0.03 & $ <0.67 $& \\
  & $K^{*-} \rightarrow K_s \pi ^-$ &  0 &  0.11 & $ <0.30 $& \\
 \end{tabular}
\end{table}
\setlength{\topmargin}{-0.5in}
\begin{table}[htb]
\caption{Detailed $\bar{B}^0$ branching ratios. Experiment: CLEO II}
\begin{tabular}{llllll}
\label{cleoii_b0}
$\bar{B}^0 $ decay & Signature & Signal & Eff. & BR [\%] & Branching ratio
[\%]\\
\hline
$\bar{B}^0 \rightarrow D^+ \pi ^-$ & & & & & $ 0.29 \pm 0.04 \pm 0.03 \pm 0.05
$ \\
 & $D^+ \rightarrow K^- \pi ^+ \pi ^+$ &  $ 80.6 \pm 9.8 $ &  0.32 & $ 0.29 \pm
0.04 $& \\
 $\bar{B}^0 \rightarrow D^+ \rho ^-$ & & & & & $ 0.81 \pm 0.11 \pm 0.12 \pm
0.13 $ \\
 & $D^+ \rightarrow K^- \pi ^+ \pi ^+$ &  $ 78.9 \pm 10.7 $ &  0.12 & $ 0.81
\pm 0.11 $& \\
 $\bar{B}^0 \rightarrow D^{*+} \pi ^-$ & & & & & $ 0.26 \pm 0.03 \pm 0.04 \pm
0.01 $ \\
 & $D^0 \rightarrow K^- \pi ^+$ &  $ 19.4 \pm 4.5 $ &  0.35 & $ 0.22 \pm 0.05
$& \\
  & $D^0 \rightarrow K^- \pi ^+ \pi ^0$ &  $ 31.9 \pm 6.4 $ &  0.14 & $ 0.30
\pm 0.06 $& \\
  & $D^0 \rightarrow K^- \pi ^+ \pi ^+\pi ^-$ &  $ 20.5 \pm 5.2 $ &  0.15 & $
0.27 \pm 0.07 $& \\
 $\bar{B}^0 \rightarrow D^{*+} \rho ^-$ & & & & & $ 0.74 \pm 0.10 \pm 0.14 \pm
0.02 $ \\
 & $D^0 \rightarrow K^- \pi ^+$ &  $ 21.9 \pm 5.2 $ &  0.12 & $ 0.71 \pm 0.17
$& \\
  & $D^0 \rightarrow K^- \pi ^+ \pi ^0$ &  $ 39.8 \pm 7.2 $ &  0.05 & $ 1.08
\pm 0.20 $& \\
  & $D^0 \rightarrow K^- \pi ^+ \pi ^+\pi ^-$ &  $ 14.6 \pm 4.6 $ &  0.05 & $
0.52 \pm 0.17 $& \\
 $\bar{B}^0 \rightarrow D^{*+} \pi ^- \pi ^- \pi ^+$ & & & & & $ 0.63 \pm 0.10
\pm 0.11 \pm 0.02 $ \\
 & $D^0 \rightarrow K^- \pi ^+$ &  $ 13.5 \pm 3.9 $ &  0.10 & $ 0.58 \pm 0.17
$& \\
  & $D^0 \rightarrow K^- \pi ^+ \pi ^0$ &  $ 21.7 \pm 5.9 $ &  0.04 & $ 0.67
\pm 0.18 $& \\
  & $D^0 \rightarrow K^- \pi ^+ \pi ^+\pi ^-$ &  $ 13.9 \pm 4.4 $ &  0.04 & $
0.65 \pm 0.19 $& \\
 $\bar{B}^0 \rightarrow D^{*+} a_1^-$ & & & & & $ 1.26 \pm 0.20 \pm 0.22 \pm
0.03 $ \\
 & $D^0 \rightarrow K^- \pi ^+$ &  $ 13.5 \pm 3.9 $ &  0.10 & $ 1.16 \pm 0.34
$& \\
  & $D^0 \rightarrow K^- \pi ^+ \pi ^0$ &  $ 21.7 \pm 5.9 $ &  0.04 & $ 1.34
\pm 0.36 $& \\
  & $D^0 \rightarrow K^- \pi ^+ \pi ^+\pi ^-$ &  $ 13.9 \pm 2.4 $ &  0.04 & $
1.30 \pm 0.38 $& \\
 $\bar{B}^0 \rightarrow D^{0} \pi ^+ \pi^- $ & & & & & $<0.16 $ \\
 & $D^0 \rightarrow K^- \pi ^+$ & $ < 10.1$ &  0.19 & $ <0.16 $& \\
 $\bar{B}^0 \rightarrow D^{**+}(2460) \pi^- $ & & & & & $<0.22 $ \\
 & $D^{**+} \rightarrow D^0 \pi^+ $ & $ < 5.6$ &  0.26 & $ <0.22 $& \\
 $\bar{B}^0 \rightarrow D^{**+}(2460) \rho^- $ & & & & & $<0.49 $ \\
 & $D^{**+} \rightarrow D^0 \pi^+ $ & $ < 5.1$ &  0.11 & $ <0.49 $& \\
 $\bar{B}^0 \rightarrow \psi K^0$ & & & & & $ 0.075 \pm 0.024 \pm 0.008 $ \\
 & $\psi  \rightarrow \mu ^+ \mu ^- , e^+ e^-$ &  $ 10.0 \pm 3.2 $ &  0.34 & $
0.08 \pm 0.02 $& \\
 $\bar{B}^0 \rightarrow \psi ' K^0$ & & 0 & & & $<0.08 $ \\
$\bar{B}^0 \rightarrow \psi \bar{K}^{*0}$ & & & & & $ 0.169 \pm 0.031 \pm 0.018
$ \\
 & $\psi  \rightarrow \mu ^+ \mu ^- , e^+ e^-$ &  $ 29.0 \pm 5.4 $ &  0.23 & $
0.17 \pm 0.03 $& \\
 $\bar{B}^0 \rightarrow \psi ' \bar{K}^{*0}$ & &$ 4.2 \pm 2.3$ & & & $<0.19 $
\\
$\bar{B}^0 \rightarrow \chi_{c1} K^0$ & & & & & $<0.27 $ \\
 & $\chi_{c1} \rightarrow \gamma \psi $ &  $   1 \pm  1 $ &  0.14 & $ <0.27 $&
\\
 $\bar{B}^0 \rightarrow \chi_{c1} \bar{K}^{*0}$ & & & & & $<0.21 $ \\
 & $\chi_{c1} \rightarrow \gamma \psi $ &  $ 1.2 \pm 1.5 $ &  0.13 & $ <0.21 $&
\\
 \end{tabular}
\end{table}


\begin{references}
\bibitem{clns}{Copies of the figures in this review are available  on request
from the authors.}
\bibitem{only4s}{The integrated luminosities quoted here are for the $\Upsilon
(4S)$ data
samples only.}
\bibitem{Artuso}{The definitions of thrust, sphericity and $s_{\perp}$
can be found in the contribution of M. Artuso in this volume.}
\bibitem{mbrange}{The resolution in beam constrained mass
has a weak dependence on the decay channel.
It varies between 2.6 MeV to 3.3 MeV.}
\bibitem{cfrac}{These are the fractions determined in the
CLEO~II analysis.
The lowest fraction is for the
mode $\bar{B^0}\to D^{*+}\rho^-$ while the highest
fraction is for the mode $B^- \to D^0 \pi^-$.}
\bibitem{ARGUSback}{In the ARGUS experiment the
beam constrained mass distributions were fitted
to a slightly different background function with a kinematic threshold.
The functional form of the background shape used was
$$f(x)= c_1x\left(\sqrt{(1-({x\over E_{beam}})^2}
\right)e^{c_2(1-({x\over E_{beam}})^2)}$$
CLEO~1.5 fitted the beam constrained mass distributions to the sum
of a gaussian signal of width 2.5 MeV (r.m.s) and a background function
of the form $w_1 ~ p + w_2~ p^2$ where $p= \sqrt{E_{beam}^2 - m^2}$
and the parameters $w_1$, $w_2$ are
determined from the $\Delta E$ sidebands.}
\bibitem{mcdd}{Two models are considered: non-resonant
 $B\to D \pi^- \pi^0$ and $ B\to D^{**}(2460) \pi^-$. Both  give very similar
$\pi^- \pi^0$ mass spectra and comparable limits on the non-rho contamination
in the signal region.}
\bibitem{mcdrho}{CLEO~II considers two models: (1) The $\pi^- \pi^0$ system
is produced non-resonantly following a phase space distribution or
(2) the $D^{*+} \pi^- \pi^0$ is produced from the decay
$D^{**} \pi^-$. Both give similar limits on the non-$\rho$ fraction.}
\bibitem{Dpi}{The requirement that the colors match gives a factor of
$1/3$ in the amplitude. Moreover, the $W$ couples to the $d \bar{d}$ part of
the $\pi^0$ wave function so there is an additional factor of $1/\sqrt{2}$ in
the amplitude. This is also  true for the $\rho^0$ and the $\omega$ mesons. The
numerical factor is different for the case of the
$\eta/ \eta^{'}$ system.}
\bibitem{wex}{These modes are also accessible via the W-exchange graph which is
expected to be small in B decay.}
\bibitem{BReta}{We use the PDG values for the $\eta$, $\eta^{'}$
and $\omega$ branching ratios.}
\bibitem{omeg}{In the mode $\bar{B^0}\to D^0 \omega$, $\omega \to
\pi^- \pi^+ \pi^0$, $\Theta_{N}$, the angle between the normal
to the $\omega$ decay plane and the $D^0$ direction calculated in
the $D^0$ rest frame, must satisfy $|\sin \Theta_{N}|>0.6$.}
\bibitem{vud}{We make the approximation that $V_{ud},\; V_{cs} \approx 0.975$}
\bibitem{qcd}{The error is due to
the uncertainty in the scale at which to evaluate the Wilson
coefficient.}
\bibitem{narrow}{A value f$_\rho=196$ MeV, often quoted in the literature, is
derived from tau decays
using the narrow width approximation i.e. by assuming the $\rho$
resonance is a delta function.}
\bibitem{width}{Since the form factor for $B\to D^* \ell \nu$ is slowly
varying, the width of the $\rho^-$ meson does not significantly modify the
result.}
\bibitem{Dstbr}{The $q^2$ spectrum has been
corrected for the new values of the
CLEO~II $D^*$ and $D$  branching ratios.\cite{AbsoD},\cite{DKpi} }
\bibitem{param}{Three parameterizations of the Isgur-Wise function
$ \xi(v v^{'})$ are considered:  (a) $ \xi ( v v^{'}) = 1 + 1/4 a (v -
v^{'})^2 (v + v^{'})^2$ with
$ a = 0.54 \pm 0.01$, (b) $ \xi ( v v^{'}) = 1 /[1- (v - v^{'})^2/w_{0}^2]$
with
$ w_0 = 1.12 \pm 0.17$, due to Rosner\cite{Rosfac} ,
 and   (c) $ \xi (v v^{'}) = exp [ b( v-v^{'})^2]$ with
$b= 0.91 \pm 0.03$.  All of these parameterizations are constrained by fitting
to the average of the CLEO~1.5 and ARGUS data on
$ d \Gamma/ dq^2 (B\to D^* \ell\nu)$.}
\bibitem{psicomment}{The branching ratio for the modes $B\to \psi K$ and
$B\to \psi K^{*}$ can be accommodated by the value $\xi \sim 0$ while
$\xi \sim 1/3$ gives a branching ratio that is about a factor of 4 too low (see
Ref.~\cite{Shifman}.}
\bibitem{volkie}{Private communication, Volker Rieckert; The result of a fit
including $R_4$ is not significantly different. We find in this case $a_2/a_1
\; = \; 0.282 \pm 0.07 \pm 0.06$.}
\bibitem{bkk}{The decay $\bar{B}^0 \to K^+K^-$ cannot be produced by a $b \to
u$
spectator diagram or a $b \to s$ penguin transition. Thus it is expected to be
small
compared to $\bar{B}^0 \to \pi^+\pi^-$ and $\bar{B}^0 \to K^-\pi^+$.}
\end{references}
\end{document}